# Numerical investigations on interactions between 2D/3D conical shock wave and axisymmetric boundary layer at $Ma = 2.2$

Yihui Weng[1], Qin Li[1a], Guozhuo Tan[1], Wei Su[2], Yancheng You[1]

1. School of Aerospace Engineering, Xiamen University, China, 361102
2. Beijing Institute of Space Long March Vehicle, Beijing, China, 100076

**Abstract** Numerical simulation and analysis are carried out on interactions between a 2D/3D conical shock wave and an axisymmetric boundary layer with reference to the experiment by Kussoy et al. [AIAA J, 1980, 18(12): 1477–1487], in which the shock was generated by a 15° half-angle cone in a tube at 15° angle of attack (AOA). Based on the Reynolds-averaged Navier–Stokes equations and Menter's SST turbulence model, the present study uses the newly developed WENO3-PRM$^2_{1,1}$ scheme and the PHengLEI CFD platform for the computations. First, computations are performed for the 3D interaction corresponding to the conditions of the experiment by Kussoy et al., and these are then extended to cases with AOA = 10° and 5°. For comparison, 2D axisymmetric counterparts of the 3D interactions are investigated for cones coaxial with the tube and having half-cone angles of 27.35°, 24.81°, and 20.96°. The shock wave structure, vortex structure, variable distributions, and wall separation topology of the interaction are computed. The results show that in 2D/3D interactions, a new Mach reflection-like event occurs and a Mach stem-like structure is generated above the front of the separation bubble, which differs from the model of Babinsky and Harvey for 2D planar shock wave/boundary layer interaction. A new interaction model is established to describe this behavior. The relationship between the length of the circumferentially unseparated region in the tube and the AOA of the cone indicates the existence of a critical AOA at which the length is zero, and a prediction of this angle is obtained using an empirical fit, which is verified by computation. The occurrence of side overflow in the windward meridional plane is analyzed, and a quantitative knowledge is obtained. To further elucidate the characteristics of the 3D interaction, the scale and structure of the vortex and the pressure and friction force distributions are presented and compared with those of the 2D interaction.

[a] Corresponding author, email: qin-li@vip.tom.com

# 1 Introduction

Shock wave/boundary layer interaction (SWBLI), a commonly observed phenomenon in supersonic and hypersonic aircraft, often occurs at fuselage–wing junctions, engine inlets, compressors of gas turbine engines, etc. SWBLI usually leads to an increase in boundary layer thickness and often to flow separation, changing the surface pressure and increasing the heat flux near the flow reattachment position. Hence, SWBLI is of great significance with regard to the aerodynamic performance, structural safety, and power efficiency of aircrafts.

SWBLI was first noticed by Ferri [2] in 1939, and subsequent studies have concerned a variety of topics, such as interaction intensity, patterns, and parametric influence. Among the different types of SWBLI, three-dimensional (3D) SWBLI is of the greatest practical importance, as well as being the most complex. The interaction between a conical shock and a boundary layer is of particular interest because of its common occurrence in engineering applications, for example, at the side of the inlet of an engine with a center body. When a cone is positioned in a flow at an angle of attack (AOA), the shock wave generated will have a conical 3D profile, and its interaction with an axisymmetric boundary layer will be quite complicated. However, most studies reported in the literature have been concerned with the interaction between shock waves and flat-plate boundary layers, with few focusing on interactions between conical shock waves and axisymmetric boundary layers. It should be pointed out here that the three-dimensionality of the flows referred to in this study is in the sense of the time average, not the instantaneous 3D coherence that occurs in turbulent flows.

There are two main types of 3D interactions caused by conical shock waves, namely, interaction with a *flat plate* boundary layer and that with an *axisymmetric* boundary layer. Regarding the former, Panov [3] experimentally studied such an interaction for an inflow Mach number of 2.87, and he presented an interaction sketch and derived a formula for the critical pressure of the boundary layer. Gai and Teh [4] studied a case with a Mach number of 2. By varying the cone angle and the height of the cone above the plate, they observed a boundary layer transition from nonseparation to separation and found that the strength of the incident shock wave decreased along the spanwise direction of the plate. Hale [5] investigated a case with a Mach number of 2.05 and obtained information about flow structures and parametric distributions through schlieren imaging, surface oil flow, and particle image velocimetry. The results indicated that the interaction structure was similar to that of a blunt fin. Zuo et al. [6] carried out direct numerical simulations related to Hale's experimental work. Subsequently [7], using the Reynolds-averaged Navier–Stokes (RANS) equations, they studied the effect of Reynolds number on the distributions of pressure, Mach number, skin friction, etc., as well as investigating predictability by various turbulence models. Zuo [8] investigated the size of the separation bubble and its relationship with the pressure rise during the interaction and derived a relationship between the rise in wall pressure and the geometric features of the separation bubble. Yang et al. [9] performed a numerical study of SWBLI interaction at a Mach number of 2 and presented a summary of the relationships between variations in flow separation and the characteristic parameters of the interaction. They also considered two-dimensional (2D) counterparts in which there were similar incident shock waves on the windward meridional plane as in the 3D cases and performed computations to provide comparisons with the 3D results.

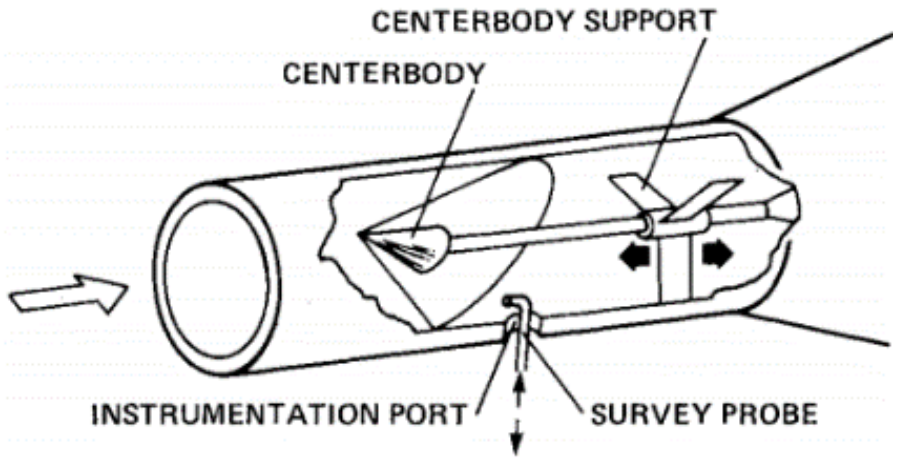

Fig. 1 Diagram of experiment by Kussoy et al. [10] with a cone at AOA = 15 °

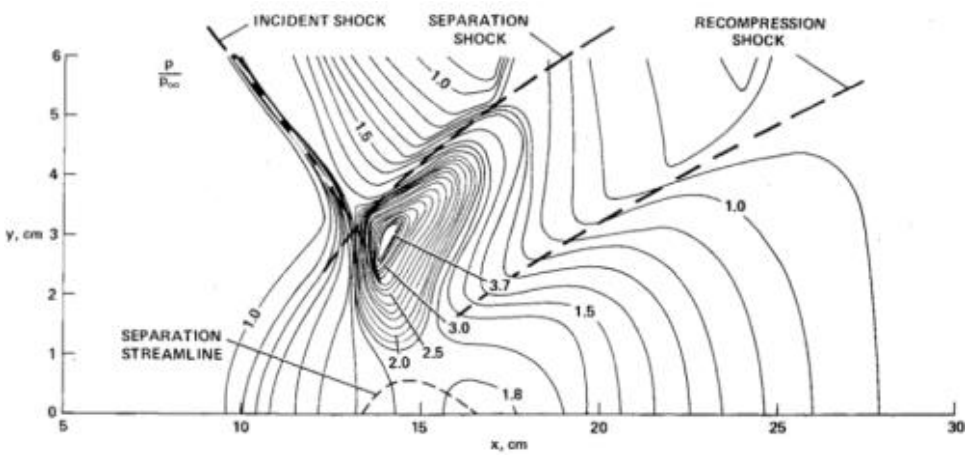

Fig. 2 Flow structures in the windward meridional plane from Kussoy et al. [10]

By comparison with the above studies of the interactions of conical shock waves with flat plate boundary layers, there have been relatively few studies of interactions with axisymmetric boundary layers. The representative work in this regard is the experimental and computational study by Kussoy et al. [10], in which a cone with 15° half-cone angle was mounted at AOA = 15° in a tube and the Mach number was 2.2. The experimental diagram is shown in Fig. 1, and the flow field structure on the windward meridional plane obtained in [10] is shown in Fig. 2. The results show the presence of typical structures such as a separation shock wave, a reattachment shock wave, and a separation bubble. In addition, Kussoy et al. [10] measured the pressure distribution on the tube surface at different azimuthal angles [see Fig. 3(a)], and they found that as the azimuthal angle increased, the conical shock wave became weaker, the pressure rise point moved backward, and the peak pressure decreased. Using an oil flow diagram [Fig. 3(b)], Kussoy et al. [10] observed that as the azimuthal angle increased, the separation streamline and the reattachment streamline gradually approached one another, finally merging into one stream trace. It is worth pointing out that although Zuo and Memmolo [11] also performed simulations of the interaction between an internal conical shock wave and an axisymmetric boundary layer in a contraction tube with Mach number of 4.0, the shock structure therein and the corresponding interaction were obviously different from those found by Kussoy et al. [10].

Although Kussoy et al. [10] carefully examined the interaction structures and variable distributions, their study suffered from a number of deficiencies:

(a) Their experiment only considered the condition AOA = 15°, and did not investigate any variations in flow structure and characteristics with AOA.

(b) They focused attention mainly on the flow structures on the windward meridional plane of the tube, paying little attention to the variations in the structures along the circumference of the tube or to analyzing the associated flow mechanisms.

(c) Their experimental study concerned only the interaction of the 3D conical shock wave with the axisymmetric tube boundary layer, with no investigation of the 2D counterpart or any comparison between 2D and 3D interactions.

(d) Except for the wall pressure distribution, which was reliably predicted, there is a lack of confidence regarding some of their other results, such as those for shock reflections and friction force distributions, owing to insufficient resolution and accuracy. Whether the structures that they obtained (as shown in Fig. 2) are correct needs further confirmation.

Therefore, it is worthwhile revisiting the subjects investigated by Kussoy et al. [10] and performing further explorations.

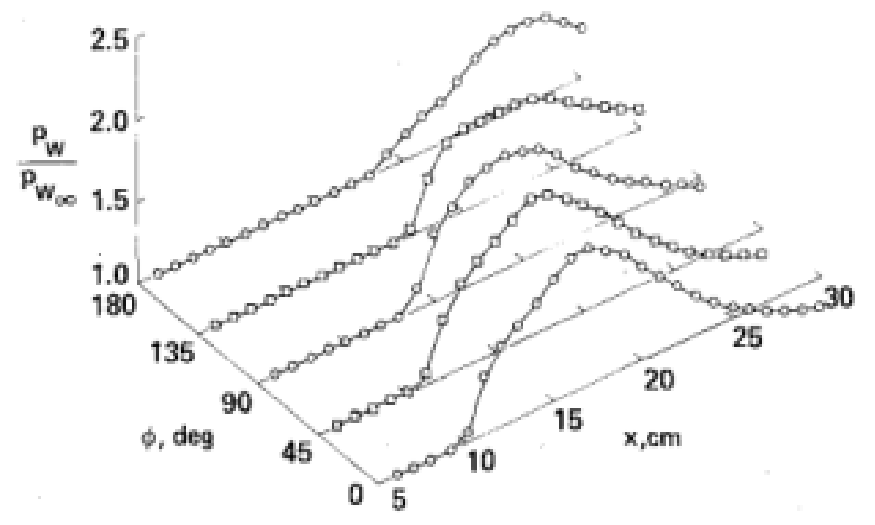

(a) Wall pressure distributions at various azimuthal angles

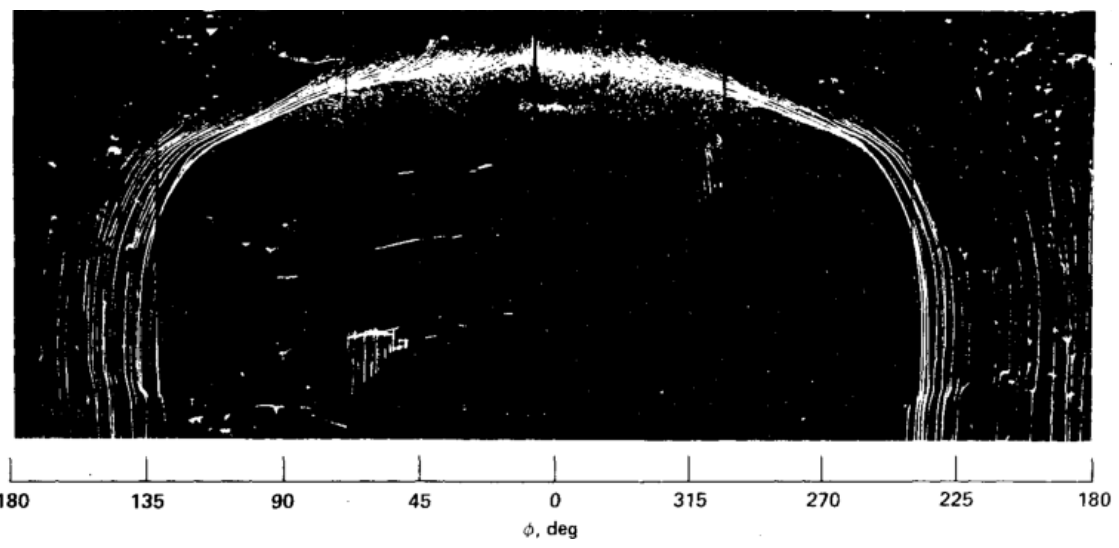

(b) Oil flow on the wall

Fig. 3 Wall pressure distributions and oil flow of the tube in the experiment by Kussoy et al. [10]

In this paper, we numerically investigate the interaction considered by Kussoy et al. [10] as well as related cases using the RANS equations and PHengLEI software [12] [13]. To achieve high resolution and accuracy when solving the interaction structures, a novel WENO-type scheme recently developed by Li et al. [1] and integrated into the software is employed. The remainder of the paper is organized as follows. The numerical scheme, turbulence model, and validations are presented in Section 2. Simulations of interactions between a conical shock wave and an axisymmetric boundary layer are performed and analyzed in Section 3. In particular, for 3D interactions, AOAs of 10° and 5° are investigated in addition to the value of 15° considered by Kussoy et al. [10]. Also, 2D axisymmetric counterparts of the 3D interactions are studied to provide comparisons with the latter for half-cone angles of 27.35°, 24.81°, and 20.96°. Finally, conclusions are drawn in Section 4.

# 2 Numerical methods

## 2.1 Governing equations

The governing equations for the computations are the RANS equations in finite volume form, namely,

$$\frac{\partial}{\partial t}\int_{\Omega} \boldsymbol{Q}\, dV + \int_{\partial\Omega}(\boldsymbol{F}\cdot\boldsymbol{n})\, dS = \int_{\partial\Omega}(\boldsymbol{F_v}\cdot\boldsymbol{n})\, dS \tag{1}$$

where $Q$ is the conserved vector, $n$ is the outward normal vector of the surface of $\Omega$, and $F$ and $F_v$ are the inviscid and viscous fluxes respectively. The expressions for $\boldsymbol{Q}$, $\boldsymbol{F}$, and $\boldsymbol{F_v}$ are

$$\boldsymbol{Q} = \begin{bmatrix} \rho \\ \rho u \\ \rho v \\ \rho w \\ \rho e \end{bmatrix}, \quad \boldsymbol{F} = \begin{bmatrix} \rho V \\ \rho uV + n_x p \\ \rho vV + n_y p \\ \rho wV + n_z p \\ \rho HV \end{bmatrix}, \quad \boldsymbol{F_v} = \begin{bmatrix} 0 \\ n_x\tau_{xx} + n_y\tau_{xy} + n_z\tau_{xz} \\ n_x\tau_{yx} + n_y\tau_{yy} + n_z\tau_{yz} \\ n_x\tau_{zx} + n_y\tau_{zy} + n_z\tau_{zz} \\ n_x\Theta_x + n_y\Theta_y + n_z\Theta_z \end{bmatrix}$$

where the variables are defined as follows:

$$\boldsymbol{V} = \vec{v}\cdot\vec{n} = n_x u + n_y v + n_z w$$

$$H = e + \frac{u^2 + v^2 + w^2}{2}$$

$$\begin{bmatrix} \Theta_x \\ \Theta_y \\ \Theta_z \end{bmatrix} = \begin{bmatrix} \tau_{xx} & \tau_{xy} & \tau_{xz} \\ \tau_{yx} & \tau_{yy} & \tau_{yz} \\ \tau_{zx} & \tau_{zy} & \tau_{zz} \end{bmatrix} \begin{bmatrix} u \\ v \\ w \end{bmatrix} + k \begin{bmatrix} \partial T/\partial x \\ \partial T/\partial y \\ \partial T/\partial z \end{bmatrix}$$

$$[\tau_{x_i x_j}] = \begin{cases} \mu\left(\dfrac{\partial u_i}{\partial x_j} + \dfrac{\partial u_j}{\partial x_i}\right) & (i \neq j) \\ 2\mu\dfrac{\partial u_i}{\partial x_i} - \dfrac{2}{3}\mu\nabla\cdot\vec{v} & (i = j) \end{cases}$$

$$\mu = \mu_l + \mu_t$$

In these equations, $e$ is the internal energy per unit mass, $k$ is the heat transfer coefficient, $\mu_l$ is the laminar viscosity coefficient, and $\mu_t$ is the turbulent viscosity coefficient. $\mu_t$ is obtained by solving the two-equation Menter's shear stress transport (SST) turbulence model [14], which will be introduced in Section 2.2. $\mu_l$ is obtained using the Sutherland formula:

$$\frac{\mu_l}{\mu_0} = \left(\frac{T}{T_0}\right)^{3/2} \frac{1+T_S/T_0}{T/T_0+T_S/T_0} \tag{2}$$

where $T_0$ = 273.16 K, $\mu_0$= 1.7161×$10^{-5}$ / (Pa · s), and $T_S$=124 K. The perfect gas model is used with equation of state $p = \rho RT$.

For time discretization, the lower–upper symmetric Gauss–Seidel (LU-SGS) scheme proposed by Yoon and Jameson [15] is used. Since the approximation of the convective term has an important influence on computational accuracy, the WENO3-PRM$^2_{1,1}$ scheme proposed by Li et al. [1] is used and will be introduced in Section 2.3.

## 2.2 Menter's SST turbulence model

In Menter's SST model, the following partial differential equations are solved to compute the turbulent viscosity coefficient:

$$\frac{\partial}{\partial t}(\rho k)+\frac{\partial}{\partial x_i}(\rho u_i k)=P-\beta^*\rho\omega k+\frac{\partial}{\partial x_i}\left[(\mu+\sigma_\kappa\mu_t)\frac{\partial k}{\partial x_i}\right] \tag{3}$$

$$\frac{\partial}{\partial t}(\rho\omega)+\frac{\partial}{\partial x_i}(\rho u_i\omega)=\frac{\gamma}{\nu_t}P-\beta\rho\omega^2+\frac{\partial}{\partial x_i}\left[(\mu+\sigma_\omega\mu_t)\frac{\partial\omega}{\partial x_i}\right]+2(1-F_1)\frac{\rho\sigma_{\omega 2}}{\omega}\frac{\partial k}{\partial x_i}\frac{\partial\omega}{\partial x_i} \tag{4}$$

where $k$ is the turbulent kinetic energy, $\omega$ is the specific turbulent dissipation rate, $\nu_t=\mu_t/\rho$, and

$$P=\tau_{ij}\frac{\partial u_i}{\partial x_j},\quad \tau_{ij}=\mu_t\left(2S_{ij}-\frac{2}{3}\frac{\partial u_k}{\partial x_k}\delta_{ij}\right)-\frac{2}{3}\rho k\delta_{ij}$$

$$S_{ij}=\frac{1}{2}\left(\frac{\partial u_i}{\partial x_j}+\frac{\partial u_j}{\partial x_i}\right),\quad \mu_t=\frac{\rho a_1 k}{\max(a_1\omega,\Omega F_2)}$$

Some parameters in the equations, denoted by $\phi$ hereinafter, are presented in the following weighted form:

$$\phi=F_1\phi_1+(1-F_1)\phi_2 \tag{5}$$

where $\phi$, $\phi_1$, and $\phi_2$ are specified in Table 1.

Table 1. Specifications of $\phi$, $\phi_1$, and $\phi_2$ in Menter's SST model

| $\phi$ | $\phi_1$ | $\phi_2$ |
|---|---|---|
| $\gamma$ | $\gamma_1=\beta_1/\beta^*-\sigma_{\omega_1}\kappa^2/\sqrt{\beta^*}$ | $\gamma_2=\beta_2/\beta^*-\sigma_{\omega_2}\kappa^2/\sqrt{\beta^*}$ |
| $\sigma_\kappa$ | $\sigma_{\kappa_1}=0.85$ | $\sigma_{\kappa_2}=1.0$ |
| $\sigma_\omega$ | $\sigma_{\omega_1}=0.5$ | $\sigma_{\omega_2}=0.856$ |
| $\beta$ | $\beta_1=0.075$ | $\beta_2=0.0828$ |

The other variables in Eqs. (3) and (4) are defined as follows:

$$F_1=\tanh(\arg_1^4),\quad \arg_1=\min\left[\max\left(\frac{\sqrt{k}}{\beta^*\omega d},\frac{500\upsilon}{d^2\omega}\right),\frac{4\rho\sigma_{\omega_2}k}{CD_{k\omega}d^2}\right]$$

$$F_2=\tanh(\arg_2^2),\quad \arg_2=\max\left(2\frac{\sqrt{k}}{\beta^*\omega d},\frac{500\nu}{d^2\omega}\right)$$

$$CD_{k\omega} = \max\left(2\rho\sigma_{\omega_2}\frac{1}{\omega}\frac{\partial\kappa}{\partial x_i}\frac{\partial\omega}{\partial x_i}, 10^{-20}\right), \qquad \beta^* = 0.09$$

## 2.3 WENO3-PRM$^2_{1,1}$ schemes

To enhance numerical precision while maintaining computational robustness, Li et al. [1] proposed third-order improvements of the WENO schemes of Jiang and Shu [16] (WENO3-JS) through the use of a new piecewise-rational polynomial mapping method (PRM). The new schemes can maintain third-order accuracy when a first-order critical point occurs, achieve numerical stability in long-period computations in the case of the scalar convection equation, and perform well in terms of accuracy, resolution, and robustness. The following is a brief introduction to these schemes.

Considering the one-dimensional hyperbolic conservation law $u_t + f(u)_x = 0$, the conservative approximation of $f(u)_x$ can be written as

$$\frac{\partial f(u)}{\partial x_j} \approx \frac{\hat{f}_{j+1/2} - \hat{f}_{j-1/2}}{\Delta x}$$

where $\hat{f}_{j+1/2}$ is some numerical flux. Supposing that *r* is the number of sub-stencils as well as the grid number of each stencil, WENO schemes of order *r* can be formulated as

$$\hat{f}_{j+1/2} = \sum_{k=0}^{r-1} \omega_k q_k^r, \quad q_k^r = \sum_{l=0}^{r-1} a_{kl}^r f(u_{j-r+k+l+1})$$

where $q_k^r$ is the candidate scheme with coefficients $a_{kl}^r$ and $\omega_k$ is the nonlinear weight. For WENO-JS, $\omega_k$ is obtained as follows:

$$\omega_k = \frac{\alpha_k}{\sum_{l=0}^{r-1}\alpha_l}, \qquad \alpha_k = \frac{d_k^r}{(\varepsilon + IS_k^{(r)})^2}$$

and

$$IS_k^{(r)} = \sum_{m=0}^{r-2} c_m^r \left[\sum_{l=0}^{r-1} b_{kml}^r f(u_{j-r+k+l+1})\right]^2$$

where $\varepsilon$ is a small quantity included to prevent the denominator from becoming zero, and the values of $a_{kl}^r$, $b_{kml}^r$, $c_m^r$, and $d_k^r$ as well as other coefficients can be found in [1]. It is known that the canonical WENO-JS scheme will suffer from order degradation at the first-order critical point. To solve this problem, Henrick et al. [17] pioneered the idea of the mapping method by introducing a mapping function for the nonlinear weight according to the following principles. For $\omega \in [0, 1]$, consider some mapping function $g_k(\omega)$ with respect to the linear weight $d_k$ satisfying $g_k(\omega) = \omega$ when $\omega = \{0, d_k, 1\}$. If there exists a constant $n$ such that $g_k^{(i)}(\omega) = 0$ and $g_k^{(n)}(\omega) \neq 0$ when $1 \leq i < n$ [i.e., if $d_k$ is an $(n-1)$th-order zero point of $g_k(x)$], then it holds in the neighborhood of $d_k$ that $g_k(\omega) = d_k + (1/n!)g_k^{(n)}(\omega - d_k)^n + \cdots$. Considering $g_k(\omega)$ as a new non-normalized weight $\alpha_k'$ about $g_k(\omega)$, then $\alpha_k' = d_k + O(\Delta x^{n\times m})$, providing $\omega - d_k = O(\Delta x^m)$, and a new nonlinear weight $\omega_k'$ will be obtained after normalization, with $\omega_k' = d_k + O(\Delta x^{n\times m})$. Hence after mapping, the original order of $(\omega - d_k)$ is improved from $n$ to $m \times n$.

After establishing the mapping function, one may employ single or piecewise functions (e.g., a piecewise-polynomial mapping, PPM [18]), with the latter often having advantages due to their good adjustability [1]. On the basis of WENO-PPM schemes [18], Li et al. [1] developed WENO-PRM schemes through the use of a piecewise-rational mapping (PRM). This new mapping overcomes the difficulty of balancing resolution and stability and has the following form:

$$\begin{cases} PRM_{n,m;m_1;c_1,c_2}^{L,n+1} = d_k + \dfrac{(\omega-d_k)^{n+1}}{(\omega-d_k)^n+(-1)^{1+n}c_2^L(\omega-d_k)\omega^{m_1}+(-1)c_1^L\omega^{m+1}} \text{ when } \omega < d_k \\ PRM_{n,m;m_1;c_1,c_2}^{R,n+1} = d_k + \dfrac{(\omega-d_k)^{n+1}}{(\omega-d_k)^n+c_2^R(\omega-d_k)(1-\omega)^{m_1}+c_1^R(1-\omega)^{m+1}} \text{ when } \omega \geq d_k \end{cases} \tag{6}$$

where superscript $L$ indicates that the function is for $\omega < d_k$, and superscript $R$ indicates that it is for $\omega \geq d_k$, and $n$ corresponds to the constant $n$ mentioned with regard to the previous $g_k(\omega)$. By adjusting the parameters $c_1, c_2,$ and $m_1$ in Eq. (6), the mapping function can have good flatness at $d_k$ and an optimized characteristic when $\omega$ approaches the endpoints $\{0,1\}$, and therefore the mapped WENO schemes have not only high accuracy and resolution, but also good robustness.

For WENO3-JS, the mapped scheme is referred to as WENO3-PRM$^2_{1,1}$ [1], where $n$ = 1 and $m$ = 1 in Eq. (6) and the other parameters are shown in Table 2. More details can be found in [1].

Table 2. Definitions of parameters ($c_1$, $c_2$, $m_1$) in the third-order WENO3-PRM$^2_{1,1}$ scheme

| | | $c_1$ | $c_2$ | $m_1$ |
|---|---|---|---|---|
| $d_k$=1/3 | L | 1 | $7\times10^7$ | 5 |
| | R | 1 | $3\times10^6$ | 5 |
| $d_k$=2/3 | L | 1 | $1\times10^5$ | 4 |
| | R | 1 | $3\times10^6$ | 4 |

## 2.4 Validating computations

The following two examples are tested to validate the numerical method and verify the turbulence model described above.

(1) Hollow cylinder–extended flare flow at $Ma$ = 11.35

The other inflow conditions are Reynolds number $Re = 3.596\times10^5$, freestream temperature $T_\infty$ = 132.8 K, and wall temperature $T_w$ = 296.7 K; the grid number is 256×128.

Fig. 4 shows the computed pressure distribution compared with that from experiment [19], and it can be seen that good agreement is achieved. Fig. 5 shows the density gradient contours and streamlines obtained at the separation zone, and it can be seen that the shock wave and vortex structure are well resolved. Thus, the high resolution of WENO3-PRM$^2_{1,1}$ is demonstrated.

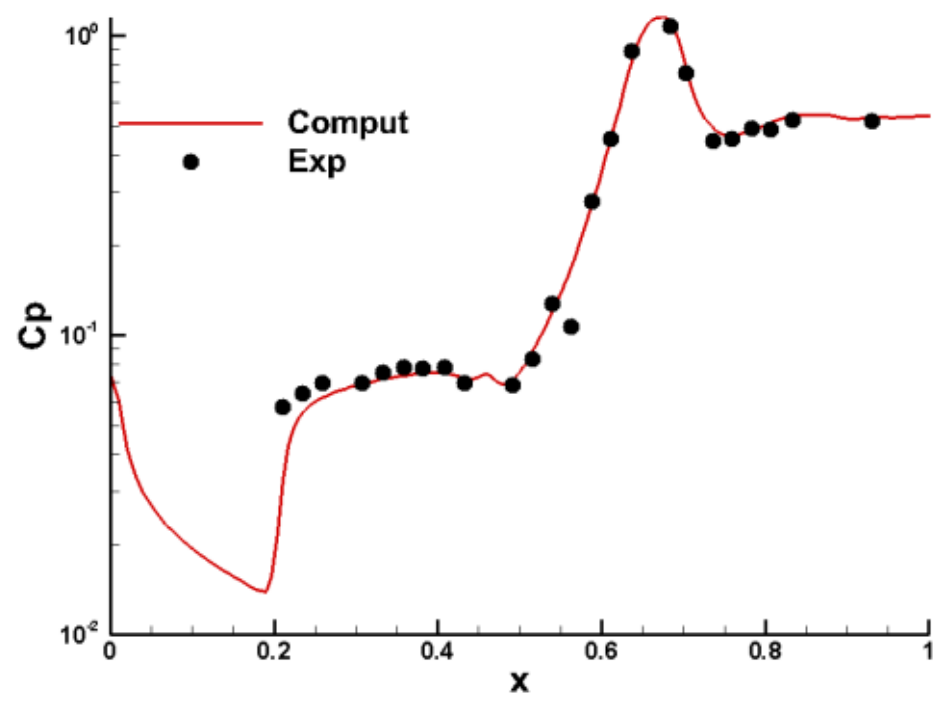

Fig. 4 Wall pressure coefficient of the hollow cylinder–extended flare flow at *Ma* = 11.35 compared with that from experiment [19]

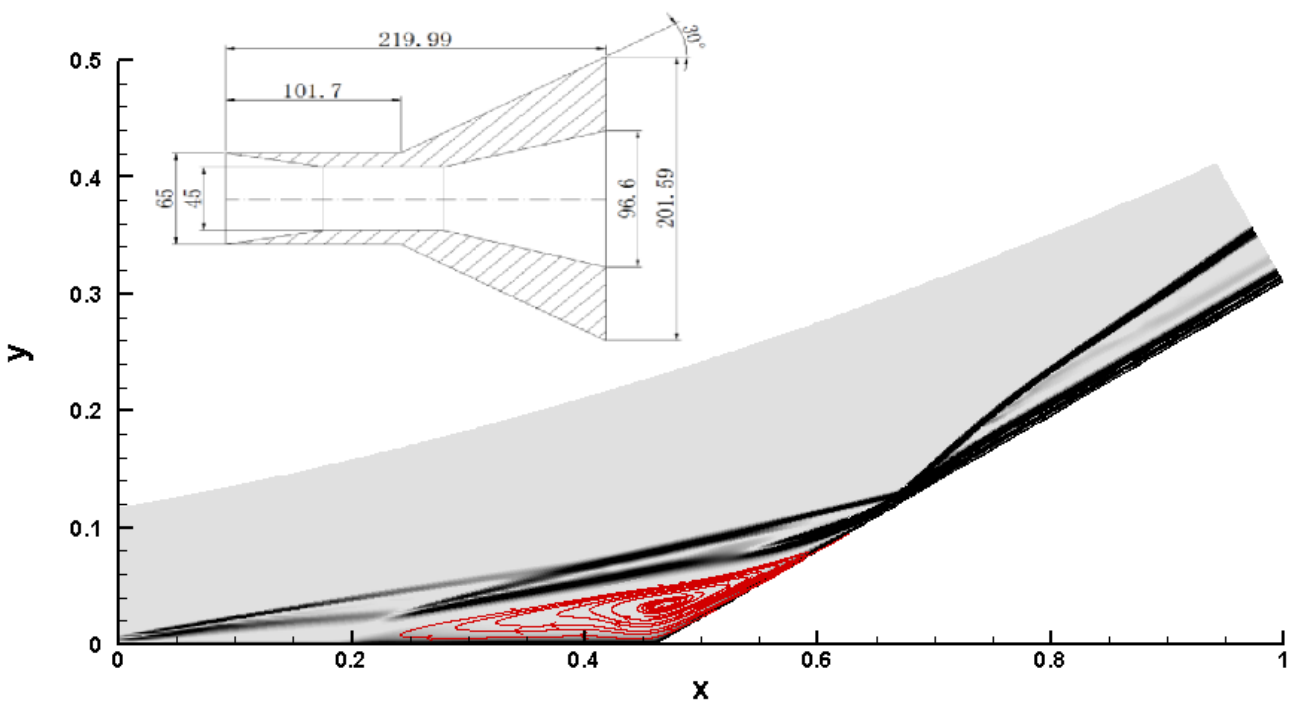

Fig. 5 Density gradient contours and streamlines at the separation zone of the hollow cylinder–extended flare flow at *Ma* = 11.35

(2) Flat plate flow at $Ma = 4.5$

The other inflow conditions are $Re$=3.81×10$^5$ and $T_\infty$ = 409.201 K, and the wall is adiabatic; the grid number is 273×193. Fig. 6 shows a comparison of the velocity profile of the boundary layer obtained from the computation with the experimental result of Coles [20]. The comparison reveals overall agreement and verifies the correct implementation of the SST model.

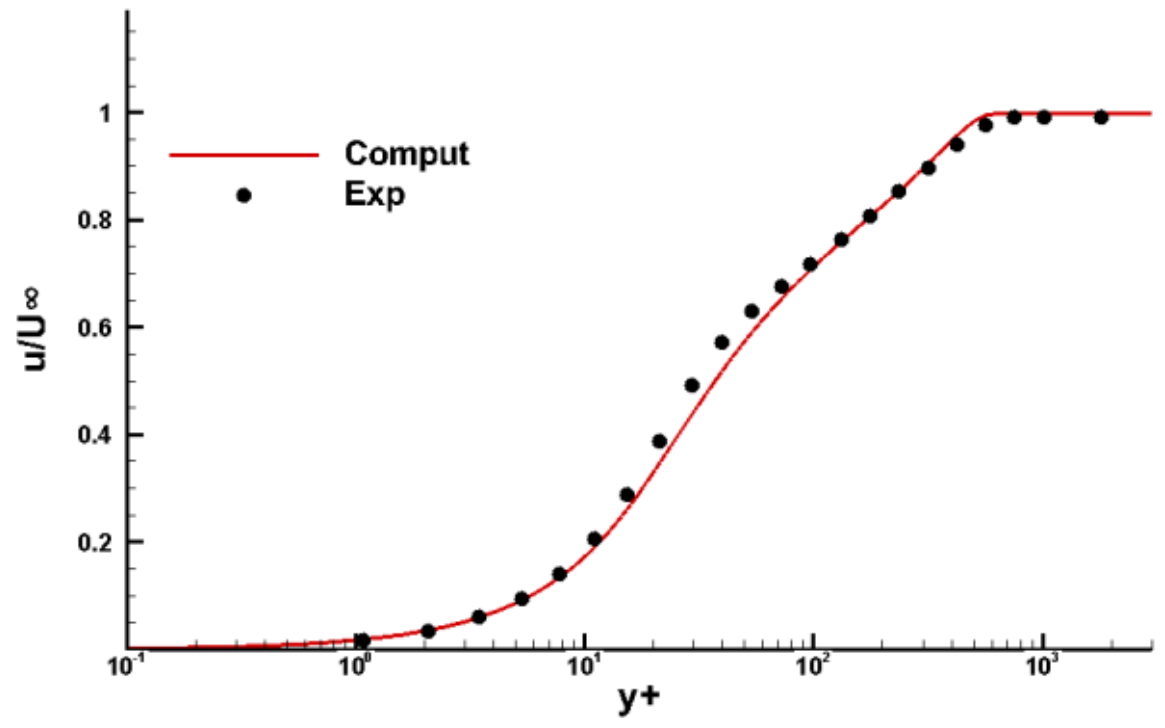

Fig. 6 Velocity profile of the boundary layer of the flat plate at *Ma* = 4.5 compared with experiment [20]

# 3 Interaction between conical shock wave and axisymmetric boundary layer at $Ma = 2.2$

As described in Section 1, Kussoy et al. [10] investigated the interaction between the conical shock wave generated by a cone with 15° half-cone angle at AOA = 15° and the boundary layer of a circular tube at $Ma = 2.2$. Given the experimental and computational capabilities at that time (1980), some of the results might have suffered from insufficient precision; besides, the investigation might not have been comprehensive enough owing to limitations on the conditions imposed. In response to these problems, this study revisits the case considered by Kussoy et al. [10] as well as investigating more related cases through simulations. Owing to the nonzero included angle between the cone axis and the incoming flow in the experiment, the conical shock wave generated is not axisymmetric. Thus, the flow is uneven in the circumferential direction of the tube, and therefore the interaction is a 3D event. To explore the influence of AOA on 3D interactions, we have performed further simulations at AOA = 5° and 10° besides the original 15° adopted in the experiment by Kussoy et al. [10]. To aid deeper understanding, it is useful to compare the 3D interaction with the appropriate 2D axisymmetric counterpart in which the AOA of the cone is zero. For this reason, we also consider a 2D axisymmetric case based on the following considerations: under the same inflow conditions as the 3D interaction, the angle of the 2D conical shock wave with respect to the circular tube is the same as that defined by the incident shock in the windward meridional plane with respect to the wall in the 3D case, which is realized by adjusting the half-cone angle. By this means, we first study the corresponding axisymmetric 2D interaction.

## 3.1 Interaction between 2D conical shock wave and axisymmetric boundary layer

The 2D inflow condition references the experiment of Kussoy et al. [10] where $Ma_\infty = 2.2$, $Re = 1.2\times10^7 m^{-1}$, and the total temperature $T_0$ and $T_w$ are both 278 K. In particular, the half-cone angles are 27.35°, 24.81°, and 20.96°, which correspond to the 3D interactions of the 15° half-cone angled cone at AOA = 15°, 10°, and 5°, respectively, in the manner previously described. A schematic of the computational configuration in the meridional plane is shown in Fig. 7(a), where the domain is virtually divided into zones I and II to facilitate later discussion. The geometric parameters of the domain are given in Table 3. The computation shows that the conical shock wave generated by the cone with 27.35° half-cone angle intersects with the tube at $x$ = 16.40 cm. To facilitate comparisons of wall pressure and friction force between cases at different AOAs, we shift the vertices of the other two cones by appropriate amounts along the negative $x$ axis, so that the intersection of the conical shock wave with the tube in three cases with different half-cone angles are at the same location.

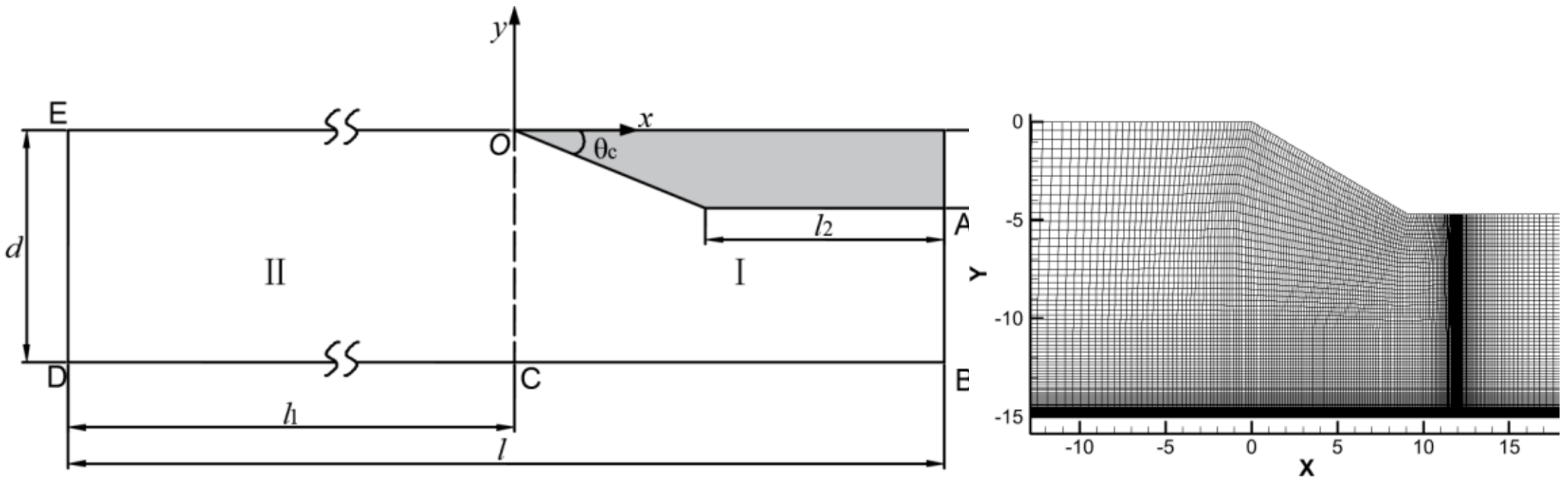

(a) Schematic of computational configuration

(b) Local view of grids near the cone apex

Fig. 7 Schematic of computational configuration and local view of grids

Table 3. Geometric parameters of 2D interactions, with lengths in centimeters, $\theta_c$ the half-cone angle, and $x_c$ the coordinate of the cone vertex

| $\theta_c$ | $l$ | $l_1$ | $l_2$ | $d$ | $h$ | $x_c$ |
|---|---|---|---|---|---|---|
| 27.35 ° | 290.0 | 270.0 | 9.9 | 15.03 | 4.8 | 0.0 |
| 24.81 ° | 290.0 | 268.4 | 10.4 | 15.03 | 4.8 | 1.6 |
| 20.96 ° | 290.0 | 265.8 | 10.9 | 15.03 | 4.8 | 4.2 |

3.1.1 Grid convergence study

To obtain suitable grids, we perform a grid convergence study using the case with a half-cone angle of 27.35°. A local view of the grids is presented in Fig. 7(b). To avoid unnecessary computational difficulties caused by the strong expansion at the bottom of the actual cone, the geometry is simplified by making an extension from the cone base to the exit of the tube as shown in Fig. 7 while otherwise preserving the configuration of the experiment. Although this changes the end angle of the expansion waves originating from the original cone base, it is known from characteristic line theory [21] that the expansion waves generated under different turning angles will yield the same solutions in the region having the same deflection angle, i.e., the flow fields therein are the same. In the grid study, assuming that the inflow in zone II remains uniform along the streamwise direction and that there is only slow variation in the boundary layer, 50 streamwise grids are assigned to zone II, and only the grid number in zone I is varied to examine grid convergence. On the basis of the above considerations, three sets of streamwise and radial grid numbers of zone I are chosen as shown in Table 4.

Figs. 8 and 9 respectively show the wall pressure and friction force distributions using different grids. It can be seen that the results with Grid2 and Grid3 basically coincide, while the positions of the separation predicted using these two grids are advanced farther upstream than the position predicted using Grid1. Hence, the results with Grid2 can be considered to reach grid convergence, and this grid is chosen for subsequent computations. In addition, when generating the grids to study the 3D interaction, the grid number of Grid2 also serves as reference.

Table 4. Three sets of streamwise and radial grid numbers of zone I in grid convergence study

| Grid1 | Grid2 | Grid3 |
|---|---|---|
| 120×120 | 200×200 | 300×300 |

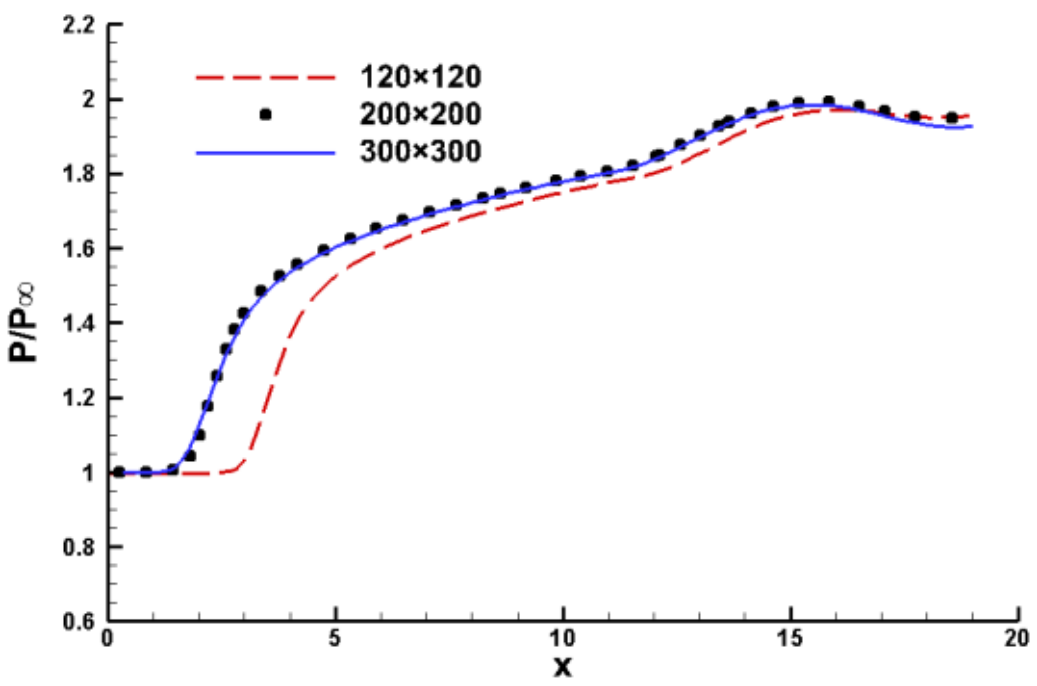

Fig. 8 Variation of wall pressure distribution with grid numbers in zone I

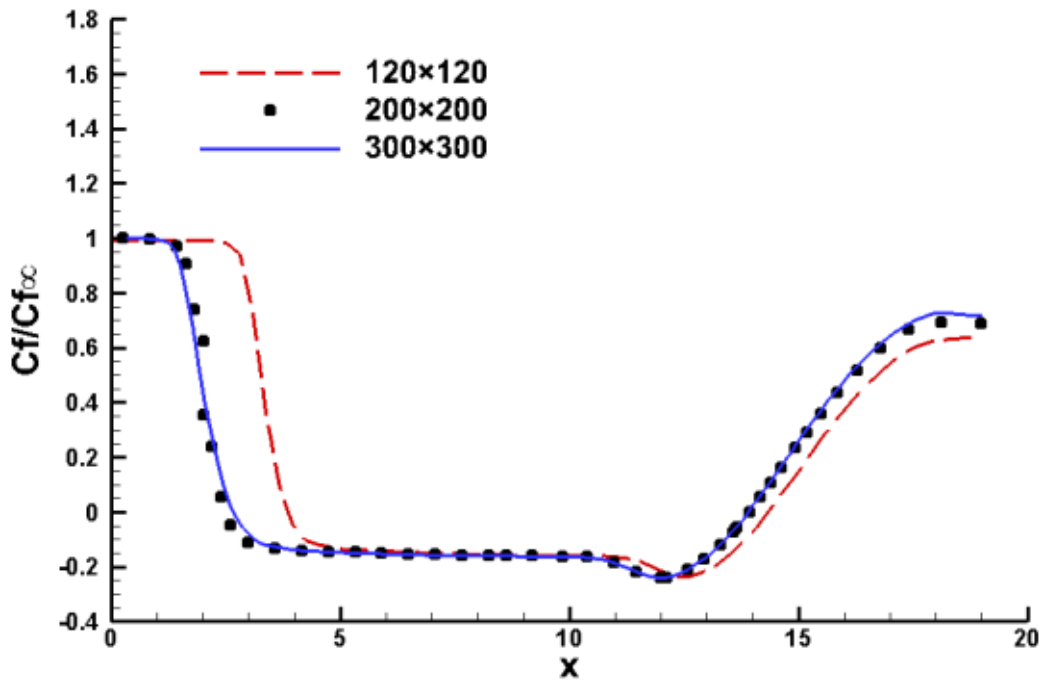

Fig. 9 Variation of friction force distribution with grid numbers in zone I

In summary, in the following computations, the grid numbers in zones II and I will be chosen as 50×200 and 200×200, respectively, with the cell closest to the wall being set as 0.0046 cm. The length ratio of adjacent cells in the stretched meshes is about 1.05 to 1. The corresponding computations and analyses will be described next.

3.1.2 Wave structures and variable distribution characteristics in flow interactions

To reveal the complexity of flow interactions, the pressure characteristics are first examined. In Fig. 10, the pressure contours are shown for different half-cone angles, with the separation bubbles also being visualized by streamlines. With the help of this visualization, the geometric relationship between the shock wave and the vortex is revealed.

Fig. 10 shows that the composition of the shock waves includes an incident shock, a separation shock, and a transmitted shock. However, by comparison with the canonical 2D interaction model of Babinsky and Harvey [22], the following distinctions exist: (a) there is no obvious reattachment shock in the flow field, and the transmitted shock ends before the top of the separation bubble (see the illustrations in Figs. 15 and 16); (b) in the cases with 24.81° and 20.96° half-cone angles, a distinct stem-like structure appears above the separation bubble, with a reflection shock being generated at the top of the stem. It is presumed that a Mach reflection-like event occurs related to the transmitted shock at this instant. To obtain the geometric relations between this structure and the boundary layer, in Fig. 10, we draw two streamlines passing through the top and bottom of the stem and mark the edge of the boundary layer with a red solid line, through which the variation of the streamlines with respect to the boundary layer can be observed. The results show that the upper streamline moves toward the wall as the half-cone angle decreases, with its location gradually shifting from the outside of the boundary layer to the inside. In addition, the stream tube formed by the two streamlines tends to shrink along the streamwise direction in the section perpendicular to the incoming flow, and it is this that is suspected to give rise to the stem.

A further examination of Fig. 10 reveals that the shock incident angle with respect to the wall increases with the half-cone angle, as a consequence of which the shock strength is amplified. Consequently, a gradient of adverse pressure is imposed, which follows the upstream movement of the separation shock and the enlargement of the separation bubble.

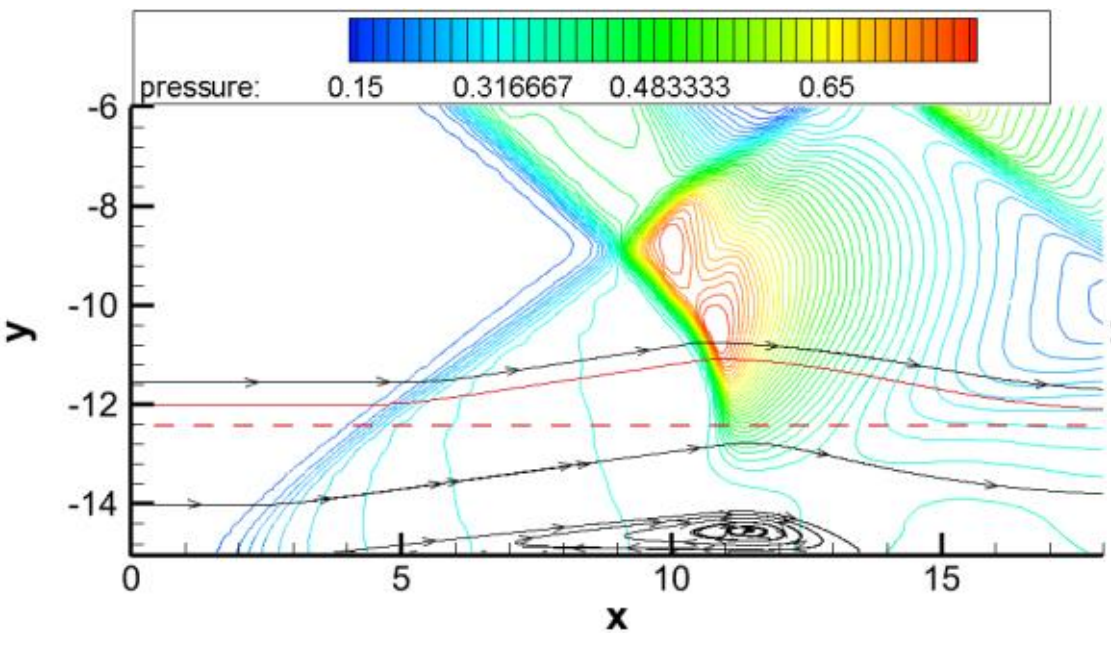

(a) $\theta_c = 27.35°$

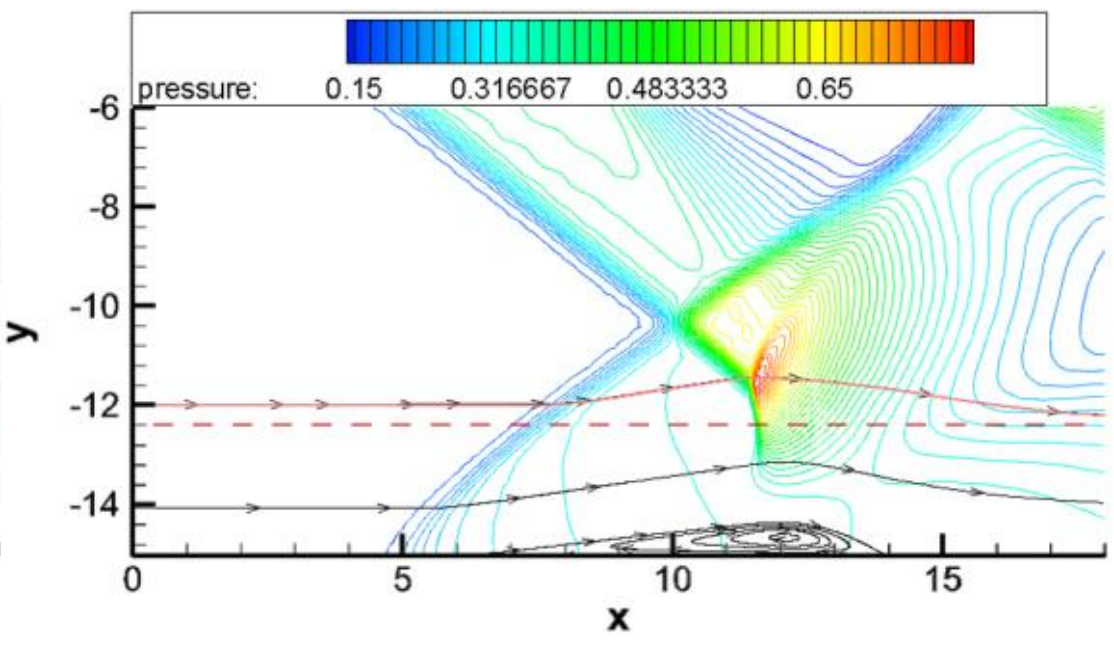

(b) $\theta_c = 24.81°$

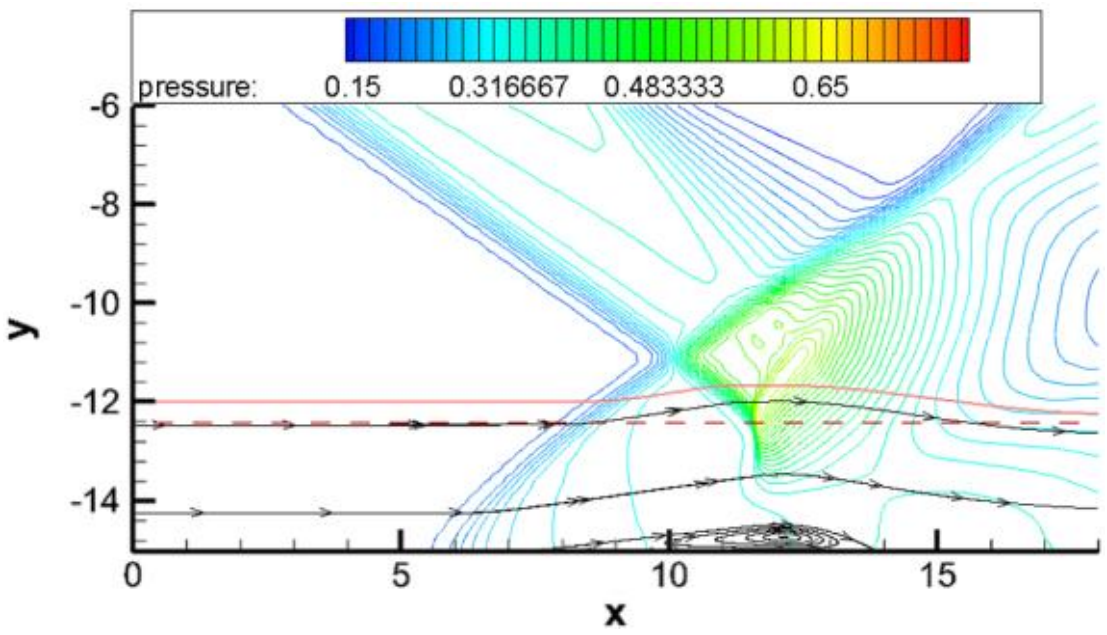

(c) $\theta_c = 20.96°$

Fig. 10 Pressure contours in zone I for three cases with different half-cone angles. The red solid lines indicate the edge of the boundary layer, the horizontal dashed lines show the variable distributions passing through the Mach stem-like structure, the upper and lower streamlines are chosen to pass through the top and bottom of the stem, and the separation bubbles located below the stem are visualized by streamlines

To show the wave system more clearly, Fig. 11 presents contours of the density gradient magnitude, with the sonic lines being drawn in red. The distribution of sonic lines indicates that a subsonic region exists behind the Mach stem-like structure, and therefore a Mach reflection-like event is considered to occur involving the transmitted shock wave. These phenomena are more evident in the cases with half-cone angles of 24.81° and 20.96°. In addition, we have not observed any obvious reattachment shock waves in any of the three cases.

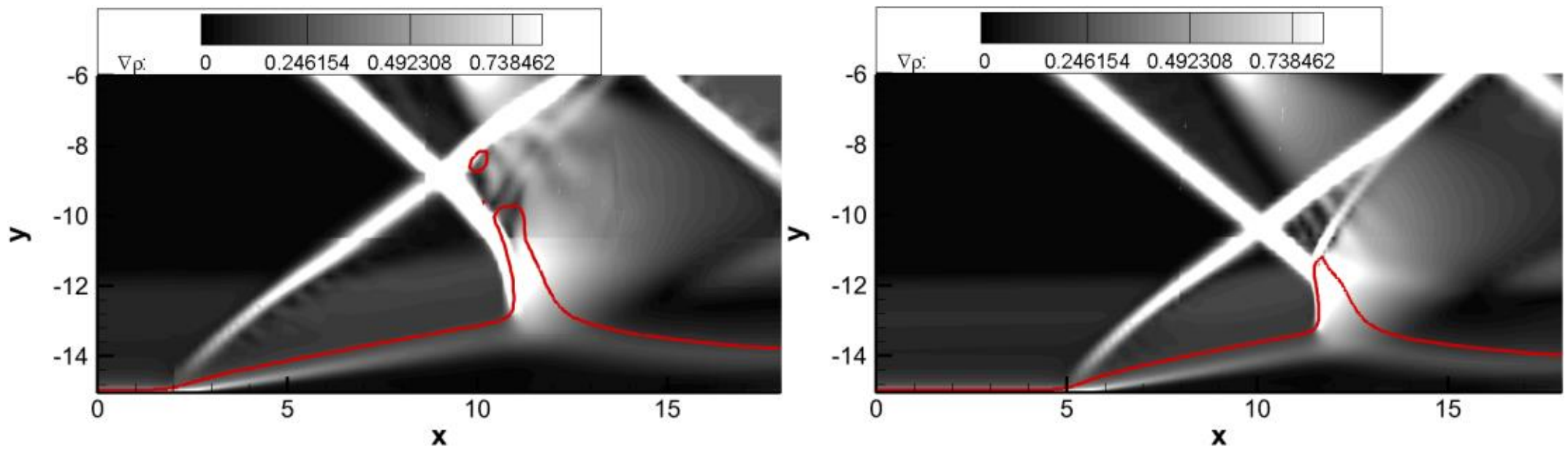

(a) $\theta_c = 27.35°$ (b) $\theta_c = 24.81°$

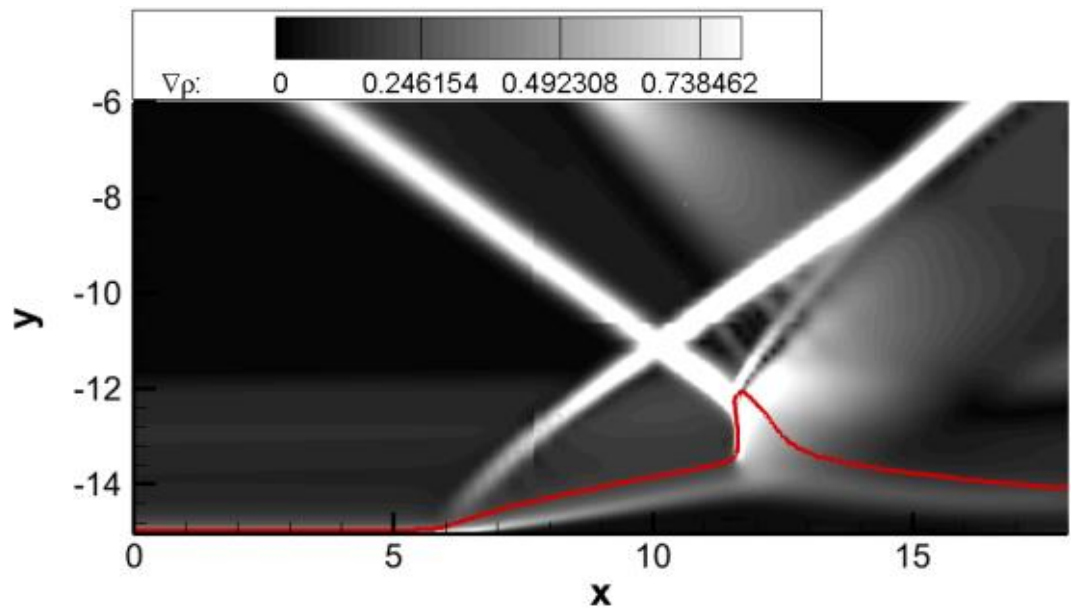

(c) $\theta_c = 20.96°$

Fig. 11 Contours of density gradient magnitude in zone I for three cases with different half-cone angles. The sonic lines are drawn in red.

To reveal the quantitative changes in the flow before and after the stem, we selected the variable distribution along the horizontal line $y = -12.4$ cm (see the red dashed lines in Fig. 10) for analysis. The Mach number and pressure distributions in the three cases with different half-cone angles are shown in Figs. 12 and 13. In Fig. 12, the following characteristics are revealed: (a) there is a discontinuity in the Mach number distribution at the stem, with the Mach number changing from 1.225, 1.286, and 1.364 before the structure to 0.854, 0.825 and 0.858 afterwards in the cases with $\theta_c = 27.35°$, 24.81°, and 20.96°, respectively; (b) undergoing expansion after the stem, the flow accelerates, with increasing Mach number, and reverts to supersonic; (c) after $x > 16$ cm, the increase in Mach number slows down, and a platform characteristic appears, which is similar to the pressure distributions shown in Figs. 10 and 13.

In Fig. 13, the pressure distributions at $y = -12.4$ cm show that a first rise occurs at the separation shock wave, next a second rise appears at the stem, and then the pressure drops rapidly from the highest point, finally approaching a plateau after $x > 16$ cm. To show the correspondence to characteristic positions such as the separation point, we take the case $\theta_c = 24.81°$ as an example and mark the positions in the distributions as shown in Fig. 14. The wall pressure distribution is displayed besides the original one at $y = -12.4$ cm, and the following characteristic locations are identified: the separation shock foot (SF), the separation shock (SS), the Mach stem-like structure (MS), the separation point (SP), and the reattachment point (RP). Fig. 14 reveals that the adverse pressure caused by separation propagates upstream through the boundary layer, which causes the wall pressure to rise before SF; the pressure saturates gradually after SP, which appears as a slow increase and a slight acceleration near RP; the pressure reaches a peak around $x \approx 11.69$ cm, then falls due to the expansion from the cone base, finally approaching a platform distribution.

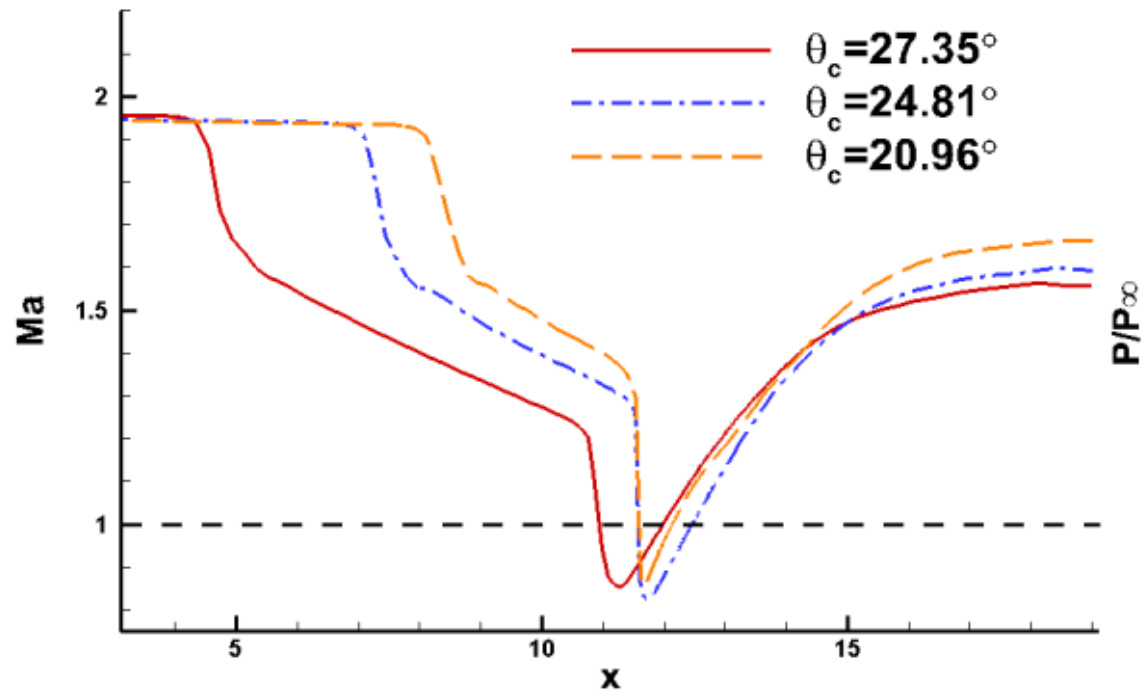

Fig. 12 Comparison of Mach number distributions along the line $y$ = -12.4 cm for three half-cone angles

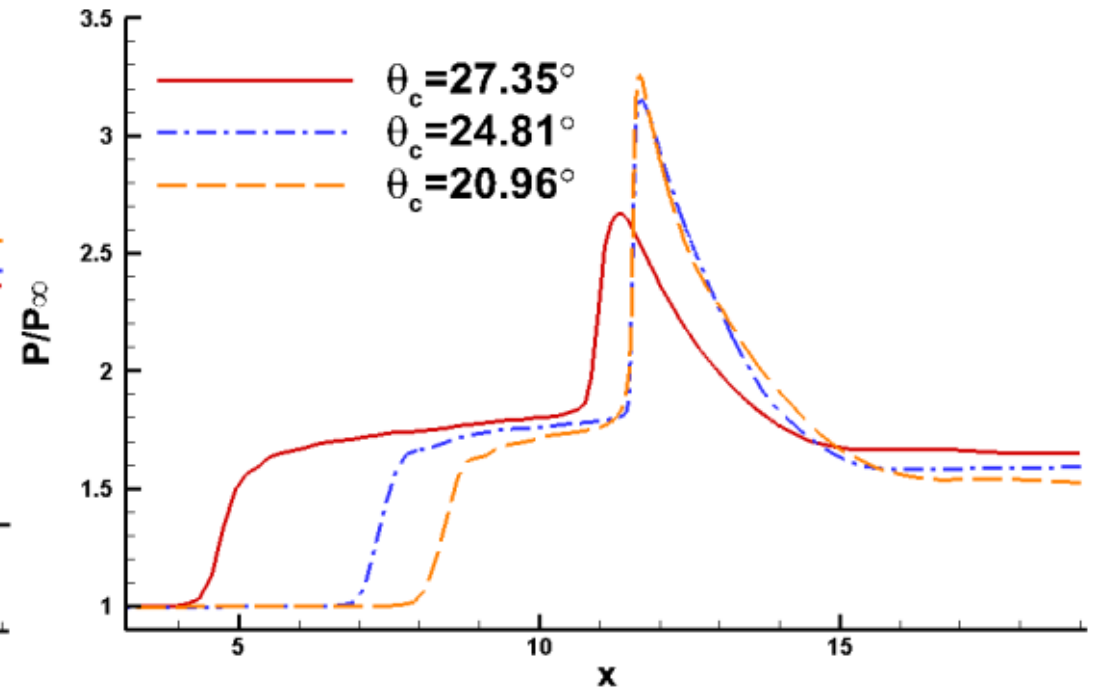

Fig. 13 Comparison of pressure distributions along the line $y$ = -12.4 cm for three half-cone angles

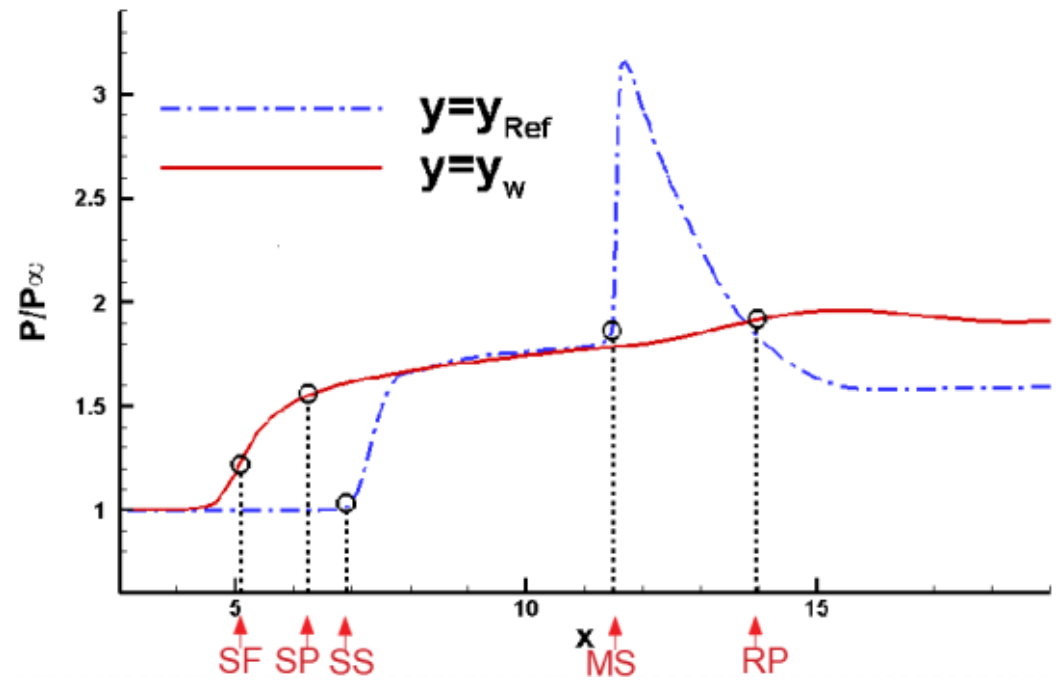

Fig. 14 Pressure distributions along the line $y$ = -12.4 cm and at the wall in the case $\theta_c = 24.81°$ with characteristic positions marked ($y_{\mathrm{Ref}}$ represents the horizontal line and $y_w$ represents the wall)

The above results show that the flow changes discontinuously at the Mach stem-like structure, which provides a qualitative indication of the occurrence of a shock wave. To reveal the characteristics of the stem, an appropriate shock wave model could be used to predict the flow behind the structure, with the prediction then being compared with the results of computation to assess the validity of the assumption that the structure is a shock wave. However, a detailed examination shows that the stem is inclined to the wall to some degree (e.g., the inclination angles with respect to the $x$ axis are found to be 77.11°, 83.02°, and 82.39° respectively), and, in addition, the flow ahead of the stem is nonuniform (e.g., quantities such as the Mach number differ at different $y$ coordinates). Owing to these complexities, it would be difficult to obtain exact shock formulas. In view of this, a simple model is employed here in which the stem is approximated as an oblique shock wave and the flow after the wave is predicted approximately by the Rankine–Hugoniot relations using the variables locally ahead of the stem. Although this model is not accurate enough for detailed calculations, it can serve to evaluate the similarity between the stem and a plane shock wave. In the process of prediction and comparison, several points are first chosen in front of the stem; then, based on the variables at these points, the Mach number $M_{2,\text{pred}}$ after the stem is predicted using the oblique shock relation; finally, the relative error $|M_2 - M_{2,\text{pred}}|/M_2$ is calculated, where $M_2$ is the Mach number from the simulation. In this procedure, the quantities in the three cases, namely, $M_1$, $M_2$, $M_{2,\text{pred}}$ and the relative errors, are acquired at chosen points as shown in Table 5. The relative errors show that $M_2$ is quite close to $M_{2,\text{pred}}$ in the middle part of the stem, while the errors are relatively large at the top and bottom of the structure. Hence, a considerable part of the stem does resemble a plane shock wave.

Table 5. Mach numbers before and after the stem from the computation, predicted Mach numbers, and relative errors at selected $y$ coordinates, with $y_{\text{stem}}$ indicating the $y$ range of the stem

| $\theta_c$ | $y_{\text{stem}}$ (cm) | $y$(cm) | $M_1$ | $M_2$ | $M_{2,\text{pred}}$ | $\frac{\lvert M_2-M_{2,\text{pred}}\rvert}{M_2}$(%) |
|---|---|---|---|---|---|---|
| 27.35 ° | [−10.7, −12.8] | −11.2 | 1.564232 | 0.846622 | 0.75539374 | 10.7755 |
| | | −11.6 | 1.452310 | 0.829762 | 0.78774661 | 5.0635 |
| | | −12.0 | 1.318956 | 0.838337 | 0.83820749 | 0.0155 |
| | | −12.4 | 1.192512 | 0.858060 | 0.90271570 | 5.2048 |
| 24.81 ° | [−11.4, −13.1] | −12.0 | 1.450708 | 0.767187 | 0.74005055 | 3.5371 |
| | | −12.4 | 1.312932 | 0.806579 | 0.79782443 | 1.0854 |
| | | −12.8 | 1.171684 | 0.850281 | 0.87666301 | 3.1027 |
| | | −13.0 | 1.105892 | 0.850805 | 0.92281661 | 8.4640 |
| 20.96 ° | [−12.1, −13.4] | -12.4 | 1.377949 | 0.850260 | 0.77225414 | 9.1743 |
| | | −12.8 | 1.240798 | 0.847901 | 0.83832239 | 1.1297 |
| | | −13.2 | 1.100226 | 0.876087 | 0.93006126 | 6.1608 |

To summarize, the so-called “Mach stem-like structure” in this study has the following characteristics: (a) the flow in front of the stem turns from supersonic into subsonic after the stem; (b) a reflection shock arises at the top of the stem; (c) the flow ahead of the stem is not directed perpendicular to the stem, nor is it uniform; (d) expansion waves occur after the stem. Hence, the structure qualitatively resembles but is not strictly identical to a Mach stem.

### 3.1.3 Vortex structures in 2D axisymmetric interaction

To reveal the vortex structure in the interaction, the streamlines are chosen for study. Because the vortex structure in 3D interactions will be investigated later and compared with the current 2D structures, we present the results for the streamlines in Fig. 31 to avoid repetition. In that figure, the locations of the separation and reattachment points are marked by “SP” and “RP”, respectively. The results show that the size of the separation bubble increases with the angle of the incident shock. Moreover, the following typical 2D characteristics are manifested: (a) the streamlines of the separation bubble form a series of closed curves around the central point; (b) the separation starts from SP and ends at RP, and the two points are connected by a single limiting streamline to form a closed boundary of the bubble. Quantitatively, we measure the length scale of the separation bubble according to the streamlines, and corresponding data are shown in Table 8.

3.1.4 Model of interaction of 2D conical shock wave with axisymmetric boundary layer

In the above discussion, we have pointed out the presence of a Mach stem-like structure and the absence of a reattachment shock wave in the interaction. In Fig. 15, the 2D planar shock wave/boundary layer interaction model of Babinsky and Harvey [22] is reproduced. The model indicates that the transmitted shock wave $C_4$ will be reflected from the upper rear of the separation bubble, with expansion waves subsequently being formed (other than the stem in this study), and a reattachment shock finally being generated. It is clear that the present axisymmetric shock wave/boundary layer interactions do not exactly match the model of Babinsky and Harvey. Based on the results and discussions above, we propose a new model of the 2D axisymmetric interactions, as shown in Fig. 16, which involves the incident shock wave S1, separation shock wave S2, transmitted shock wave S3, Mach stem-like structure MS4, reflection shock wave S5, and expansion waves E6. The flow mechanism is as follows:

(1) The adverse pressure gradient caused by the incident shock wave causes the flow to separate. (In the 2D axisymmetric situation, the separation point is connected to the reattachment point by the same singular axis, while the connection is not closed in the 3D case considered later.).

(2) A separation shock is generated ahead of the separation through convergence of compression waves and intersects the incident shock to yield two transmitted shocks. The lower transmitted shock wave S3 undergoes a Mach reflection-like event, and a stem is generated with its foot located some distance from the upper front of the separation bubble.

(3) A reflection shock is generated at the head of the stem and coalesces with the upper

transmitted shock wave, and the flow downstream is influenced by the expansion waves from the cone base.

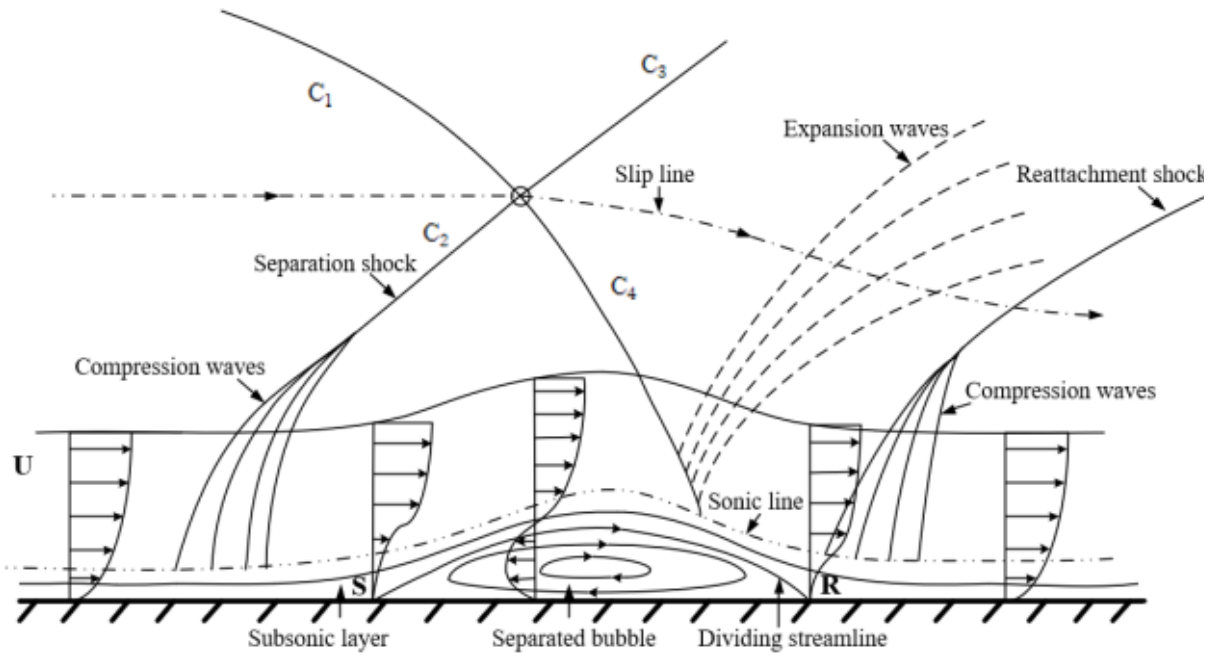

Fig. 15 Reproduction of 2D planar shock wave/ boundary layer interaction model of Babinsky and Harvey [22]

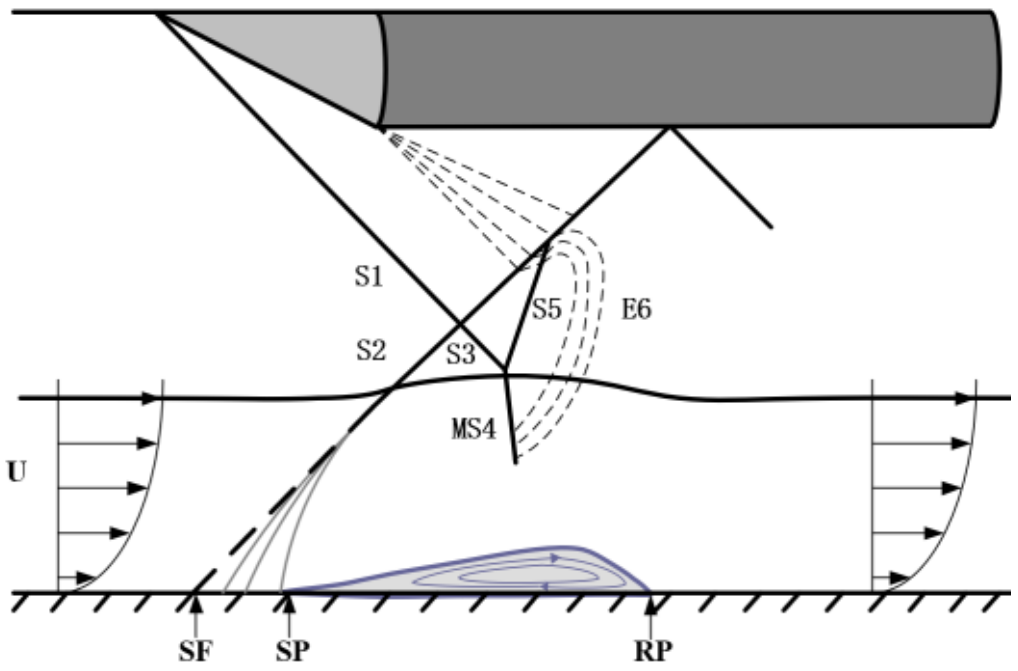

Fig. 16 New model of interaction between 2D conical shock wave and axisymmetric shock boundary layer. SF denotes the separation shock foot, SP the separation point, and RP the reattachment point

## 3.2 Interaction between 3D conical shock wave and axisymmetric boundary layer

In the above study of 2D axisymmetric interactions, we have pointed out the differences between the present interaction structure and the model of Babinsky and Harvey [22] and we have proposed a new model based on numerical results. Our purpose here in studying the 2D axisymmetric interaction is to provide a reference and comparison for the 3D counterpart in which the cone has various AOAs, including that adopted in the experiment by Kussoy et al. [10]. Therefore, in this section, considering the conditions of that experiment, for three different AOAs, we investigate the 3D characteristics of the conical shock wave from a cone interacting with an axisymmetric boundary layer, and we compare the results obtained with those from the 2D simulations discussed in Section 3.1.

We first study the interaction corresponding to the experiment by Kussoy et al. [10]. As shown in Fig. 1, a cone with half-cone angle as 15° and base diameter 6.10 cm is placed in a $\phi 24.13$ cm tube at AOA = 15°. The cone- is supported by a stick, and the center of the cone base lies on the axis of the tube. Obviously, the supersonic flow will yield a conical shock more that is more intense on the windward side than on the leeward side, and the interaction of the shock wave with the axisymmetric boundary layer is fully 3D. The shock is apparently strongest at an azimuthal angle of $\varphi$ = 0°, where its intersection with the tube is $x$ = 16.40 cm. Since Kussoy et al. only investigated the case AOA = 15°, to study the effects of varying AOA on the interaction, it is necessary to consider different AOAs. In this section, we consider three AOAs of 5°, 10°, and 15° while keeping the cone otherwise the same as that of Kussoy et al.. The computational configuration on the symmetric meridional plane is sketched in Fig. 17, where the center of the cone base is always located on the axis of the tube and the coordinate origin is taken as the apex of the cone at AOA = 15°. To avoid unnecessary difficulties in grid generation and computation, we apply a similar geometric simplification to that adopted for the 2D interaction, namely, extending the cone base to the exit of the tube and using the resulting elliptical cylinder for grid generation. According to characteristic line theory [21], although the angle at which the expansion waves end differs from that for the original cone without the extension, the solutions for the expansion waves within a domain confined by the same deflection angle are the same. Hence, as long as the domain covers the region that can influence the interaction, the simplification that we have adopted will not yield solutions that differ from those for the original interaction. To facilitate comparison of the results for different AOAs, we shift the cone vertices along the tube axis so that the location at which the shock wave is incident for AOA = 10° and 5° is the same as the location for AOA = 15°, namely, $x$ = 16.40 cm. The configurations of the cone vertices at different AOAs are illustrated in Fig. 17, and the main dimensions are given in Table 6. To facilitate later discussion, we divide the meridional plane into four regions according to the location of the cone vertices for reference, namely, zones I–IV as shown in the figure. Taking the $y$ axis to be aligned with the radial direction in the meridional plane, we do not distinguish strictly between the use of coordinate $y$ or $r$ in the following discussion.

The inflow conditions remain those of the experiment by Kussoy et al., namely, $M_\infty = 2.2$, unit $Re = 1.2 \times 10^7\ \mathrm{m}^{-1}$, and $T_0 = T_w = 278$ K.

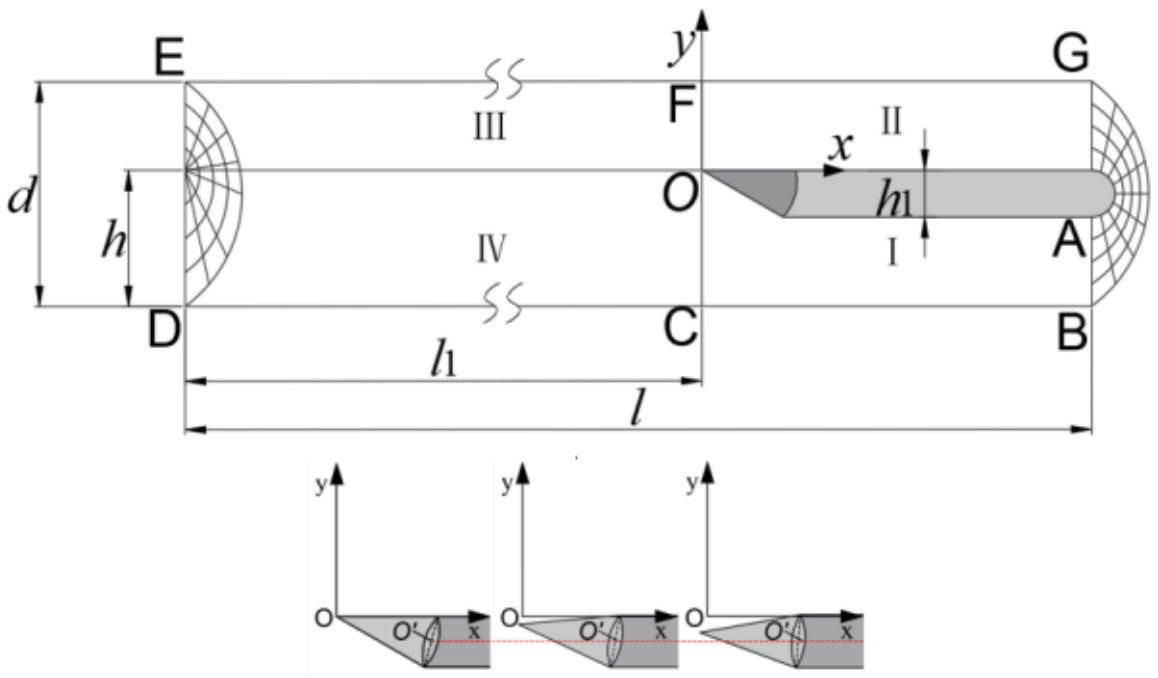

Fig. 17 Schematic of computational configuration in the meridional plane and illustration of cone vertices with respect to the coordinate origin in three cases (the bottom surfaces of the cones have been extended to the outlet of the tube)

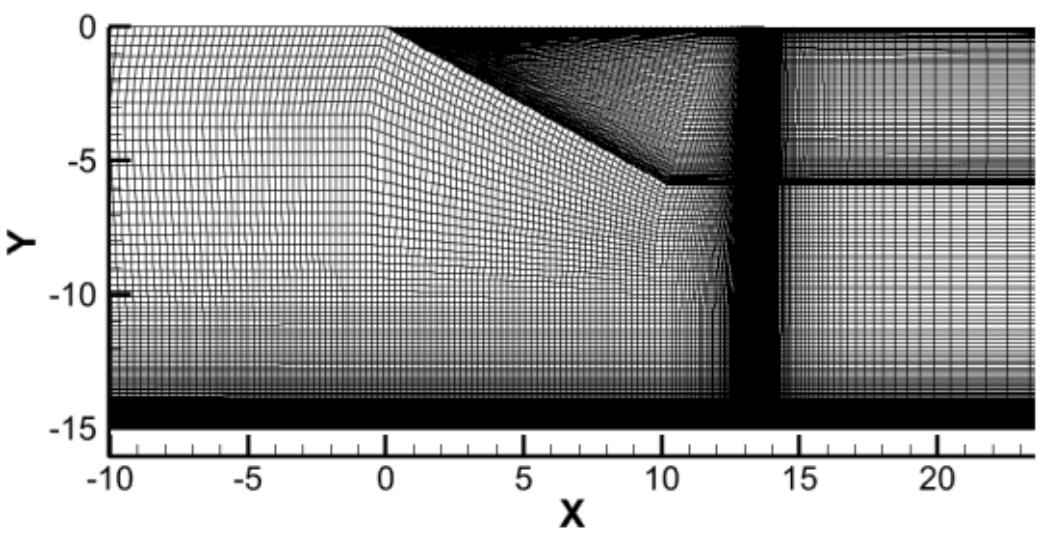

Fig. 18 Local view of grids in zone I with grid refinement used around the interaction region

Table 6. Geometric parameters of 3D interactions with lengths in centimeters

| AOA | $l$ | $l_1$ | $d$ | $h$ | $h_1$ | Coordinates of cone apex | |
|---|---|---|---|---|---|---|---|
| | | | | | | $x$ | $y$ |
| 15 ° | 310.0 | 270.0 | 9.9 | 15.03 | 4.8 | 0.0 | 0.0 |
| 10 ° | 310.0 | 269.6 | 10.4 | 15.03 | 4.8 | 0.4 | 1.0 |
| 5 ° | 310.0 | 268.5 | 10.9 | 15.03 | 4.8 | 1.5 | 2.0 |

Considering the symmetry of the configuration and the zero sideslip of the inflow, we only generate half-meshes with respect to the meridional plane to save computational resources. Since a 2D grid convergence study has already been performed in Section 3.1.1, we use the obtained outcomes to generate the streamwise and radial meshes. Specifically, we set up grids with numbers $n_{\text{streamwise}} \times n_{\text{radial}} = 200 \times 200$ in zones I and II and with the numbers $n_{\text{streamwise}} \times n_{\text{radial}} = 50 \times 200$ in zones III and IV. A local view of the grids in zone I is shown in Fig. 18. Around the interaction region, the streamwise grids are refined, the cell length closest to the wall is 0.0046 cm, and the growth rate of the adjacent grid is 1.05. To define the circumferential distribution, we perform a grid study using a different circumferential number $n_c$. Figs. 19 and 20 respectively show the wall pressure and friction distributions at three azimuthal angles, and it can be seen that the distributions at $n_c$ = 20, 40, and 80 basically coincide with each other and therefore grid convergence has been achieved. In view of the present focus on the 3D details of interactions, more circumferential grids would be favorable to higher accuracy, and so $n_c$ = 80 is employed in this study.

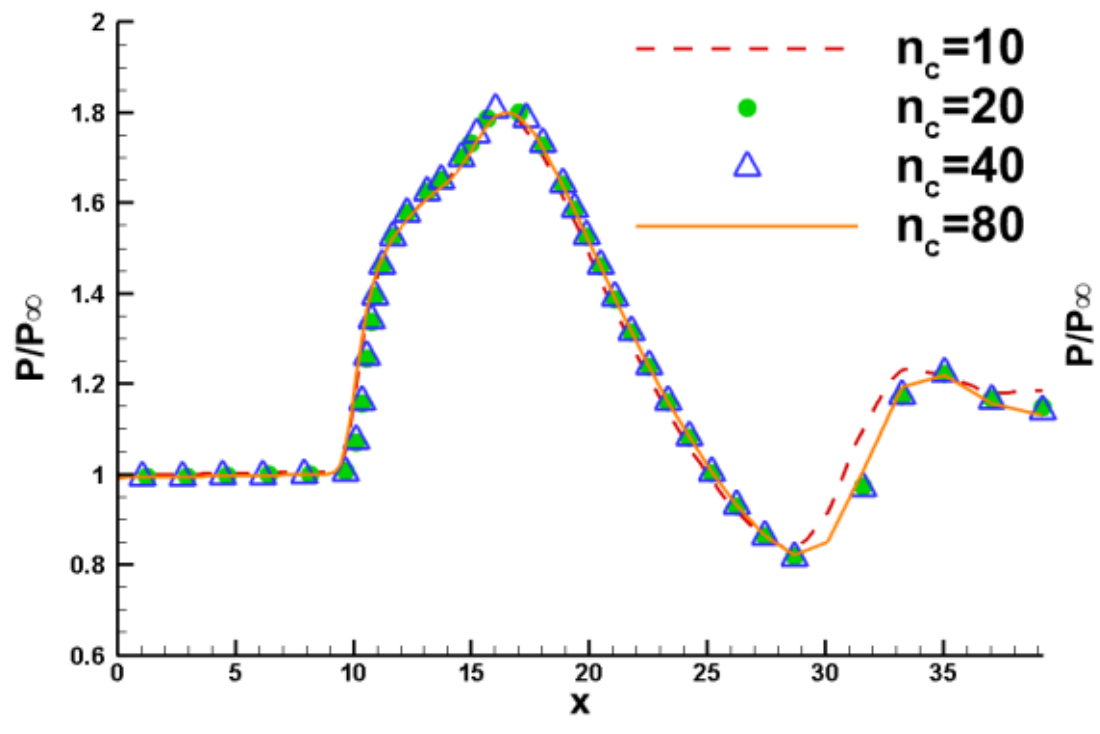

(a) $\varphi$=30°

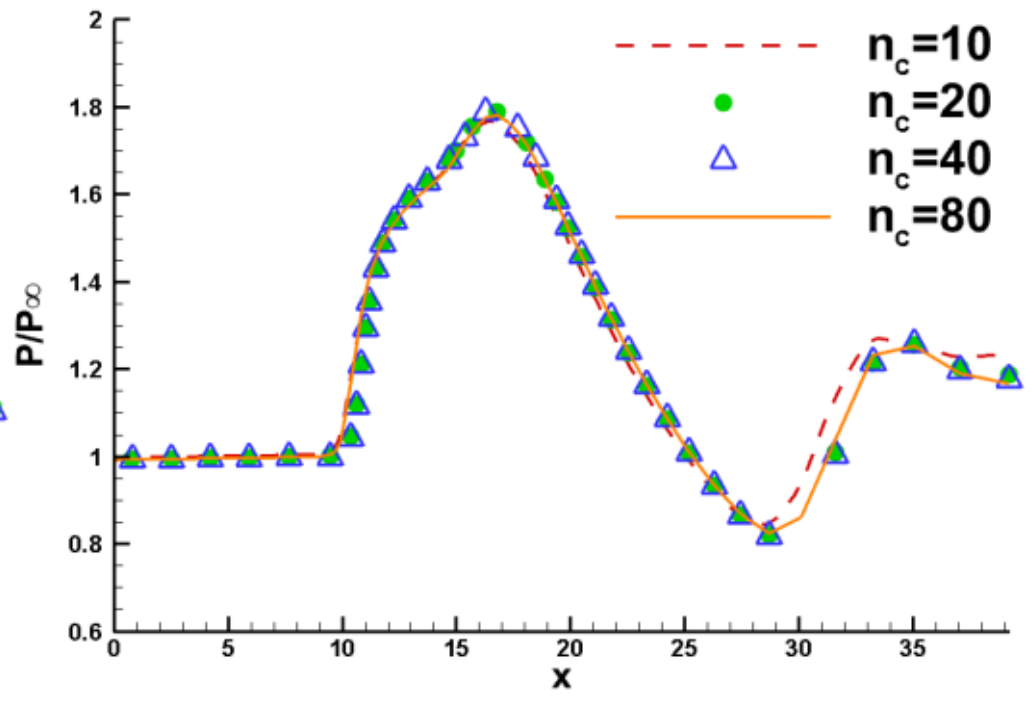

(b) $\varphi$=45°

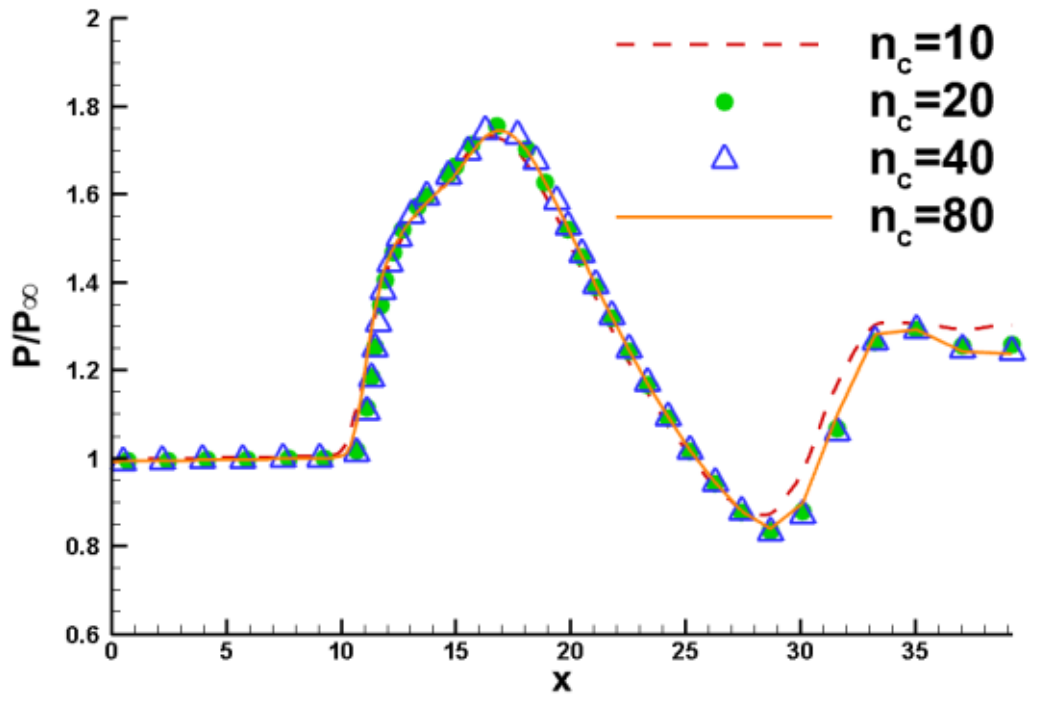

(c) $\varphi$=60°

Fig. 19 Wall pressure distributions at three azimuthal angles using four grids with different circumferential numbers $n_c$

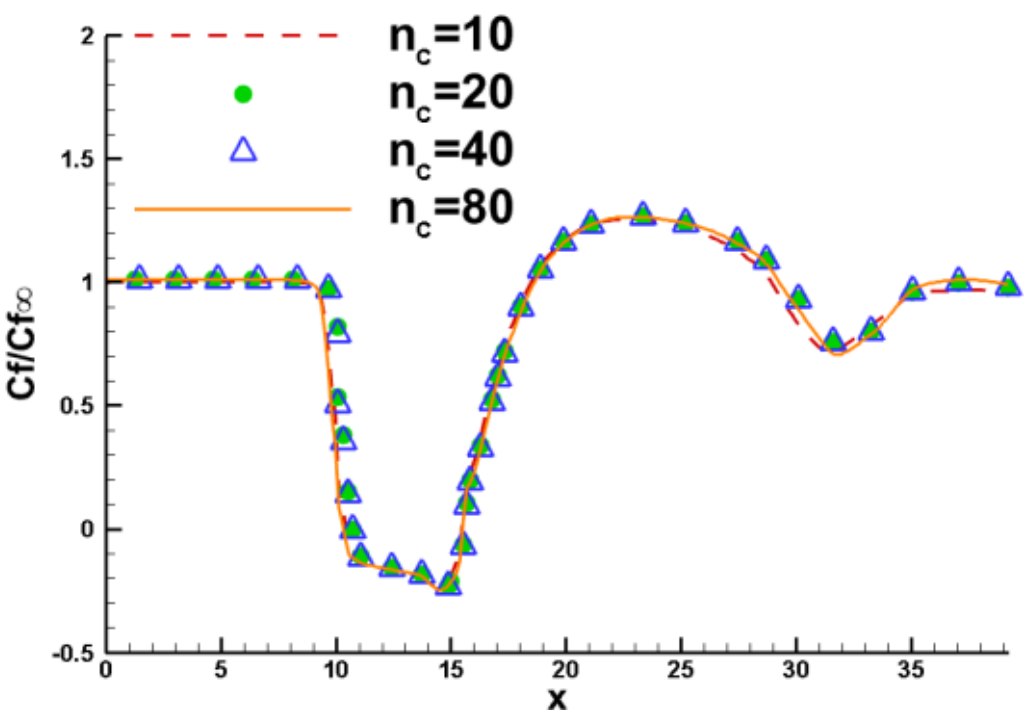

(a) $\varphi$=30 °

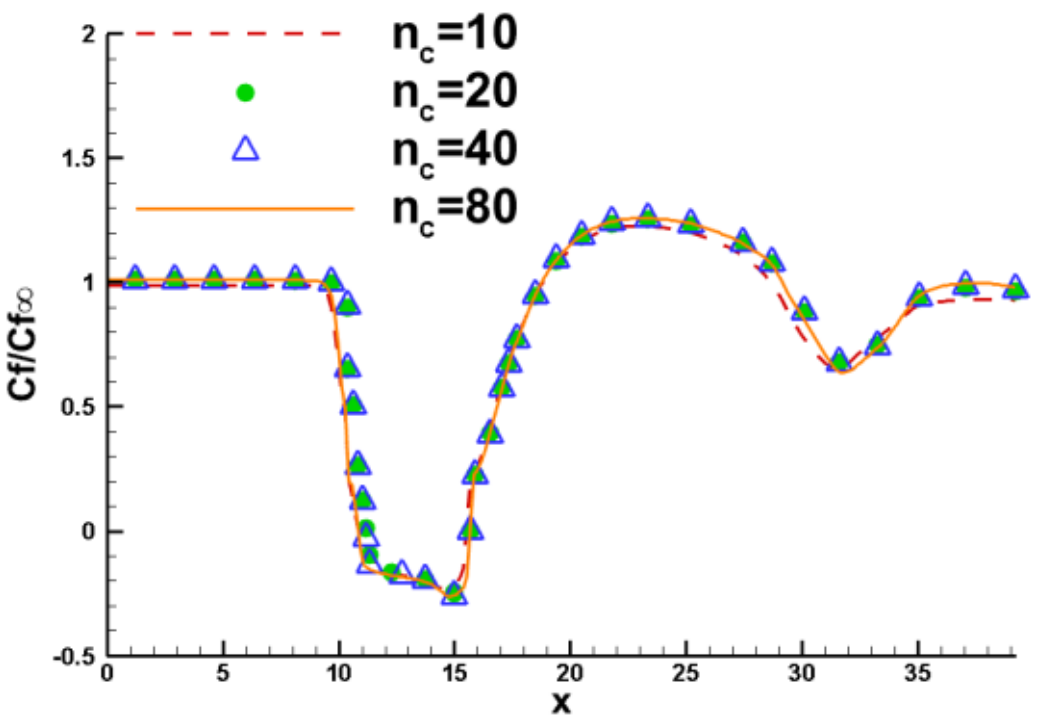

(b) $\varphi$=45 °

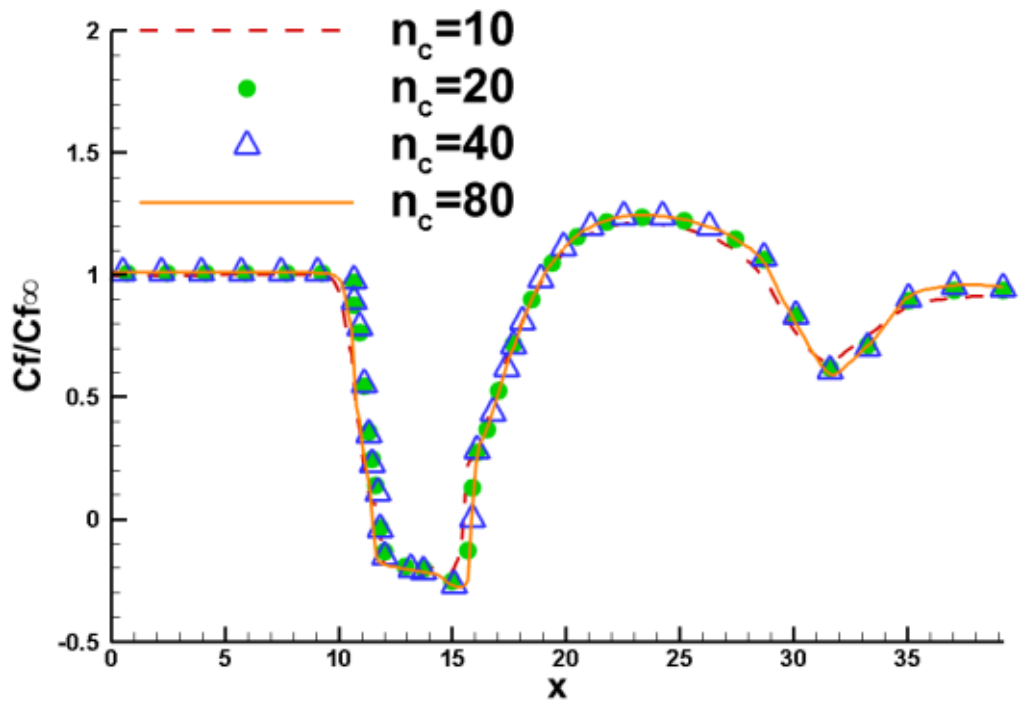

(c) $\varphi$=60 °

Fig. 20 Friction force distributions at three azimuthal angles using four grids with different circumferential numbers $n_c$

Before any further analysis of the complexity of 3D interactions, we first perform computations for the case studied experimentally by Kussoy et al. [10], which enables us to make comparisons, reveal more details, and evaluate the suitability of our approach.

3.2.1 Computation and analysis concerning the experiment by Kussoy et al. at AOA = 15°

Kussoy et al. [10] presented quantitative results for distributions such as those of wall pressure and friction force, as well as flow mechanisms such as surface separation patterns. These can be used for comparison with our computations. Fig. 21 shows a comparison of the computed pressure distributions at three azimuthal angles with experiment results, and it can be seen that good agreement is achieved. The following features should also be highlighted: the separation shock causes an initial pressure jump, after which a local platform-like distribution appears owing to flow separation, which is most obvious at $\varphi$ = 0° and almost disappears at $\varphi$ = 90°; then a secondary pressure rise occurs, which is caused by the flow reattachment; finally, a continuous pressure drop takes place owing to the expansion wave at the base of the cone, but after that no further platform is observed as the one shown in see Fig. 13. In addition, the variation of pressure with increasing azimuthal angle exhibits the following features: the peak pressure drops from 1.84 to 1.78 and finally to 1.66; the location of the pressure rise moves downstream from $x$ = 9.21 cm to $x$ = 10.17 cm and finally stays at $x$ = 11.85 cm; the extent of the pressure platform shrinks until it finally disappears, indicating a decrease in the separation scale along the circumference. These characteristics suggest that the intensity of the conical shock decreases with increasing azimuthal angle, and accordingly the adverse pressure gradient also decreases as flow separation becomes weaker.

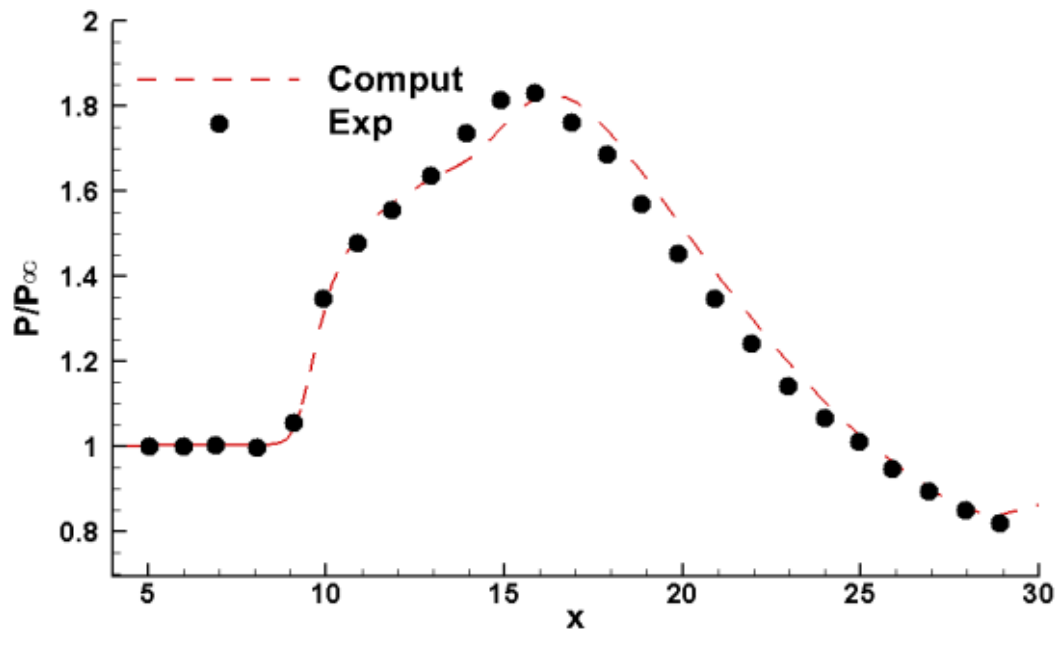

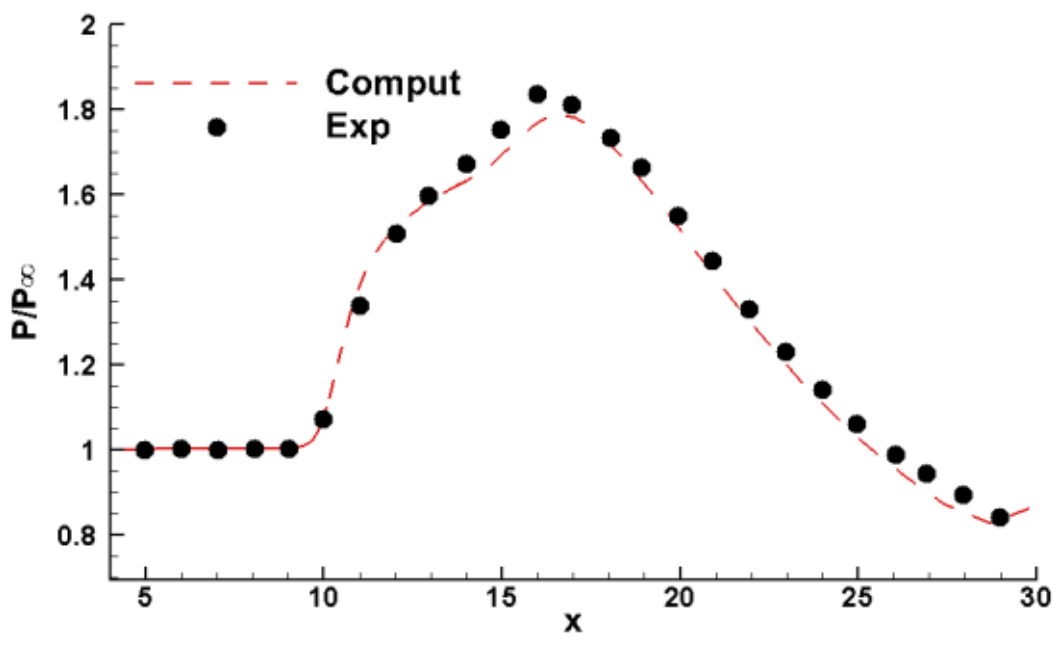

(a) $\varphi$=0° (b) $\varphi$=45°

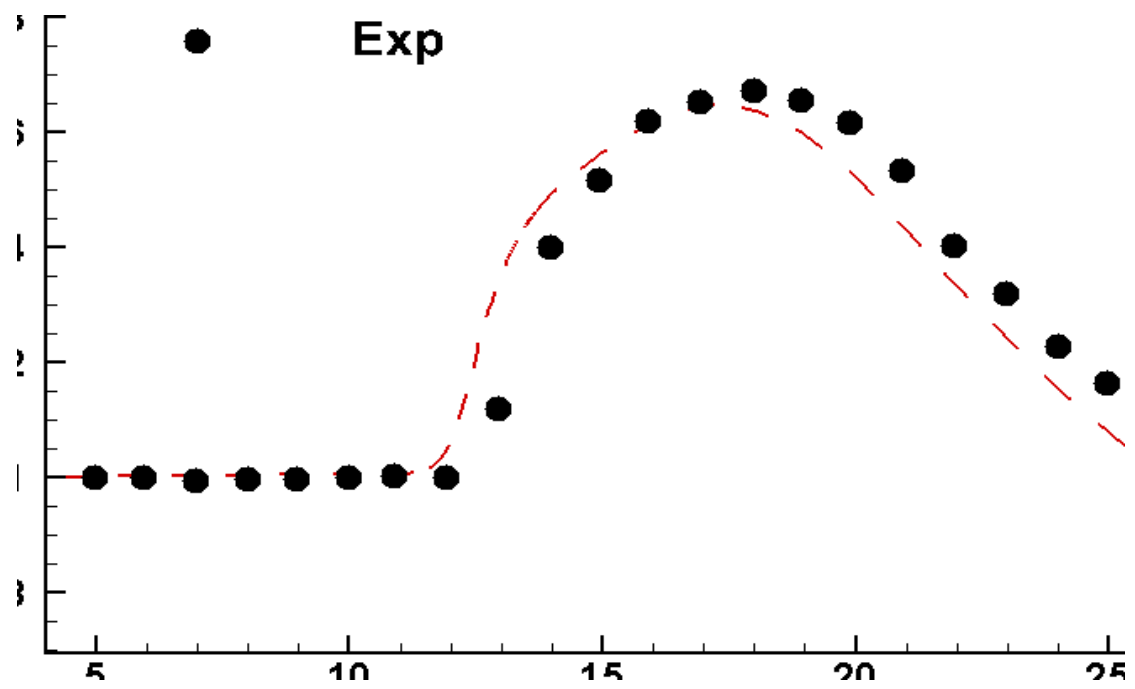

(c) $\varphi$=90°

Fig. 21 Wall pressure distributions in comparison with those from experiment at three azimuthal angles in the case AOA = 15 °

Fig. 22 shows a comparison between computation and experiment for the friction force distributions at three azimuthal angles. It can be seen that (1) although the presence of flow separation is indicated by the negative friction force in both studies, the predicted separation zone is clearly larger than that found experimentally and (2) the computed friction force distribution shows a steeper fall than is found experimentally, and the computed start of separation ($C_f$ = 0) is advanced farther upstream. In addition, in Fig. 22(c), it can be seen that a separation zone is present in the computed result at $\varphi$ = 90°, whereas there is no separation according to the experimental result. However, in Fig. 3(b), the oil flows from the experiment do show such a separation. This inconsistency in the experimental results suggests that the techniques for measurement of friction force available when Kussoy et al. [10] performed their experiment might not have been sufficiently sensitive. In short, then, the lack of agreement here between computation and experiment cannot be regarded as indicating a lack of accuracy of the former.

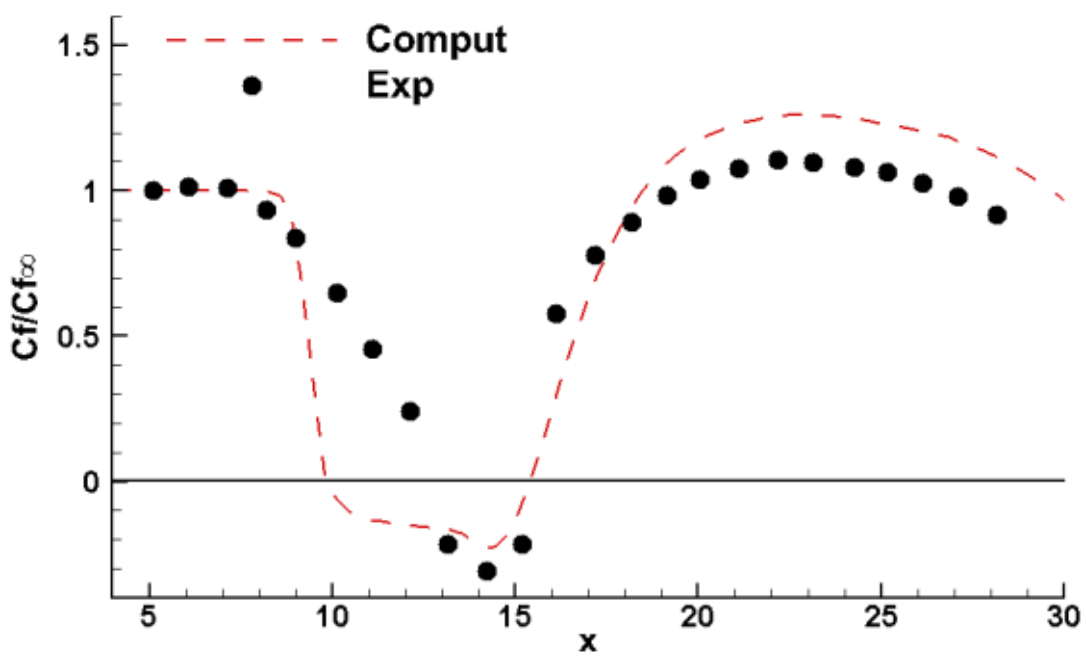

(a) $\varphi$=0°

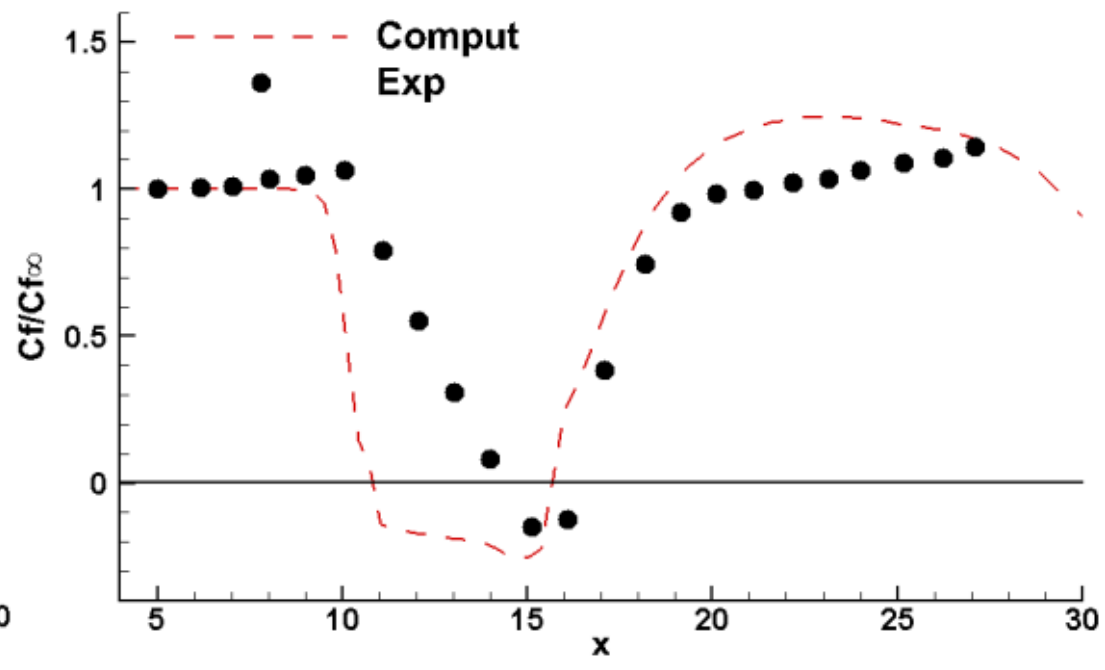

(b) $\varphi$=45°

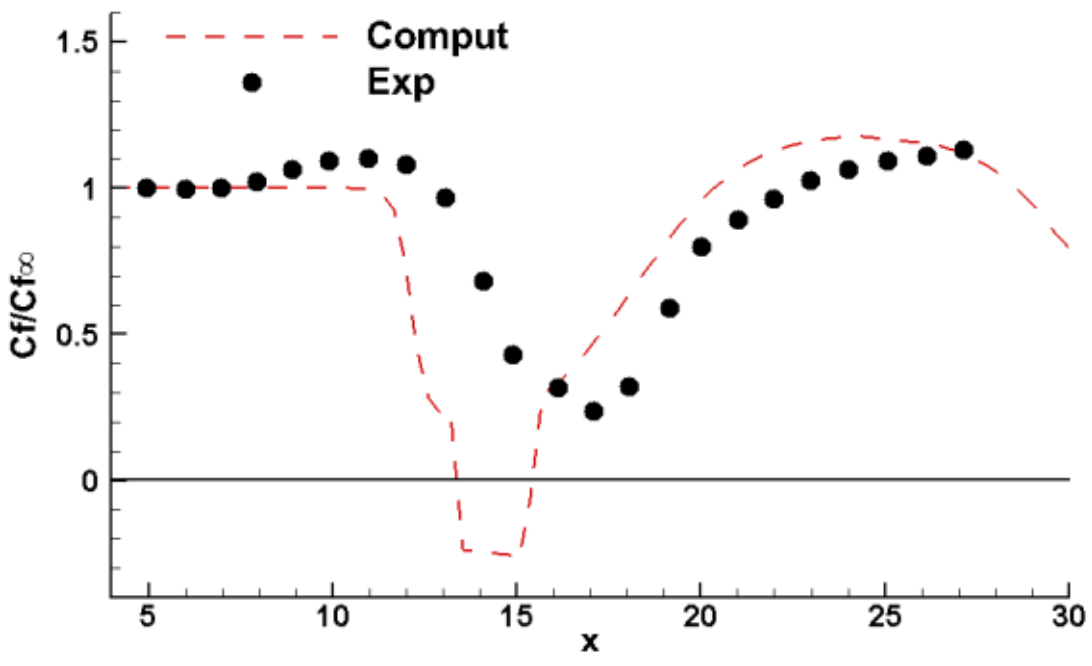

(c) $\varphi$=90°

Fig. 22 Computed friction force distributions in comparison with those from experiment at three azimuthal angles in the case AOA = 15 °

In addition to pressure and friction force distributions, we have also computed the surface separation to compare with the experimental oil flows shown in Fig. 3(b). The limiting streamlines at the wall for AOA = 15° are shown in Fig. 33(a) below. The comparison shows that the computed and experimental streamline distributions are qualitatively similar, such as in the extension of separation and reattachment streamlines along the circumference and in the convergence into a single stream trace. Despite this qualitative similarity, there are differences: for example, the separation and reattachment streamlines in the experiment converge at approximately $\varphi$ = 135° and 235° respectively, which are not symmetric with respect to the location $\varphi$ = 0° and thus violate the symmetry of the flow, whereas according to the computation, the two types of streamlines converge at $\varphi$ = 125° or 235° due to the employment of half configuration with respect to the meridional plane. Hence, the disagreement regarding the convergence of streamlines at $\varphi$ = 135° in the experiment are not sufficient to invalidate the computation.

3.2.2 Wave structures in windward meridional plane and 3D perspectives of interactions

In Section 3.1, a new flow structure was found in the 2D interaction, namely, a Mach-like stem structure in the upper front of the separation bubble without a reattachment shock during flow reattachment. Whether such a stem is also present in the 3D interaction and whether there might also be other new flow structures are worthy of investigation. In view of this, we first analyze and compare the pressure and density gradient contours in zone I at different AOAs to examine how the wave structure varies with the strength of the incident shock. We then perform a qualitative study of the pressure and Mach number distributions in the windward meridional plane. Finally, we analyze the wave structures in 0°, 45° and 90° azimuthal sections at AOA = 15° to obtain some understanding of how the flow evolves along the circumference of the tube.

(1) Flow structures in the windward meridional plane at three AOAs

Fig. 23 shows contours of the pressure and density gradient in zone I of the windward meridional plane at three AOAs, which are quite similar to the 2D counterparts in Figs. 10 and 11. Therefore, the schematic model of the 2D interaction (Fig. 16) should still be appropriate for describing the wave structure in the meridional plane considered here. In Fig. 23, we again draw the streamlines passing through the top and bottom of the stem, as well as identifying the edge of the boundary layer by a red solid line in the pressure contours. Compared with the 2D interactions, the top of the stem here is closer to the wall, while the streamlines passing through the top overall depart farther from the edge of the boundary layer and approach the wall with decreasing AOA, and the stream tube formed by the two streamlines at the front of the vortex indicates a similar contraction.

In addition, the pressure contours indicate that as AOA decreases, the incident angle of the shock with respect to the wall decreases, the separation bubbles become smaller, and the location of the separation shock moves downstream. In Fig. 23, the streamlines are also depicted with the pressure contours, and a qualitative comparison with Fig. 10 shows that the 3D separation bubble is smaller than its 2D counterpart.

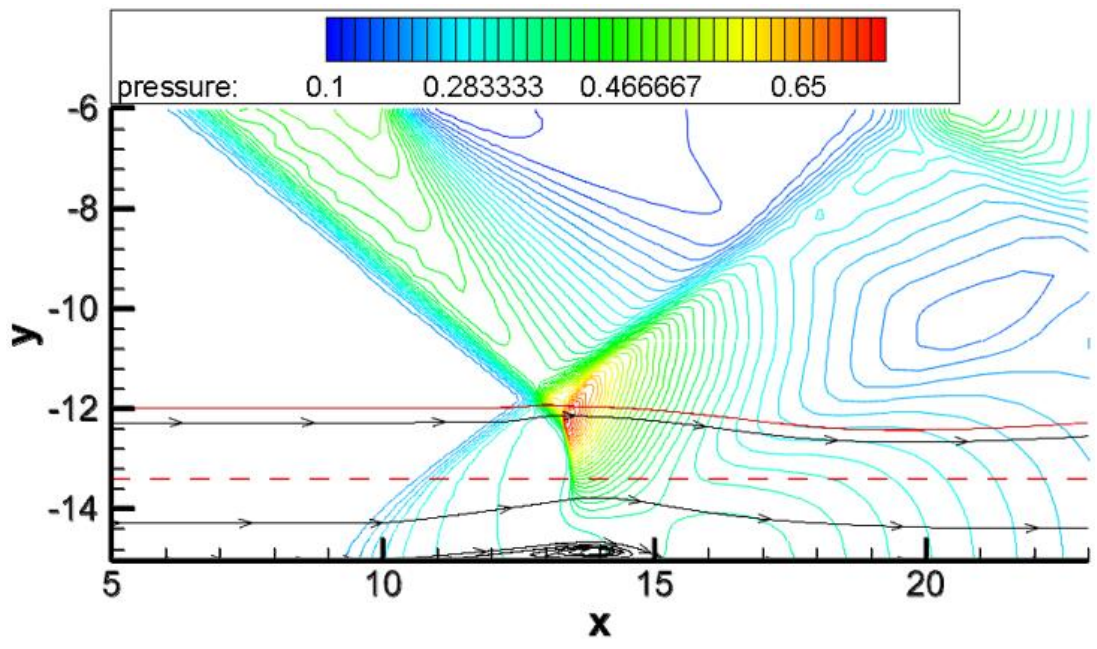

(a) Pressure contours at AOA = 15 °

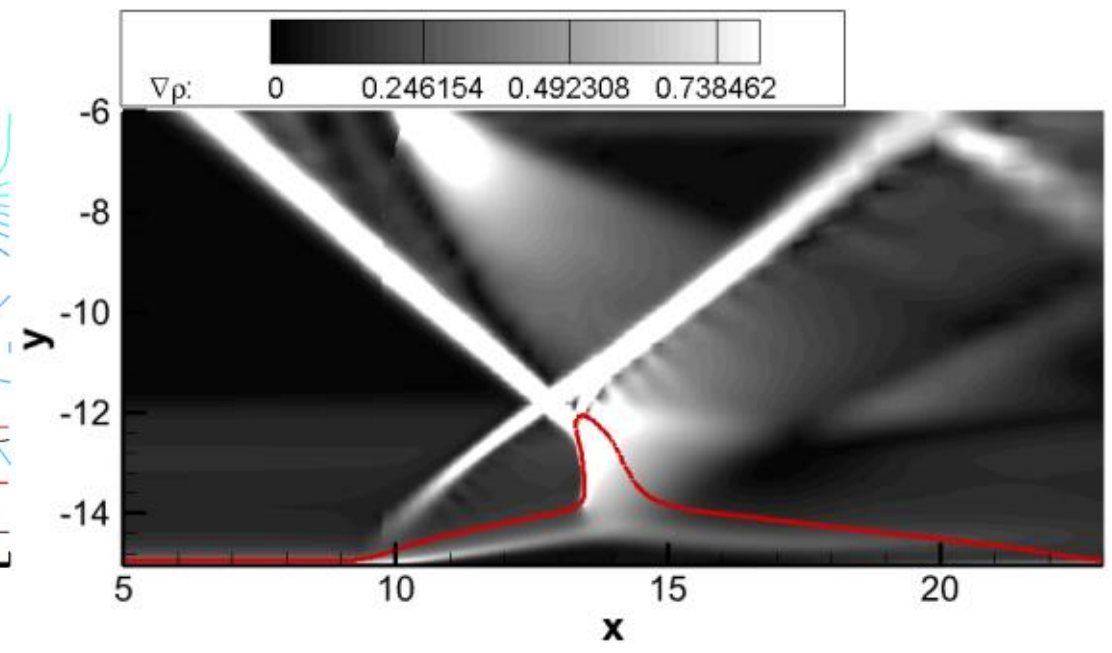

(b) Density gradient contours at AOA = 15 °

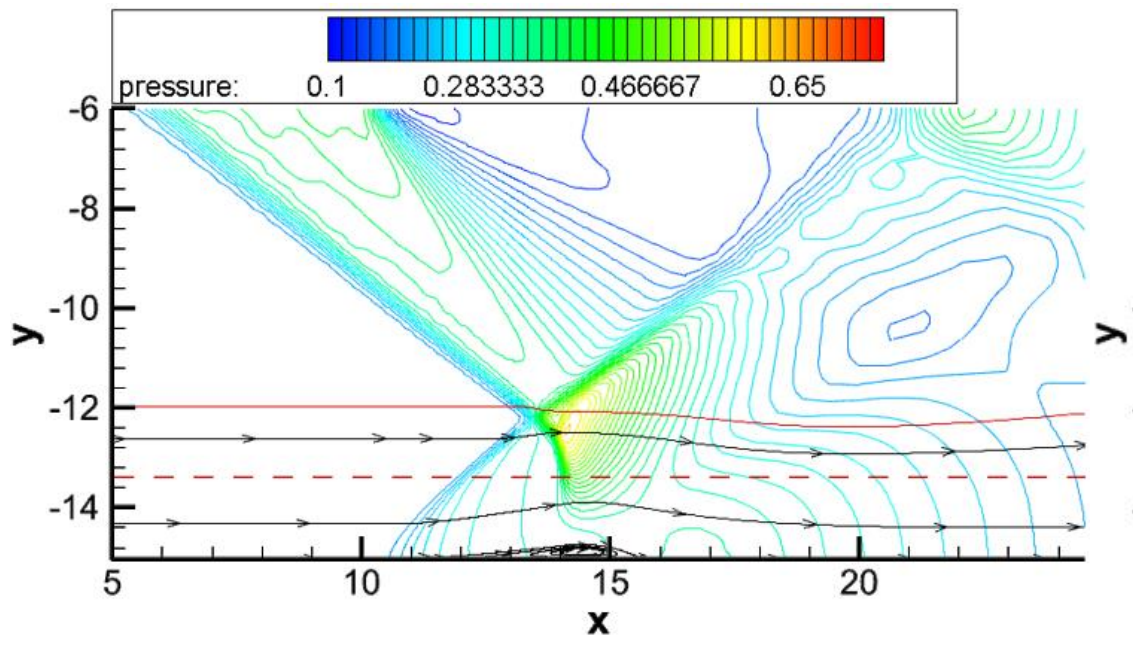

(c) Pressure contours at AOA = 10 °

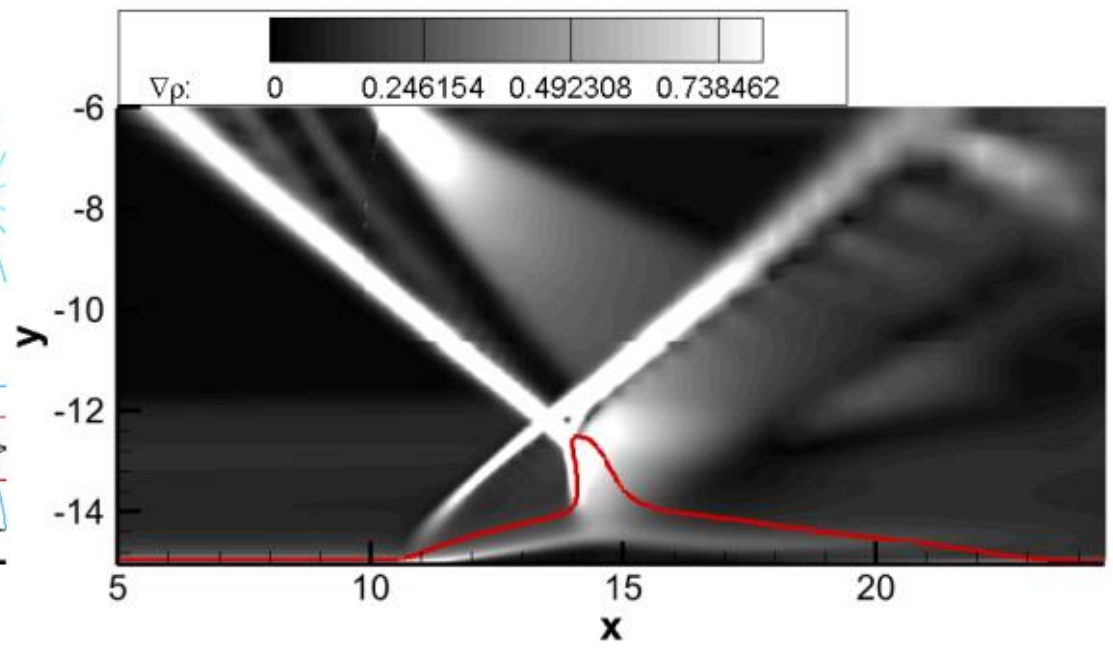

(d) Density gradient contours at AOA = 10 °

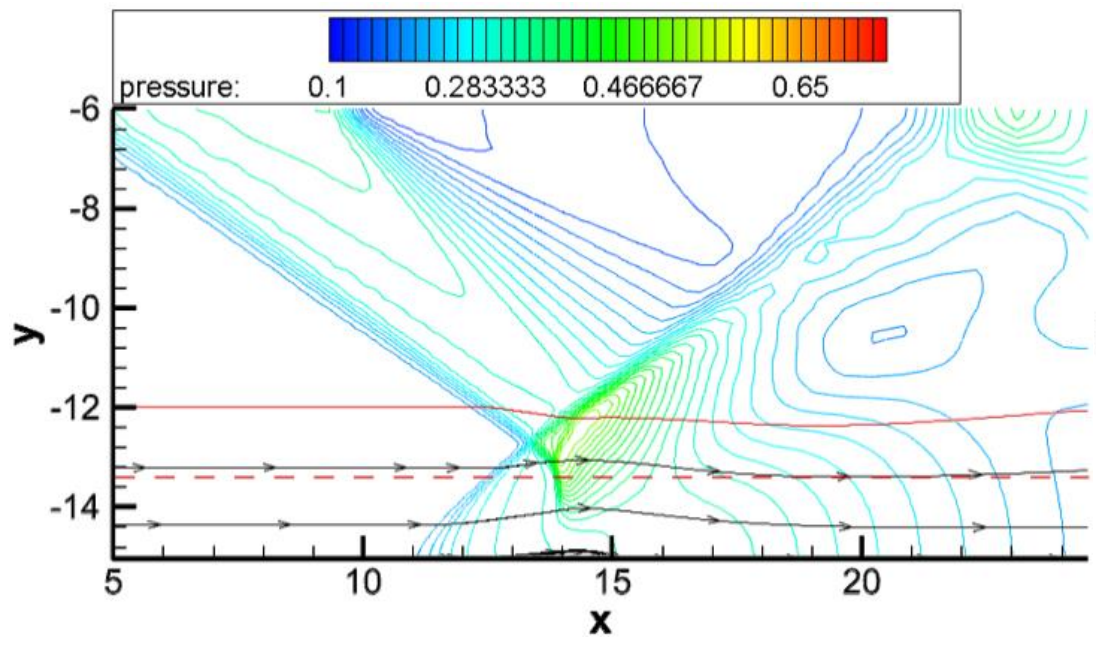

(e) Pressure contours at AOA = 5 °

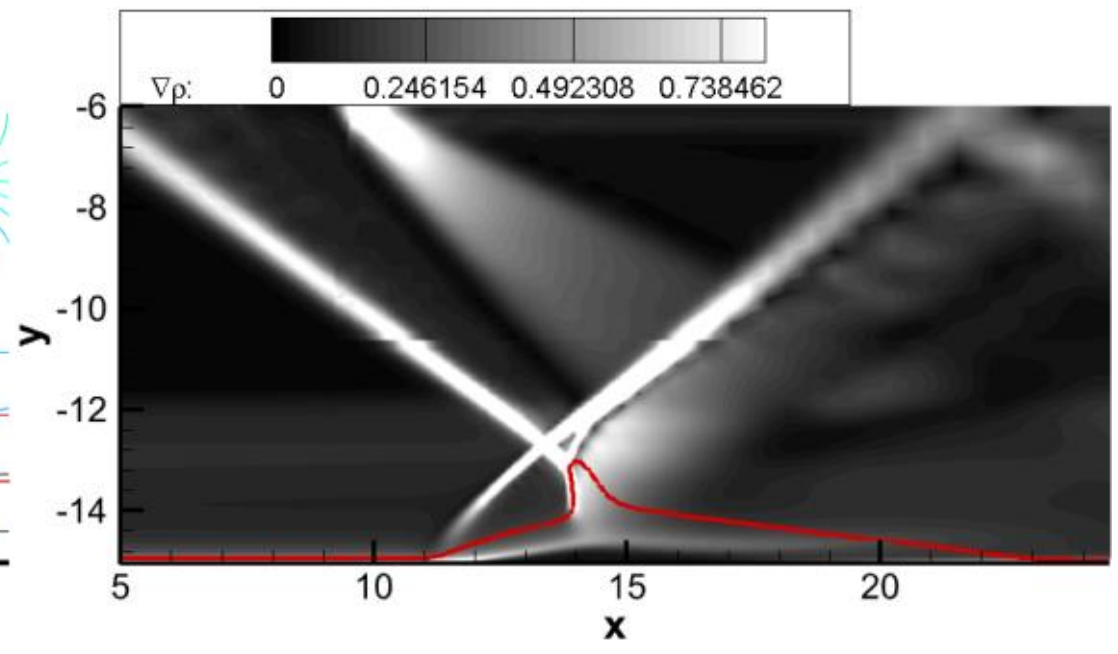

(f) Density gradient contours at AOA = 5 °

Fig. 23 Pressure and density gradient contours in the windward meridional plane in three cases with different AOAs. In (a), (c) and (e), red solid lines denote the boundary layer edge, dashed red lines show the variable distributions at $y = -13.4$ cm passing through the Mach stem-like

structure, the two upper and lower streamlines are chosen to pass through the top and bottom of the stem, respectively, and streamlines are drawn to indicate the separation bubbles. In (b), (d), and (f), the sonic lines are drawn in red

(2) Variable distributions at $y = -13.4$ cm in the windward meridional plane at three AOAs

The Mach number distributions at $y = -13.4$ cm in the meridional plane at three AOAs are shown in Fig. 24. It is known that the intensity of an incident conical shock will decrease as the AOA decreases, and the figure shows similar distributions to those found for the 2D interactions (see Fig. 12); i.e., the Mach number suddenly drops when the flow encounters the stem, after which a quick recovery occurs due to the expansion wave. To better understand the Mach number distribution, we mark the positions of the separation shock wave (SS) and the Mach stem-like structure (MS) in Fig. 24 in the case AOA = 15°. A careful comparison with the corresponding 2D case shows that the distance between the separation shock and the stem here is significantly shorter and the Mach number changes more abruptly; in addition, the Mach number continues to increase throughout the range of computation, in contrast to the evolution toward a platform distribution in the 2D interaction.

The pressure distributions at $y = -13.4$ cm are shown in Fig. 25. It can be seen that two jumps exist: the first is due to the separation shock wave, and the second arises from the stem and is accompanied by a rapid drop. In contrast to the subsequent platform profile in the 2D interaction (Fig. 13), the downstream pressure in 3D decreases continuously under the influence of expansion waves. In Fig. 26, we take the case AOA = 15° as an example and display the pressure distribution at $y = -13.4$ cm together with the wall pressure to establish possible correlations. Similarly, with the aim of revealing correlations, we mark positions such as that of the separation shock foot (SF) in the figure, as we did in Fig. 14. In the region from 0 to $x = 13.4$ cm, the pressure increases in a similar way for both curves, but, owing to the effect of the stem, the pressure at $y = -13.4$ cm increases significantly afterward and exhibits a peaked profile. Unlike the 2D cases in Fig. 14, the two distributions downstream of separation gradually coincide as they decrease, indicating their approximate consistency.

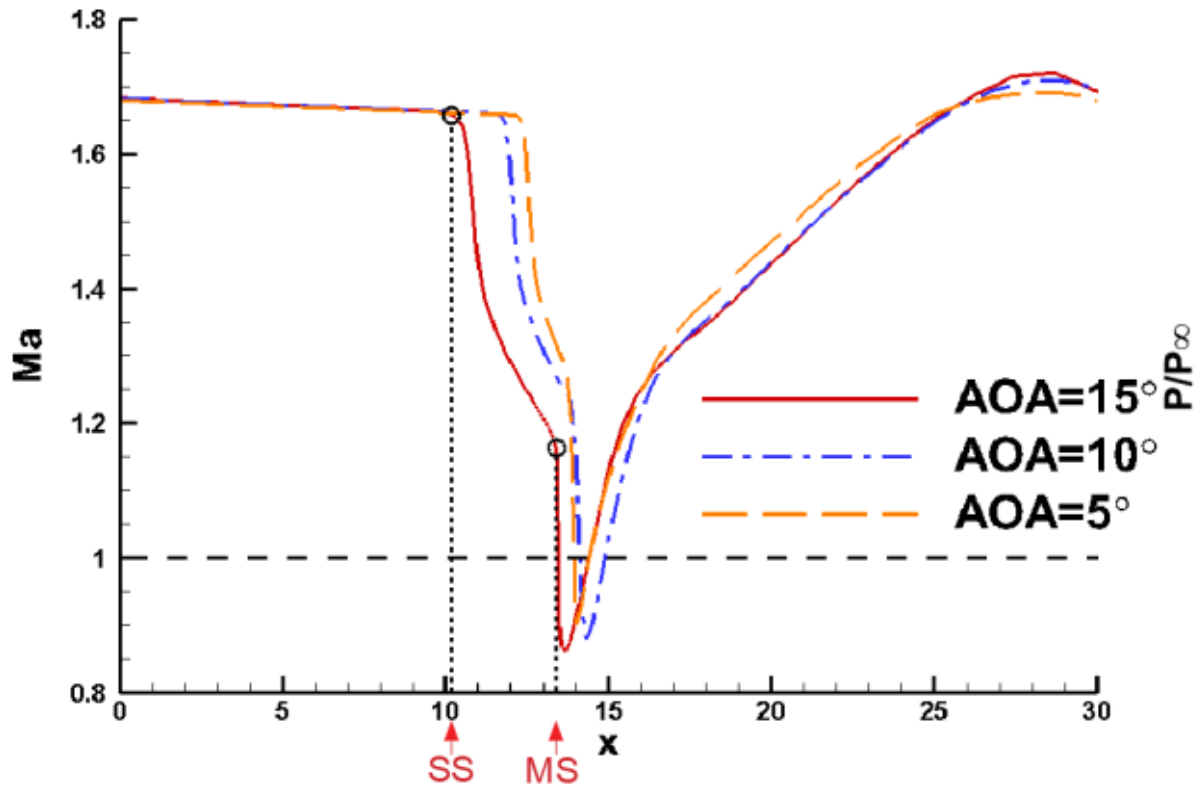

Fig. 24 Mach number distributions along the line $y = -13.4$ cm at three AOAs with characteristic positions marked

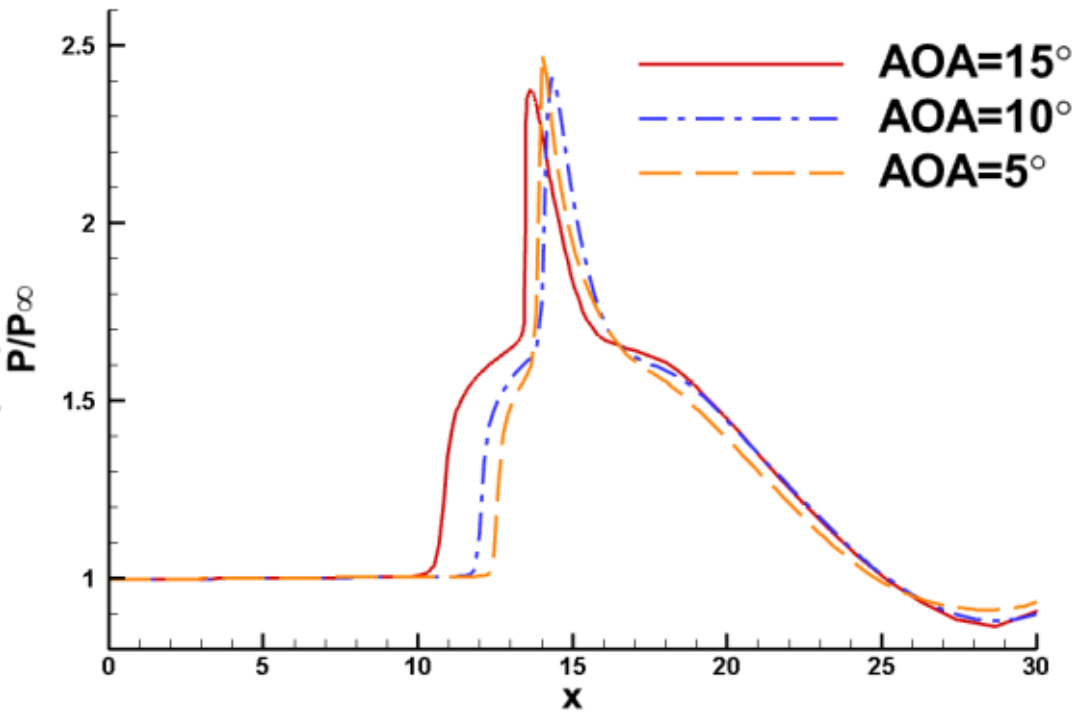

Fig. 25 Pressure distributions along the line $y = -13.4$ cm at three AOAs

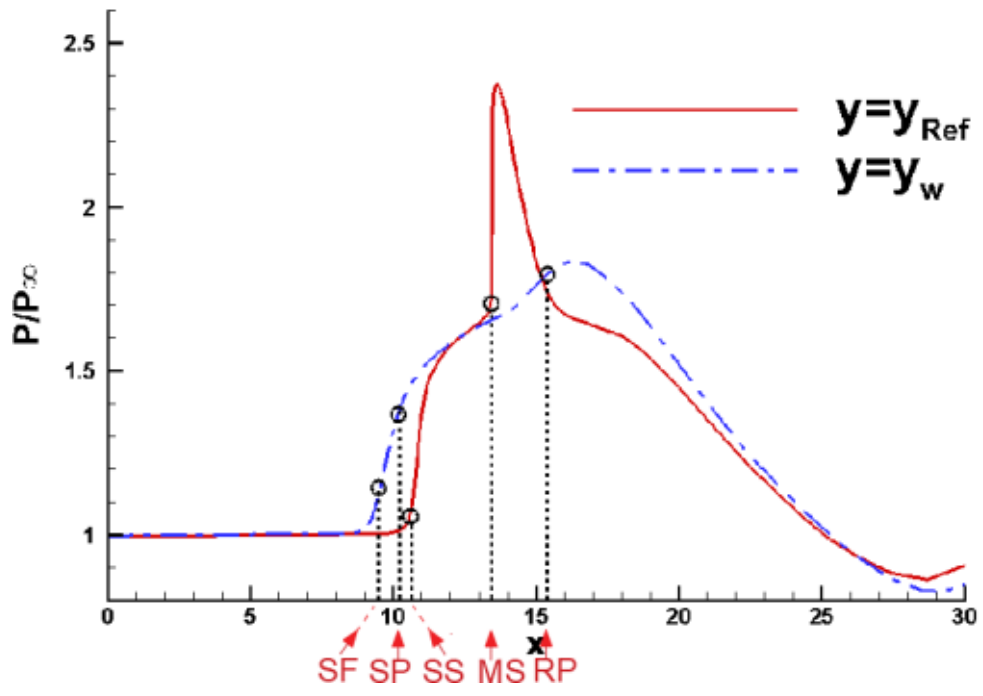

Fig. 26 Comparison of the pressure distributions at the line $y = -13.4$ cm and the wall at AOA = 15 °with characteristic positions marked

(3) Flow structures in various radial sections with respective circumferential angles at AOA = 15°

To understand the circumferential variations of the flows, two other sets of the pressure and density gradient contours are extracted and analyzed on radial sections with azimuthal angles $\varphi$ =

45° and 90° as representatives, and the results are shown in Fig. 27. From a comparison of the windward meridional results in Fig. 23(a) and (b) with those in Fig. 27, it can be seen that the strength of the shock wave on the radial section decreases with increasing $\varphi$, as do the sizes of the separation shock and the stem. Detailed examination shows that in sections where $\varphi$ = 0° and 45°, a clear Mach reflection-like event is revealed by the sonic line in the density gradient contours, where the apex formed by the sonic line coincides with the triple point at the top of the reflection; at $\varphi$ = 90°, the reflection is not obvious according to the sonic line, and the flow after the stem-like structure is not subsonic but supersonic. In the figure, the streamlines are drawn under the stem, which shows that although separations appear on the wall, the conventional winding behavior of the vortex is not observed.

In Figs. 28 and 29, we show the Mach number and pressure distributions along the line at $y$= −13.4 cm in sections with $\varphi$ = 0°, 45°, and 90° in the case AOA = 15°. These figures reveal that the peak pressure and minimum Mach number at $\varphi$ = 0° and 45° exhibit small differences, and the variable distributions are similar; however, the distributions of the Mach number and pressure at $\varphi$ = 45° and 90° are quite different, which reveals the existence of an obvious 3D effect at large $\varphi$.

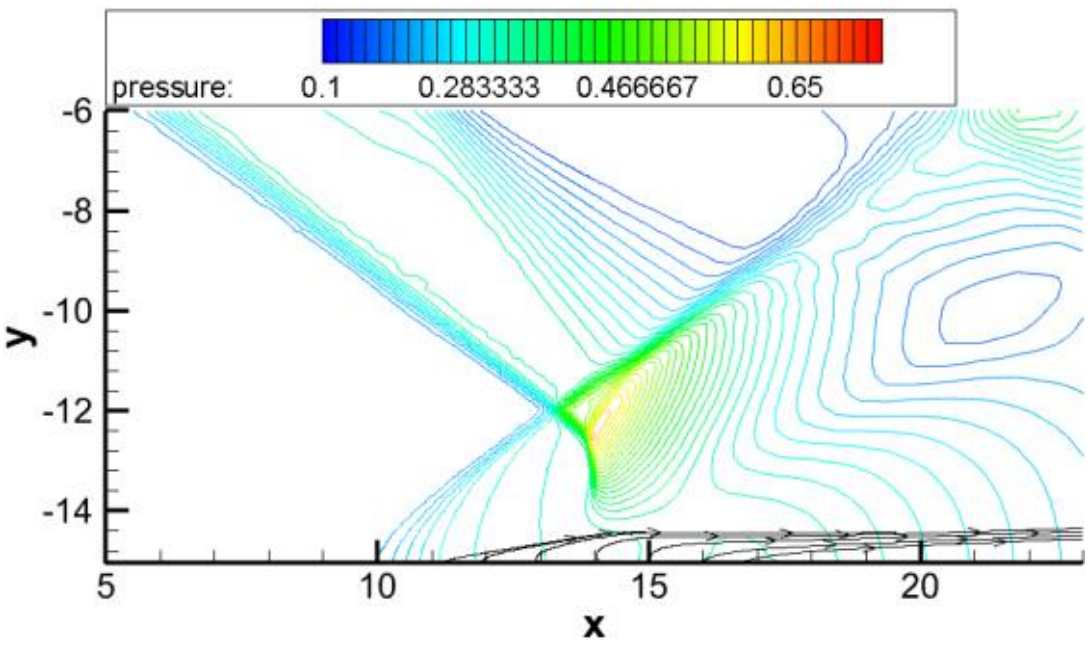

(a) Pressure contours at $\varphi$ = 45°

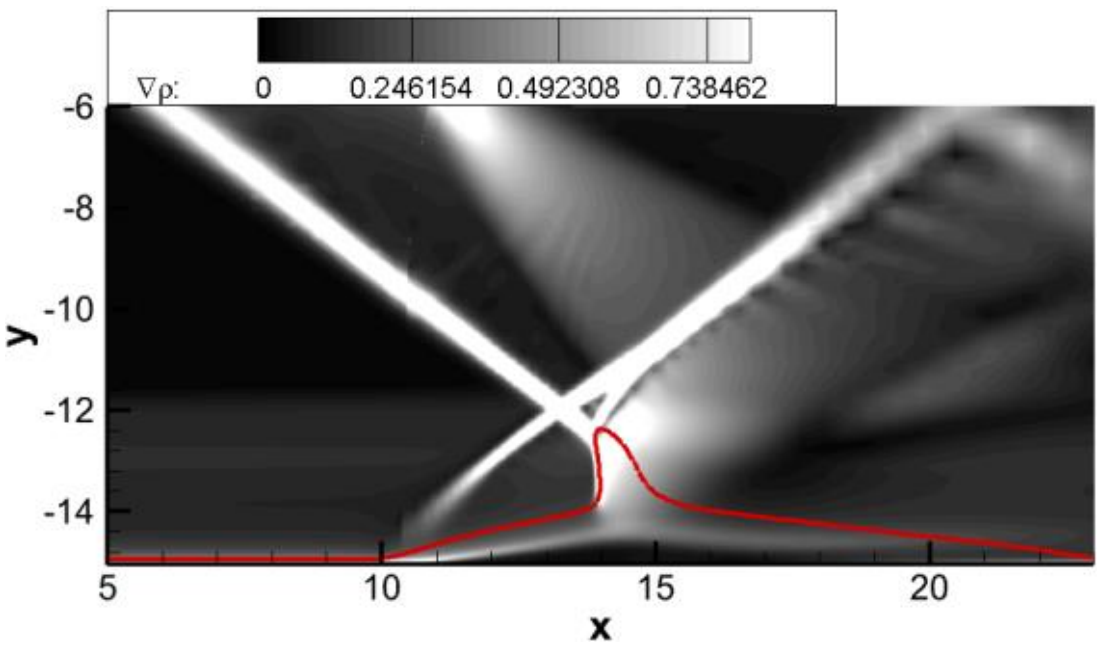

(b) Density gradient contours at $\varphi$ = 45°

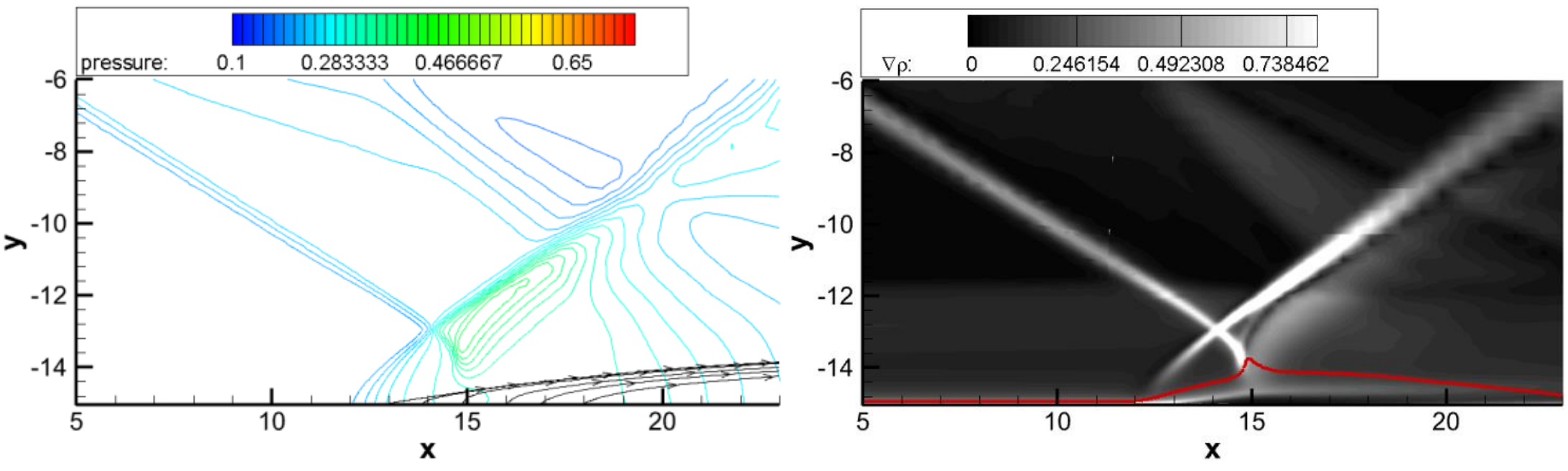

(c) Pressure contours at $\varphi = 90°$ (d) Density gradient contours at $\varphi = 90°$

Fig. 27 Contours of pressure and density gradient at different azimuthal angles in the case AOA = 15 °. On the pressure contours, streamlines are shown in the interaction region; on the density gradient contours, the sonic lines are drawn in red

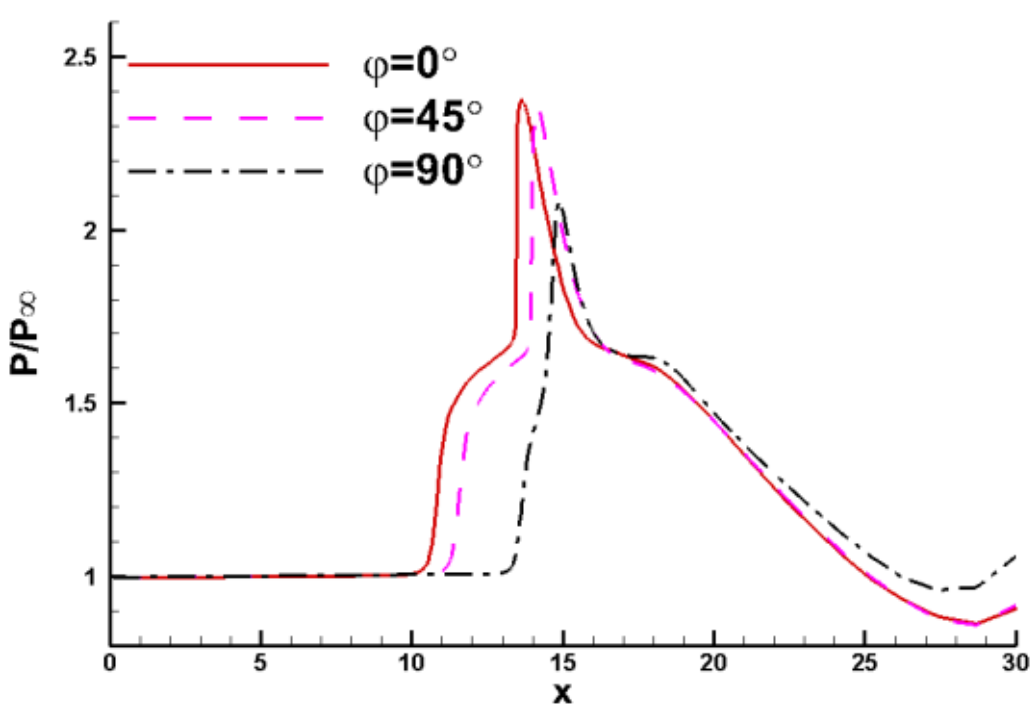

Fig. 28 Pressure distributions at the line $y = -13.4$ cm at different azimuthal angles in the case AOA = 15 °

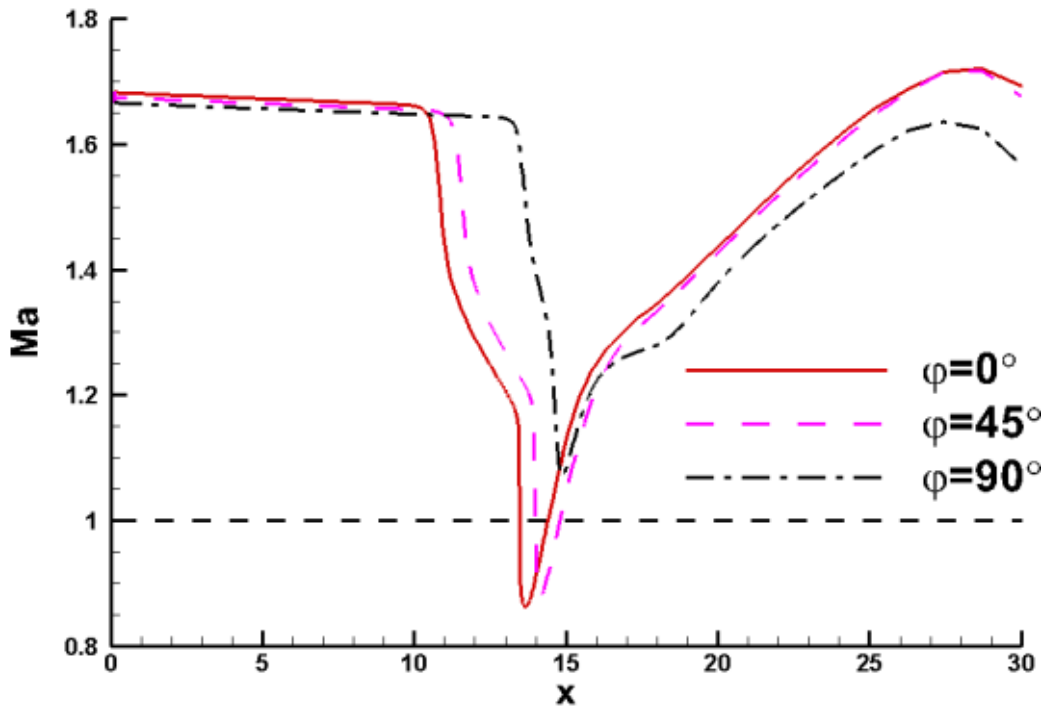

Fig. 29 Mach number distributions at the line $y = -13.4$ cm at different azimuthal angles in the case AOA = 15 °

(4) Flow structures circumferentially unfolded in a torus at $r = 10.45$ cm at different AOAs and 3D appearance of Mach stem-like structure in meridional plane

From our above investigation of the stem in the windward meridional plane, we have found that the stem has a 3D configuration that appears as a curved surface circumferentially. To analyze the 3D appearance further, an annulus with a certain radius $r$ is chosen in which the flow structures are visualized using pressure contours. To facilitate this analysis, the annulus is further expanded into a plane with the circumferential angle as abscissa, and the results are shown in Fig.

30, where $r$ = 10.45 cm. In this figure, “MS” and “SS” denote the stem and the separation shock. With regard to the structure near $\varphi = 0°$, it can be seen that the separation shock is curved and is not axisymmetric in the circumferential direction; in addition, a very distinct shock wavefront is exhibited within a range of azimuthal angles (e.g., [−120°, 120°] in the case of AOA = 15°). By comparison, the curvature related to the stem is rather mild near $\varphi = 0°$; in particular, the stem appears straighter with decreasing AOA, which corresponds to the existence of axisymmetry in 3D space. Fig. 30 also shows that the stem is most fully intact at $\varphi = 0°$, gradually becomes obscured with increasing $|\varphi|$, and finally disappears after meeting the separation shock wave. In short, during their existence, the separation shock structure exhibits a curved geometry in the circumferential direction, while the stem has a relatively axisymmetric distribution, which implies the 3D nature of the former and the 2D nature of the latter.

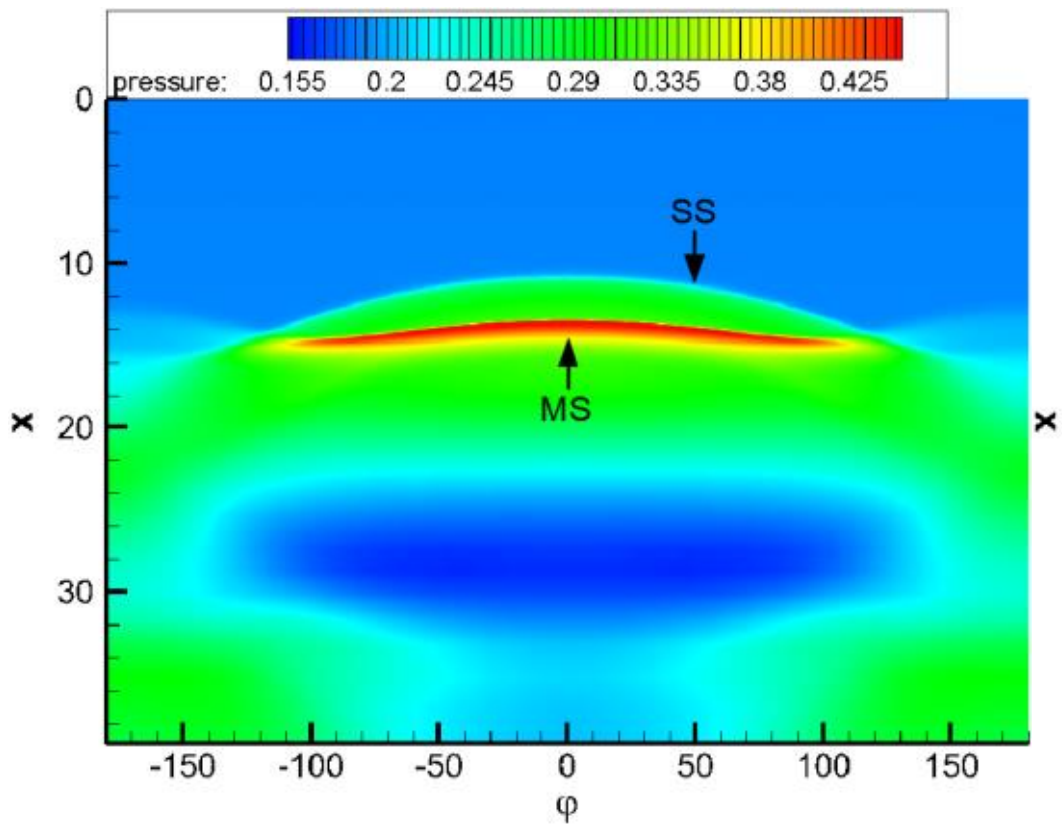

(a) AOA = 15 °

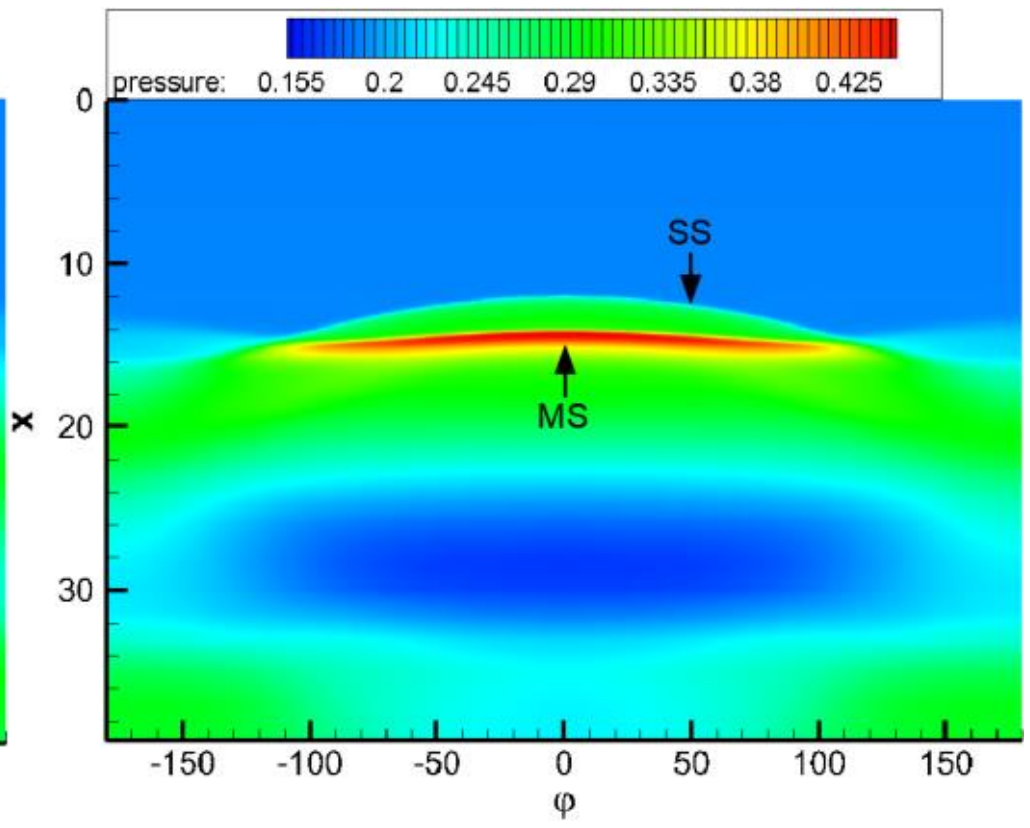

(b) AOA = 10 °

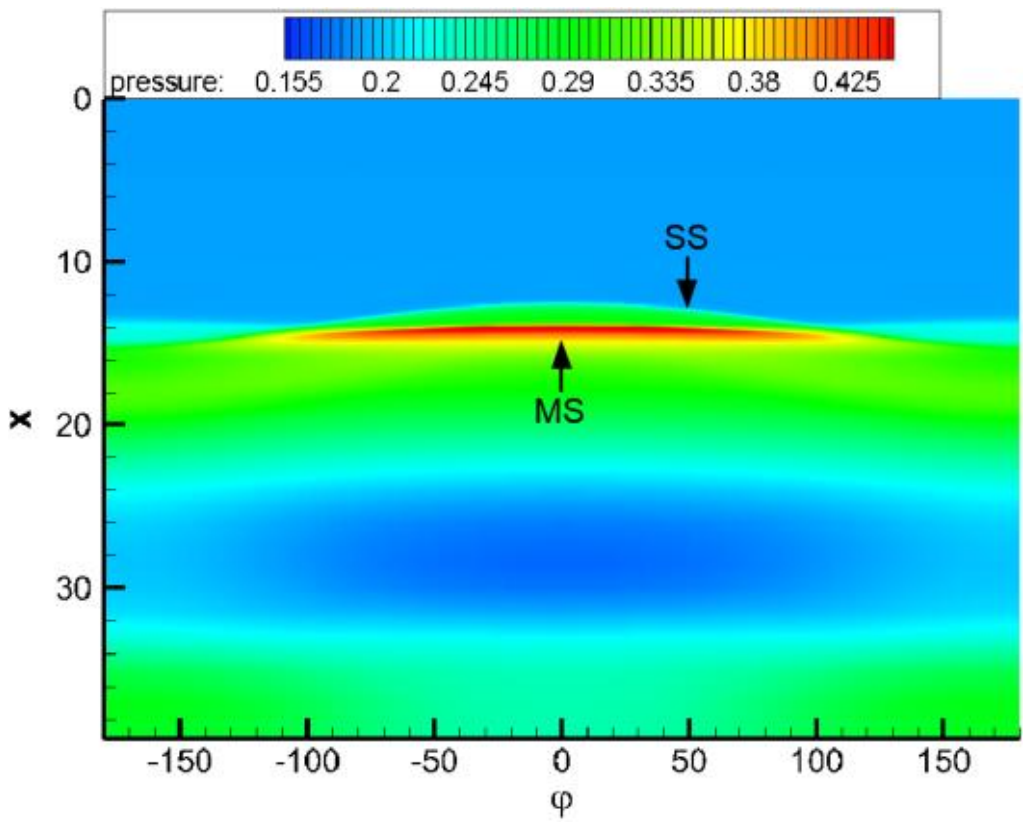

(c) AOA = 5 °

Fig. 30 Pressure contours in a torus with $r$ = 10.45 cm for three different AOAs. "SS" denotes the separation shock and "MS" the Mach stem-like structure

Based on the above discussion and recalling the discussion in Section 3.1.2, we suspect that the stem, especially around $\varphi$ = 0°, will be close to being a local planar oblique shock, and the degree to which it departs from such a shock can be estimated by comparing the computed result with that predicted by the Rankine–Hugoniot formulae. Concretely, considering the stem in the windward meridional plane, the variables such as the Mach number at selected locations after the structure are predicted from the flow in front of it. For selection of these locations and for theoretical prediction, the ranges of $y$ coordinates of the stem at different AOAs are found to be $y \in [-12.1,–13.8]$ at AOA = 15°, $y \in [-12.5, -13.9]$ at AOA = 10° and $y \in [-13.1, -14.0]$ at AOA = 5°. The predicted Mach numbers at the selected locations after the stem are shown in Table 7, together with the relative error in the computation. The results show that for these locations, the $M_2$ from the computation is very close to the predicted $M_{2,\mathrm{pred}}$, which indicates that the chosen part of the stem quite closely resembles a planar shock wave.

Table 7. Mach numbers before and after the stem from the computation, predicted Mach numbers, and relative errors at selected locations in 3D interactions, with $y_{\mathrm{stem}}$ indicating the $y$ range of the stem

| AOA | $y_{\mathrm{stem}}$ (cm) | $y$ (cm) | $M_1$ | $M_{2,\mathrm{pred}}$ | $M_2$ | $\frac{\lvert M_2 - M_{2,\mathrm{pred}}\rvert}{M_2}$ (%) |
|---|---|---|---|---|---|---|
| 15 ° | [−12.1, −13.8] | −12.6 | 1.445402 | 0.769358 | 0.775991 | 0.8548 |
| | | −13.0 | 1.305897 | 0.825534 | 0.810547 | 1.8490 |
| | | −13.4 | 1.166025 | 0.902197 | 0.862019 | 4.6609 |
| | | −13.6 | 1.095891 | 0.951393 | 0.866140 | 9.8425 |
| 10 ° | [−12.5, −13.9] | −13.0 | 1.357561 | 0.789287 | 0.866670 | 8.9288 |
| | | −13.4 | 1.217275 | 0.859140 | 0.879806 | 2.3489 |
| | | −13.8 | 1.093205 | 0.942329 | 0.876862 | 7.4661 |
| 5 ° | [−13.1, −14.0] | −13.6 | 1.200235 | 0.871397 | 0.903139 | 3.5146 |
| | | −14.0 | 1.068889 | 0.963839 | 0.889870 | 8.3123 |

The investigations above regard flow structures on an $r$ = 10.45 cm torus together with their 3D characteristics. Naturally, one wonders what would happen on another typical annulus, namely, the wall, and this will be discussed in Section 3.2.3 (3).

3.2.3 Vortex structure from different perspectives and separation topology of 3D interaction

(1) Flow separation and vortex structure in the windward meridional plane

Investigations are first carried out in the windward meridional plane. Fig. 31 shows the separation bubbles at various AOAs, with the separation point SP and reattachment point RP marked. It can be seen that the bubble becomes smaller with decreasing AOA, the streamlines in the separation zone rotate and converge toward the center, and the separation streamline (red) originating from SP lies outside the reattachment streamline (green) ending at RP without them meeting. The consequent separation bubble is not closed, which indicates that the flow between the two lines will spill out laterally and lead to a typical 3D effect. In contrast to this behavior, SP and RP in the 2D interaction are connected by a single limiting streamline and an enclosed bubble is formed, as a consequence of which no mass leakage occurs and the separation bubble is accordingly larger.

Table 8 lists the dimensions of the separation bubbles in the 2D and 3D interactions, where the streamwise sizes can be measured separately using the friction force distribution and streamlines, and where the circumferential size and the height of the separation bubble are measured by streamlines. Because the two streamwise dimensions are quite similar, only that from the streamlines is shown. From the table it can again be seen that although the height and streamwise dimension decrease with AOA/half-cone angle, the scales in the 3D interactions are smaller than those in their 2D counterparts. This might be related to lateral overflow, which will be discussed in detail in Section 3.2.4.

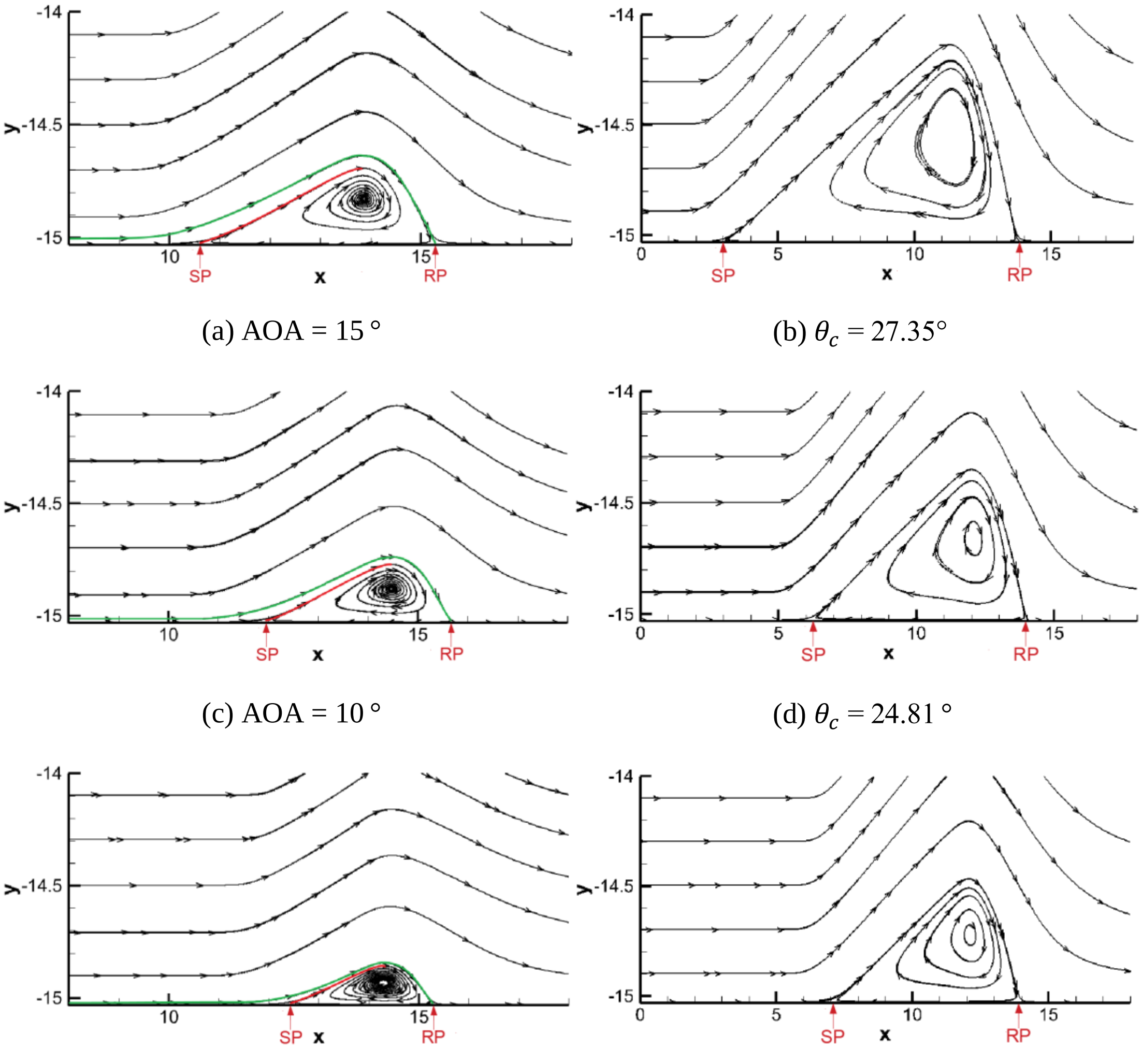

(a) AOA = 15 ° (b) $\theta_c$ = 27.35°

(c) AOA = 10 ° (d) $\theta_c$ = 24.81 °

(e) AOA = 5 ° (f) $\theta_c$ = 20.91°

Fig. 31 Structures of separation bubble shown by streamlines in the meridional plane at various AOAs/half-cone angles. (a), (c), and (e) are for 3D interactions, with separation streamlines in red and reattachment streamlines in green. (b), (d), and (f) are for 2D interactions

Table 8. Dimensions of separation bubble at various half-cone angles/AOAs in 2D/3D interactions

| | Dimensions of separation bubble (cm) |
|---|---|

| Half-cone angle/AOA | Streamwise at $\varphi = 0^0$ | Circumference | Height |
|---|---|---|---|
| $\theta_c$ = 27.35° (2D) | 11.4551 | — | 0.8903 |
| $\theta_c$ = 24.81 °(2D) | 8.5300 | — | 0.6815 |
| $\theta_c$ = 20.91° (2D) | 7.5823 | — | 0.5639 |
| AOA = 15 °(3D) | 5.3548 | 51.89 | 0.3959 |
| AOA = 10 °(3D) | 4.5195 | 53.99 | 0.2952 |
| AOA = 5 °(3D) | 3.4636 | 67.49 | 0.1868 |

(2) 3D structure of vortex

To understand the fully 3D architecture of the vortex rather than just that of the 2D separation bubble in the meridional plane, the commonly used $Q$ criterion is employed, and a $Q$-isosurface is shown in Fig. 32 for visualization of the vortex structure. To illustrate the flow behavior from a different angle, the insets in the figure show the 3D streamlines, the origins of which lie in a plane near the windward meridian. It can be seen that the $Q$-isosurface extends along the circumference and gradually tapers until it disappears, which indicates that the vortex and the associated separation weaken with increasing azimuthal angle. In addition, with decreasing AOA, the circumferential span of the vortex increases, which implies that the separation region in the wall is expanding in circumference. This implication is verified in the following examination of the behavior of the circumferential length of separation with variations in AOA.

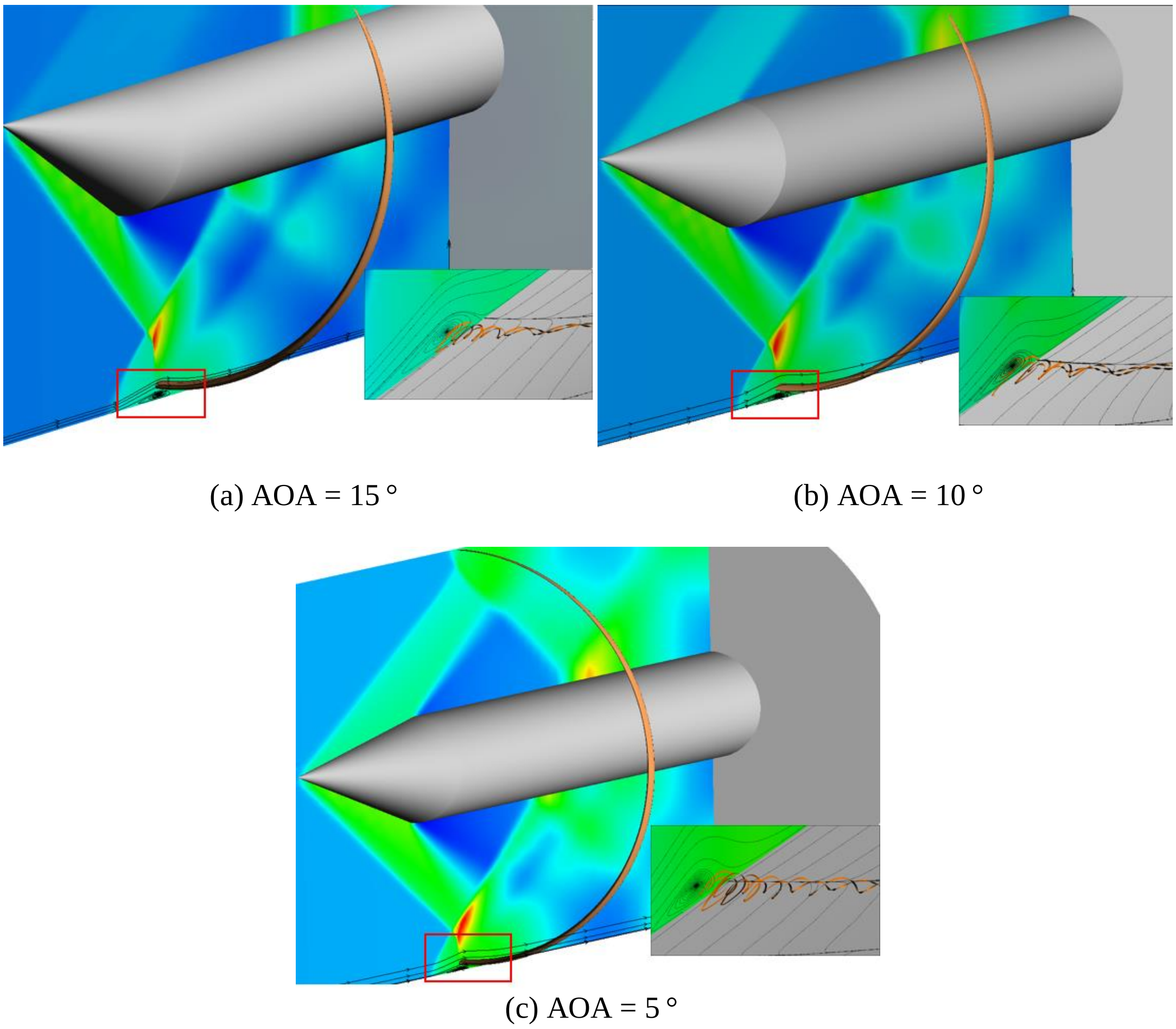

Fig. 32 Vortex structure visualized by $Q$-isosurface. The insets show 3D streamlines in local views at three AOAs

(3) Surface separation topology and circumferential scale

Examination of surface separation topology is an effective way to unveil the complexity of vortex structures. Following this approach, we unfold the annular wall into a plane with the azimuth as abscissa so that the original circumferential topology can be shown more clearly and in greater detail. In Fig. 33, the limiting streamlines on the wall are drawn for three different AOAs, and it can be seen that the separation and reattachment streamlines first extend circumferentially and then approach each other, finally converging into one stream trace. Such an evolution is qualitatively consistent with the experimental result shown in Fig. 3(b). With decreasing AOA, the angle of the conical shock wave as well as its strength decrease in the azimuthal region of 0°–90°, which leads to attenuation of the adverse pressure gradient, a weakening of flow separation, and a shortening of the streamwise separation length; on the other hand, in the azimuthal region of 90°–180°, the incident angle of the conical shock wave increases, the pressure behind the wave rises, and the adverse pressure gradient increases accordingly, which provokes an increase in separation circumferentially. The scales of separation at different AOAs can be found in Table 8. From the above discussion, it can be seen that as the AOA decreases, the unseparated region confined by the two merged streamlines shrinks and finally disappears. Thus, it can be inferred that there is a critical AOA at which the merged streamlines just connect and the circumferential scale of the separation is just equal to the tube circumference.

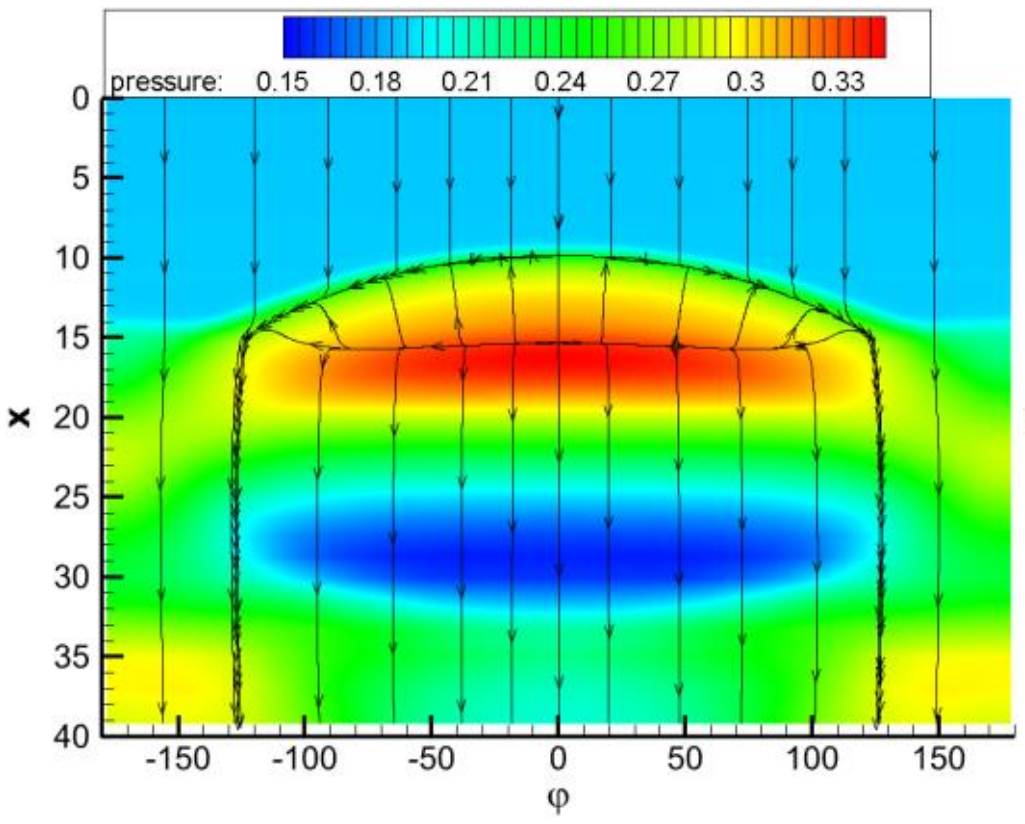

(a) AOA = 15 °

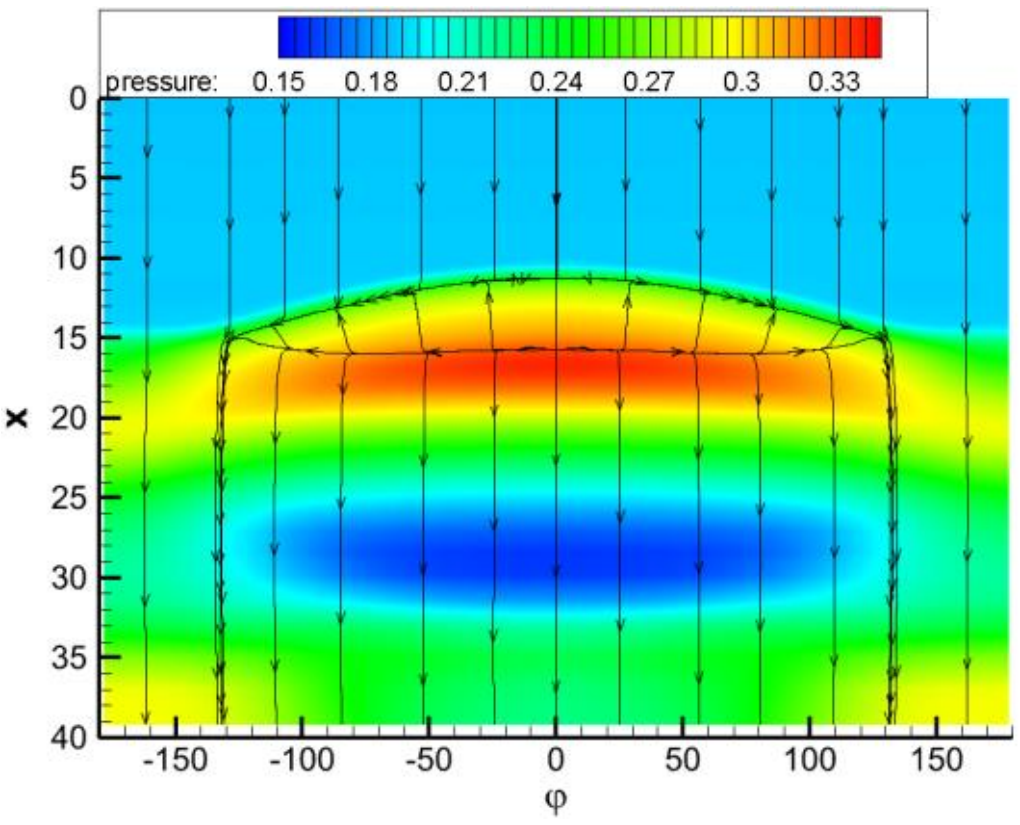

(b) AOA = 10 °

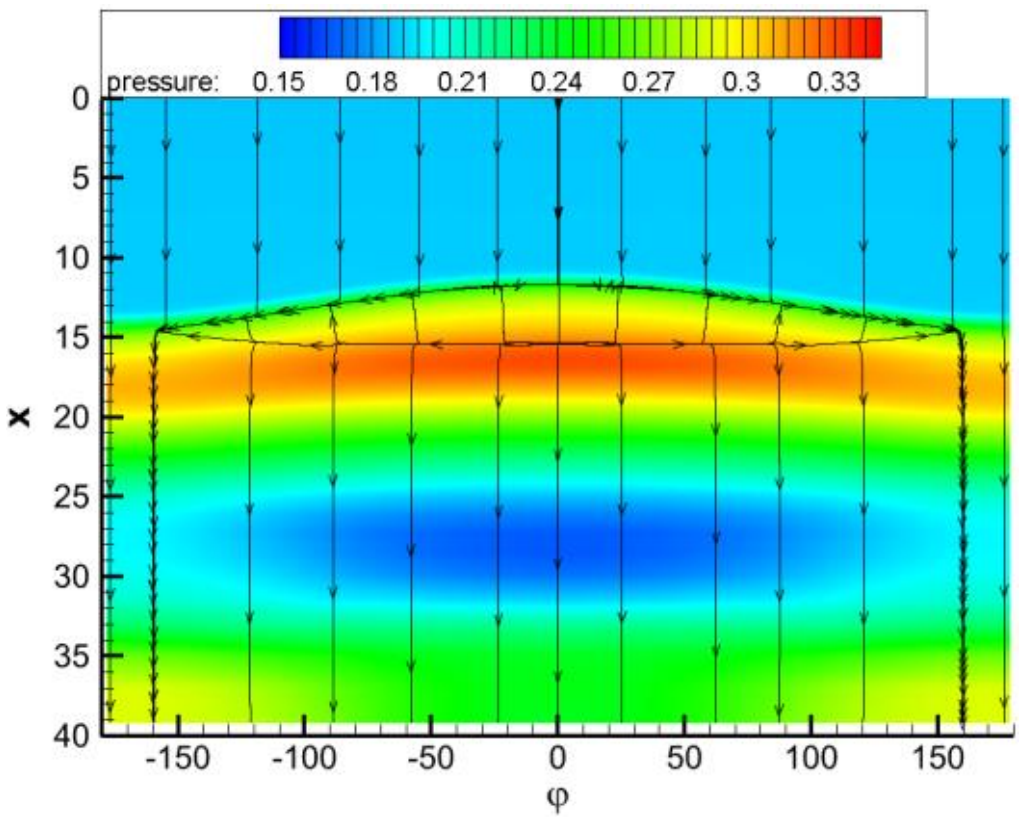

(c) AOA = 5 °

Fig. 33 Surface separation topology as illustrated by limiting streamlines against a background of isobar contours at three AOAs

To obtain the corresponding critical angle, a simple model is established based on the circumferential length $L$ of the unseparated region at three AOAs. Specifically, taking $L$ to be a function of AOA, the following empirical fitting can be obtained:

$$L = \frac{26.91AOA - 119}{AOA - 3.158} \tag{7}$$

From this equation, the critical AOA can be found to be approximately 4°. Based on this solution, we generate the corresponding 3D grids and carry out numerical simulations, from which the surface separation is acquired as shown in Fig. 34. It can be seen from this figure that at this time there is a quite small piece of undivided separation that is about to vanish near $\varphi = 180°$ with no fusion of the separation and reattachment streamlines, which indicates a critical state of separation topology.

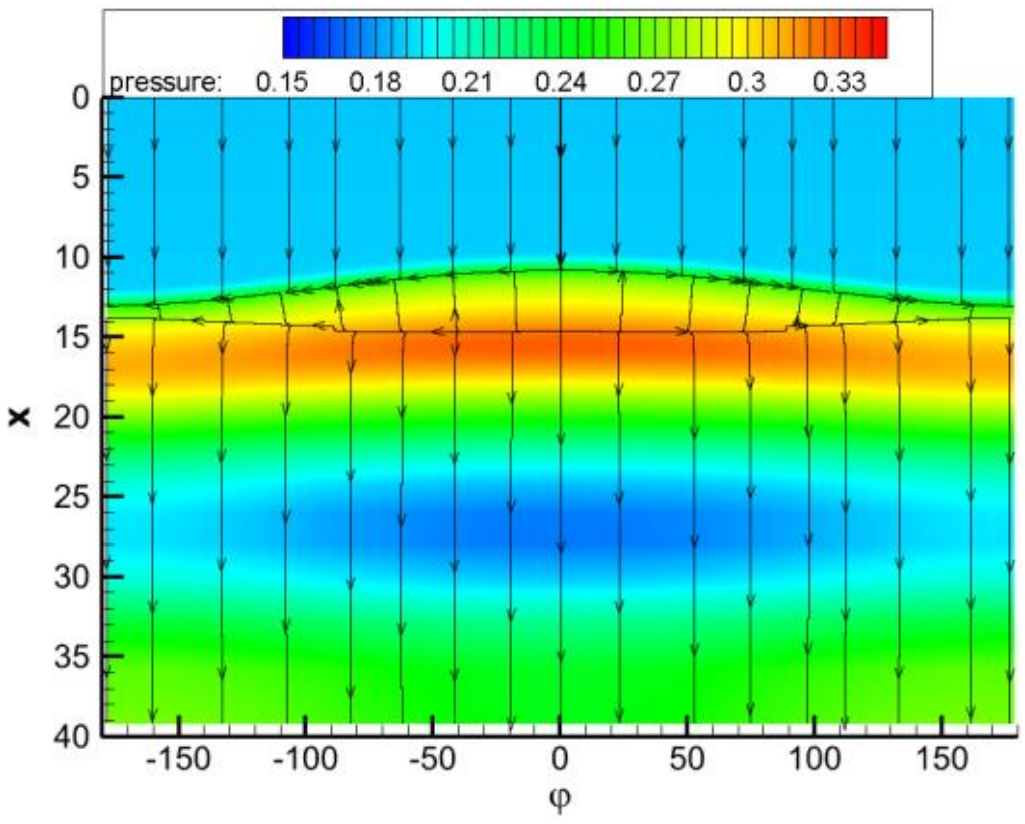

Fig. 34 Limiting streamlines against a background of isobar contours at AOA = 4 °

In Section 3.2.2 (4), the wave structure at a torus with $r$ = 10.45 cm was studied, and here a similar analysis is performed for the pressure distributions on the wall. In Figs. 33 and 34, the pressure contours indicate the following: in the streamwise direction, the pressure rises adversely owing to separation and reaches a maximum downstream of the nearby reattachment streamline, whereas in the circumferential direction, the pressure decreases from a maximum at $\varphi = 0°$ with increasing azimuth, and therefore a positive pressure gradient exists circumferentially and contributes to alleviating the streamwise pressure rise near the meridional plane.

3.2.4 Quantitative analysis of side overflow at the meridional plane in the case AOA = 15°

In Figs. 31–33, the separation on the windward meridian indicates that the flow between the separation and the reattachment streamlines will overflow laterally and exhibit typical 3D behavior. To study this phenomenon, the flow at AOA = 15° is chosen as a representative example and the windward meridian plane is partitioned by series of streamlines vertically equidistant in the upstream of interaction, after which the mass flow rate of each divided section is calculated at four streamwise locations, enabling the variation in mass flow to be examined. Fig. 35 shows a schematic of the partitioning of the flow field, in which the space within the boundary layer is divided vertically into five sections [see Fig. 35(b)] and four characteristic streamwise locations are selected [see Fig. 35(a)]. The $x$ coordinates of these four locations correspond approximately to locations before the separation point, at the separation point, at the top of the bubble, and at the reattachment point, respectively. In particular, location 1 is ahead of the separation shock wave, and the boundaries of section 1 are chosen to be the reattachment and separation streamlines; the other sections at location 1 have the same vertical interval as that of section 1. From Fig. 35, it can be seen that streamlines equidistant from location 1 are distributed in a parallel manner upstream of the interaction, which implies that there is no lateral overflow in that region.

Based on this spatial partition, the mass flow rates of each section at the four streamwise locations are listed in Table 9, from which it can be seen that (1) the reduction in the mass flow rate between locations 1 and 4 in different sections indicates the presence of an overall side overflow, (2) the side overflow in sections close to the wall is more pronounced than that in sections farther away from the wall, and (3) as the height of a section increases the mass flow rates at the four locations increase.

The differences in mass flow rate between locations 1 and 4 in each section, $\Delta\dot{m} = \dot{m}_1 - \dot{m}_4$, are calculated to evaluate the overall side overflow. The results are shown in Fig. 36, from which it can be seen that on going from section 1 to section 5, the curve gradually decreases and tends to $\Delta\dot{m} = 0$, which shows that the overflow attenuates farther away from the wall. It is to be expected that at a certain $y$ position, $\Delta\dot{m}$ will be near zero and side overflow will be negligible.

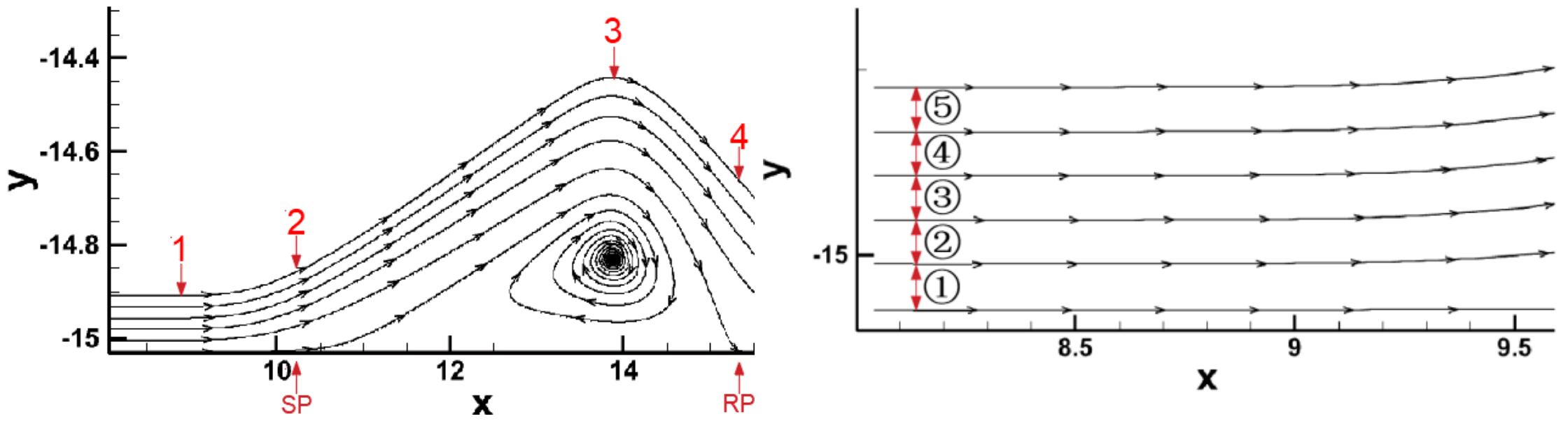

(a) Four selected streamwise locations (b) Five partitioned sections at location 1

Fig. 35 Schematic of spatial partition of the flow in the windward meridional plane at AOA = 15 °

Table 9. Mass flow rates at four streamwise locations in five sections

| Section | Mass flow rates at four streamwise locations $(\mathrm{kg/(m^2 \cdot s)} \times 1000)$ | | | |
|---|---|---|---|---|
| | 1 | 2 | 3 | 4 |
| 1 | 0.7377144 | 0.7106874 | 0.5914243 | 0 |
| 2 | 1.1487160 | 1.1406800 | 1.0703970 | 0.9379403 |
| 3 | 1.2968100 | 1.2932570 | 1.2645130 | 1.1765790 |
| 4 | 1.3939970 | 1.3930170 | 1.3873460 | 1.3146670 |
| 5 | 1.4676640 | 1.4683690 | 1.4730960 | 1.4127730 |

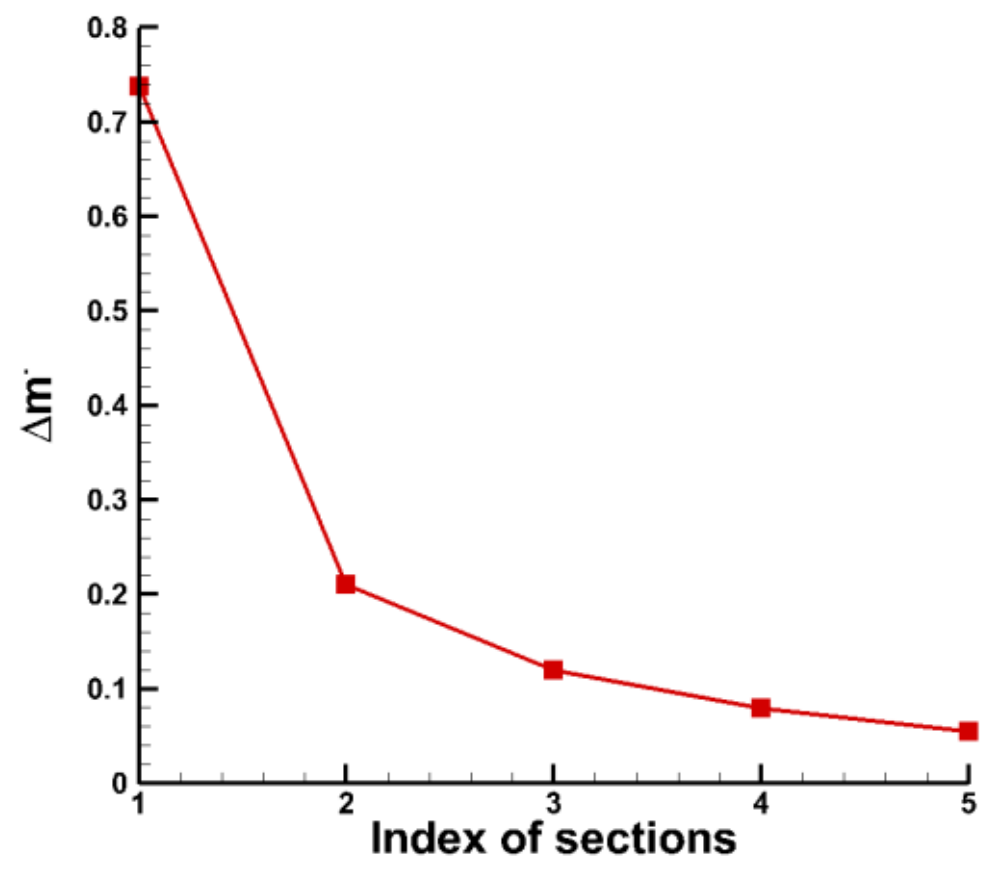

Fig. 36 Difference in mass flow rate between locations 4 and 1 with respect to section number

### 3.3 Comparisons of wall pressure and friction force distributions for 3D interaction at $\varphi = 0°$ and 2D interaction

In Section 3.2, a qualitative comparison was made between the 2D axisymmetric interaction and the 3D interaction. In the present subsection, a quantitative examination is made of the wall pressure and friction force distributions of the 3D interaction at $\varphi = 0°$ and the 2D interaction.

Figs. 37 and 38 show the pressure and friction force distributions, respectively. The results for the 3D and 2D interactions are displayed using the same line type and color, with the following correspondence between the AOA in 3D and the half-cone angle in 2D: $15°(3D) \leftrightarrow 27.35°(2D)$, $10°(3D) \leftrightarrow 24.81°(2D)$, $5°(3D) \leftrightarrow 20.96°(2D)$. The results in Fig. 37 show that the 2D separation point is located farther upstream than that for the 3D interaction at $\varphi = 0°$, and the peak value of the wall pressure is also higher for the 2D interaction, which indicates the stronger adverse pressure gradient in the latter case. A possible reason for this is that in the 3D interaction, the separation bubble is no longer closed, owing to the existence of side overflow, and a positive pressure gradient arises in the circumference (see Fig. 33), yielding a weaker adverse pressure gradient along the wall and thus reduced flow separation.

From the friction force distributions in Fig. 38, the length of the separation can be obtained by measuring the lengths of the negative friction zones, and this again confirms the smaller separation in 3D interactions than in the 2D case. Fig. 38 also shows that although the starting positions of the 3D and 2D separations differ greatly, the differences between their end positions are small.

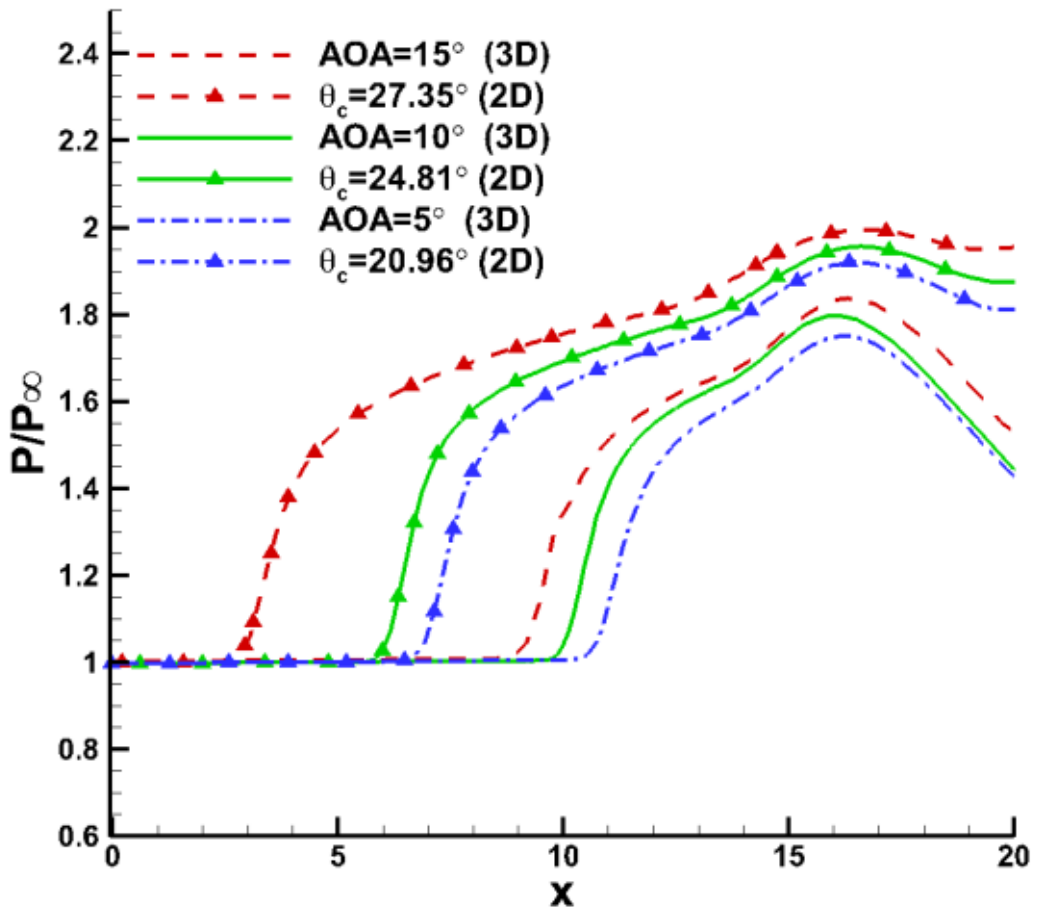

Fig. 37 Comparison of wall pressure distributions between 2D and 3D interactions

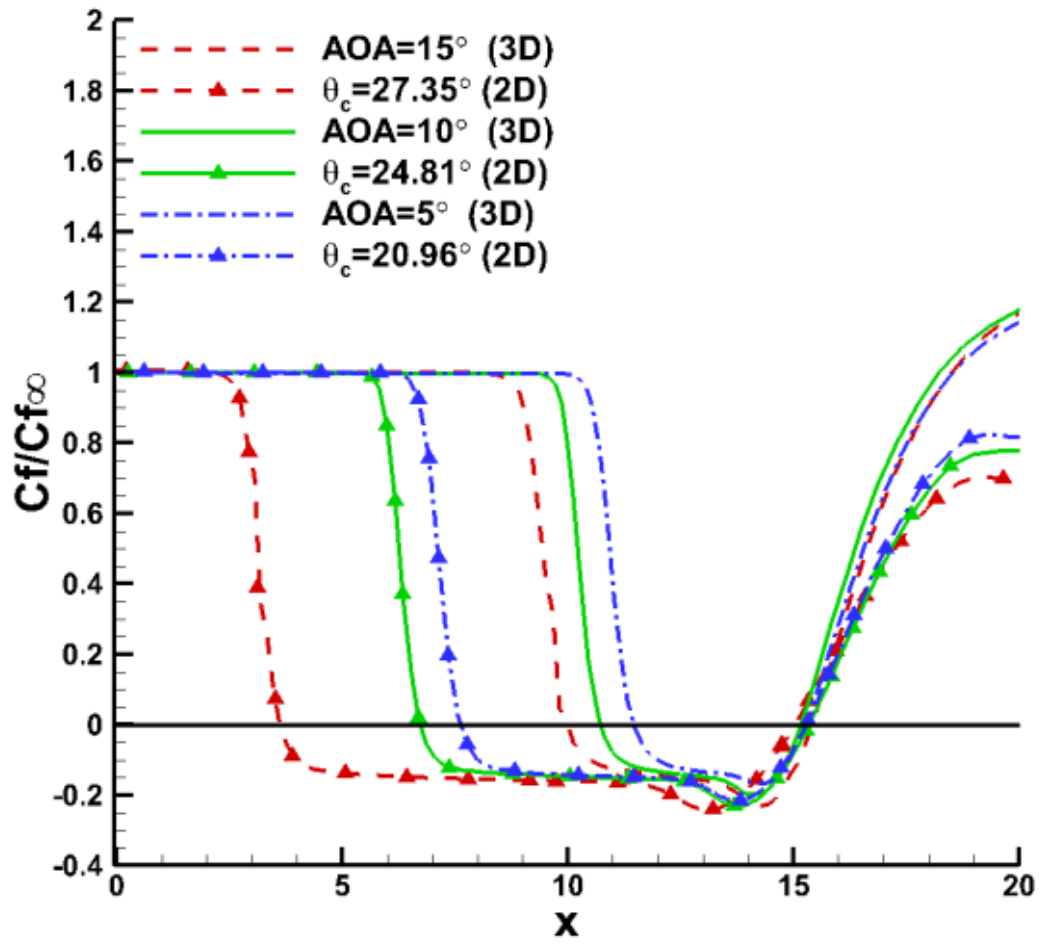

Fig. 38 Comparison of friction force distributions between 2D and 3D interactions

# 4 Conclusion

In this paper, with reference to the experiment by Kussoy et al. [10], we have studied the interaction between a 3D conical shock wave and an axisymmetric boundary layer for a 15° half-angle cone at AOA = 15° and $Ma = 2.2$ based on RANS simulations using the WENO3-PRM$^2_{1,1}$ scheme and Menter's SST model. We have also investigated further conditions of AOA = 10° and 5° that were not considered in the experiment, as well as axisymmetric 2D interactions for comparison in which the cones have half-cone angles of 27.35°, 24.81°, and 20.96°. Our analyses of the flow structures and variable distributions in 2D and 3D interactions under different conditions, and the resulting conclusions, can be summarized as follows:

(1) To enable comparisons with the experimental results of Kussoy et al. [10], we considered a 2D axisymmetric cone with half-cone angle of 27.35°, taking account of the fact that the shock angle in a 2D interaction is the same as the angle in the windward meridional plane in the corresponding 3D interaction. For this 2D cone, a grid convergence study was carried out and appropriate grids were obtained. The half-cone angles of other 2D axisymmetric cones were also determined corresponding to the cone used by Kussoy et al. but at AOA = 10° and 5°. The results of the calculations for the 2D axisymmetric shock wave/boundary layer interaction reveal that a Mach reflection-like event occurs in the flow field and a Mach stem-like structure arises, while no reattachment shock wave appears during flow reattachment. A new interaction model was established to describe this behavior.

(2) A comparison of the results of computations corresponding to the experiment by Kussoy et al. [10] with their results showed good agreement with regard to wall pressure distributions, although there were differences in the friction force distribution and the oil flow diagram. However, given that in the experimental results the separation inferred from the friction force at 90° azimuthal angle was inconsistent with that from the oil flow, the differences might have arisen from inadequacies in the experimental measurement techniques.

(3) Computations were performed for 3D interactions at three AOAs to investigate the wave structure in the windward meridional plane, the vortex structure, and the separation topology. In particular, the results of these computations confirm the presence of similar Mach reflection-like events and Mach stem-like structures, indicating the validity of the new interaction model. The computed results for the separation topology on the wall and the 3D structure of the vortex provide an improved understanding of the 3D interaction. Based on the numerical results, side overflow was also studied, and its quantitative characteristics were obtained. The existence was established of a critical AOA at which the length of the circumferentially unseparated region in the tube is zero.

(4) A comparison of the results for the 2D and 3D interactions shows that the size and height of the 3D separation bubble in the windward meridional plane are less than those of the 2D counterpart, and the position of separation is postponed. This behavior can be explained by the presence in the 3D interaction of side overflow in the windward meridional plane, while there is a positive pressure gradient circumferentially on the wall, as a result of which the adverse wall pressure gradient in the streamwise direction on the wall is weakened and flow separation is thereby reduced.